\documentclass{aastex631}

\usepackage{enumitem}
\usepackage{verbatim}
\usepackage{float}
\usepackage{makecell}

\newcommand\pmra{$\mu^*_\alpha$}
\newcommand\pmdec{$\mu_\delta$}
\newcommand\parallax{\overline{\omega}}
\newcommand\mfirst{$M_1$}
\newcommand\msecond{$M_2$}
\newcommand\mthird{$M_3$}
\newcommand\mfour{$M_4$}

\begin{document}

\title{Seeking Kinematic Association of Known FU Orionis Stars with Young Clusters in Cygnus}

\author[0009-0003-9906-2745]{Tamojeet Roychowdhury}
\affiliation{Department of Electrical Engineering, Indian Institute of Technology Bombay, Mumbai, 400076, India}

\author{Lynne A. Hillenbrand}
\affiliation{Department of Astronomy, California Institute of Technology, Pasadena, CA, 91125, USA}
\email{lah@astro.caltech.edu}

\begin{abstract}

Kinematic studies of young stars and star clusters have increased our understanding of the process of star formation and evolution in the Milky Way. FU Orionis objects are a specific class of young stellar object notable for their extremely high disk-to-star accretion rates. 
We use parallax and proper motion information from the \textit{Gaia} astrometric survey to study five known FU Ori stars towards the Cygnus clouds, in the distance range $\sim$500--900 parsecs, and seek evidence of their kinematic association with proximal stellar groups or clusters. We develop multiple search criteria within the \textit{Gaia} datasets to look for nearby stellar aggregates and to reliably isolate their likely member stars. We show that V1057 Cygni and HBC 722 are kinematically consistent with the 3D locations as well as the inferred proper motion fields of the North America Nebula cluster. We show a similar association of V1515 Cygni with NGC 6914 in the Cygnus-X region, and of V2494 Cygni with stars in the dark cloud LDN 1003 and Braid Nebula. Further, we find that V1735 Cygni is consistent in both position and proper motion with the streamer structure of IC 5146, and we trace the streamer's similar proper motions to the main cluster.  
Color-magnitude diagrams of all identified clusters show the presence of pre-main-sequence populations, strengthening the likelihood of a physical association between the young FU Ori stars and their respective nearby clusters.
\end{abstract}


\section{Introduction } \label{sec:intro}

The \textit{Gaia} DR3 astrometric survey \citep{Gaia_2023} has enabled the kinematic study of various features in the Milky Way over a wide range of scales. Of these, star clusters and associations are of particular interest in better understanding structure and evolution in the Galaxy \citep{review_cantat2024}. Multiple new star clusters have been identified using 6D phase space information \citep[e.g.][]{new_clusters_2019, new_clusters_2021}. Young clusters further help decipher the process of star formation in molecular clouds and nebulae as a whole, including their gradual dispersion into the galaxy. Several earlier works have also looked at expansion and rotation in young clusters \citep{Kuhn_2019} as well as their tidal tails and disruption \citep{Risbud_2025,Semyeong_2020}. 

Star forming clusters are also often traced by young O- and B-type stars located in what are known as OB associations. Recent works have used machine-learning based models on the \textit{Gaia} astrometric data of these stars to find structures in known nearby OB associations \citep[e.g.][]{Quintana_2023, Quintana_2025}. This has resulted in the discovery of multiple groups and sub-groups in some of them, and enabled trace-back to their kinematic origin in known spiral arms of the Galaxy.  \textit{Gaia} astrometric data  has also been used to improve the estimates for star formation rates and for core-collapse supernovae rates, both locally and on the scale of the Milky Way.  Other studies have used various methods to assess membership probabilities and to catalog young stars in known star-forming regions \citep{Kuhn_NAP_2020,Armstrong_2025}, identifying new kinematic properties (such as clustering in subgroups) using the full 6D information. Such analyses help us build a more complete picture of star forming clouds and their evolution.

A specific class of young stellar objects (YSOs) are FU Ori objects \citep{HERBIG1966}. These are characterised by extreme brightening ($\sim4-6$ mag) over timescales of a few months to a few years. This unpredictable brightening is attributed to a large (factor of tens to hundreds) increase in the disk accretion rate \citep{Hartmann-Kenyon1996}. Being some of the youngest YSOs, these stars would be expected to still be embedded in the gas cloud or young cluster where they have formed. However, many or most of the known FU Ori stars seem to be located (in projection on the sky) on the peripheries of known star-forming nebulae or clusters, rather than well inside them. Due to spatial proximity in RA--Dec, they are often assumed in the literature to be physically associated with these nearby regions. Stronger evidence for true association, such as phase space proximity, is usually absent in older literature, primarily due to a lack of (reliable) velocity information.

With newly available astrometric data from \textit{Gaia} DR3, we can now look for this physical association in a larger phase space. Parallax information combined with photometry further helps to constrain absolute luminosities and spectral energy distributions. This can provide useful insights into key properties of the system including mass, age and accretion rate -- helping us probe the physical processes in young stellar systems and circumstellar disks better, such as mass assembly and the evolution of disks to planets \citep{Fischer_2023,Molyarova2018}. Apart from parallax, \textit{Gaia} also contains proper motion data in RA (\pmra) and Dec (\pmdec) for a large portion of its sample, allowing us to spatially select stars close to the FU Ori target and/or the nearby star-forming region, and look for kinematic consistency in the proper motion space.

In our previous work \citep{TR_LAH_2024}, we studied two of the closest FU Ori type sources and demonstrated their kinematic association with two clusters forming in the Orion molecular cloud complex at $\sim$400 pc
(FU Ori with the $\lambda$ Ori cluster and V883 Ori with the Orion Nebula Cluster). 
In this work, we extend our analysis to the next closest region known to host outbursting FU Ori sources, the Cygnus clouds, which  together contain six such objects but span a range in distance from $\sim500-900$ pc.

In Section~\ref{sec:methods} we describe our methodology to select our star sample for finding physical association, and how we determine the kinematic consistency between the FU Ori target and any nearby young clusters or nebulae. In Section~\ref{sec:application} we apply it to our sample of well-known FU Ori stars in Cygnus. In Section~\ref{sec:results} we provide overall results concerning the kinematics of the clusters/stellar groups identified as physically close to the FU Ori targets, as well as to what extent they are consistent with each other in both position and velocity space.

\section{Data and Analysis} \label{sec:methods}

There are six known FU Ori stars located towards the Cygnus region of the Galaxy. Five of these that have associated \textit{Gaia} DR3 \citep{Gaia_2023} records are listed in Table~\ref{tab:object-list}. The sixth, V2495 Cygni, does not have a \textit{Gaia} DR3 entry and hence can not be studied astrometrically and used for reliable kinematic analysis. 
The main aim of this work is to identify stars that are kinematically related to the five sources in Table~\ref{tab:object-list}, 
with a focus on testing star-forming regions and young clusters in the vicinity of each object. 
Our analysis differs from that of \citealt{Kuhn_2019} (Appendix A), in that we use a later data release from \textit{Gaia} with improved parameters, and we base our search specifically in the vicinity of the FU Ori targets and find closest kinematic components from the vector point diagrams of the samples, instead of targeting individual clusters or subclusters and studying bulk kinematics.

To identify potential stellar neighbours, we first obtain astrometric information for the target FU Ori stars, namely their right ascension (RA), declination (Dec) and parallax ($\parallax$).  In addition to the Gaia-sourced distances for the individual objects, we also consider the distance estimates from previous literature.  
Proper motions in RA (\pmra, declination-corrected) and Dec (\pmdec), as well as their errors, were also obtained from \textit{Gaia} DR3. While relying on astrometry from single sources can be misleading, we show later that both proper motion and parallax uncertainties are low compared to the intrinsic dispersion within the clusters that we identify.

Regarding astrometric quality,
two of our sources (HBC 722 and V1735 Cygni) have \textit{Gaia} RUWE (renormalised unit weight error) $< 1.2$, indicating a good astrometric solution \citep{Lindegren_2021}. Two more sources (V1057 Cygni and V1515 Cygni) have RUWE $\sim 1.42$, just outside the widely accepted upper limit for good astrometry at 1.4. However, as shown by \cite{RUWE_2022}, astrometry for sources with RUWE as large as 8 may be used if reported uncertainties are scaled up by a correction factor of $ 1.5-2$ for observed magnitude $9<G<13$, which contains both these sources. Moreover, unlike unresolved binaries where high RUWE indicates astrometric unreliability \citep{Lindegren_2018, Belokurov2020}, YSOs with disks are known to have elevated RUWE owing to photometric variability. The astrometry starts becoming unreliable due to potential stellar multiplicity at RUWE $\gtrsim 2.5$ in such systems \citep{Shannon_2022}. Only one source, V2494 Cygni, has $G\sim 17$ and RUWE $\sim 1.8$, which is still well within the threshold of 2.5.

For each of the FU Ori stars, we then search for stars that are both located physically near, and co-moving with the object.

\begin{table}[ht]
\begin{center}
\begin{tabular}{cccccc} \hline
FU Ori Star & Parallax Distance & Photogeometric & Nearby Young Cluster & Lit. Distance & Reference \\
& [parsecs] & Distance [parsecs] & & [parsecs] & \\\hline
HBC 722 & $763 \pm 8$ & $757 \pm 8$ & North America Nebula &  $794 \pm 25$ & \cite{Kuhn_NAP_2020} \\
V1057 Cygni & $907 \pm 19$ & $891 \pm 20$ & North America Nebula &  $794 \pm 25$ & \cite{Kuhn_NAP_2020} \\
V1515 Cygni & $902 \pm 18$ & $900 \pm 14 $& Cygnus -- X (NGC 6914) &  $\approx 1050$ & \cite{Racine_1968} \\
V1735 Cygni & $690 \pm 36$ & $663 \pm 35$ & IC 5146 &  $783 \pm 25$  & \cite{Kuhn_2019} \\
V2494 Cygni & $533 \pm 75$ & $580 \pm 82$ & Cyg OB7 (Braid Nebula) &  $800$  & \cite{Magakian2013} \\

\hline
\end{tabular}
\end{center}
\caption{List of our FU Ori targets,
distances from Gaia DR3 \citep{Gaia_2023}, and
photogeometric distance estimates from \cite{Bailer_Jones_2021}that are also based on Gaia DR3 data products.
Also given are proximate young cluster or star forming regions and their available literature distances.
} 
\label{tab:object-list}
\end{table}

\subsection{Selecting Associated Stars}

Our selection criteria require measurements for $\parallax$, \pmra, and \pmdec, in addition to the spatial positions RA and Dec. 
We impose that the Gaia renormalized unit weight error (RUWE) is $<1.4$, denoting a good astrometric solution \citep{Lindegren_2021}.
We also retrieved radial velocities (RVs) from \textit{Gaia}, but the numbers are limited and the measurements have high fractional errors.
Additional queries of SDSS/APOGEE radial velocities \citep{APOGEE_dr17} also had sparse returns for the regions we probe, and are not considered in our analysis. 

We use several different methods with different criteria for selecting stars that are potentially associated with a given FU Ori star.  These are defined as \mfirst, \msecond, \mthird, and \mfour, in order of increasing number of restrictions placed on the retrieved stars. In assessing candidates for kinematic association,
we let $(\alpha, \delta, d)$ denote the coordinates and distance of an individual star, with $d=1/\parallax$ in units of pc, and $\sigma_{\parallax}$ the error in its parallax. 
We also let $(\alpha_0, \delta_0, d_0)$ denote the RA, Dec and distance of the center of a nominal cluster or nebula, as described in the subsections below on the individual sources.

\begin{itemize}
    \item \mfirst: 
    We initially select stars satisfying the simple fixed criteria:

    \begin{enumerate}
        \item $(\alpha - \alpha_0)^2\cos^2\delta_0 + (\delta-\delta_0)^2 \leq (25/d_0)^2$
        \item $|d-d_0| \leq 25$
        \item $\parallax / \sigma_{\parallax} \geq 5.$
    \end{enumerate}
    This set of criteria is identical to those we used earlier for FU Ori and V883 Ori \citep{TR_LAH_2024}, searching a cube of 25 pc on a side. The choice of 25 pc is motivated by the typical scale of $r_{50}$ (the radius containing 50\% of stars within the tidal radius) to be $\sim 2-5$ pc, and the tidal radius itself being $\sim 5-20$ pc for a large segment of the open cluster population of the Milky Way \citep{Hunt_Reffert_2023, Cantat-Gaudin2020}.

    \item \msecond: 
    We account for the much higher errors in the third parallax dimension, as compared to the two dimensions in the plane of the sky RA and Dec, especially at the larger distances ranging up to $\sim 1$ kpc. 
    A relaxation on \mfirst\ is to use an adaptively chosen distance cut $D$.
    The selection criteria are the same as in \mfirst\ but allow $D$ to vary from 25 pc, as follows:

    \begin{enumerate}
        \item $(\alpha - \alpha_0)^2\cos^2\delta_0 + (\delta-\delta_0)^2 \leq (25/d_0)^2$
        \item $|d-d_0| \leq D$
        \item $\parallax / \sigma_{\parallax} \geq 5.$
    \end{enumerate}

    \item \mthird: We allow both the angular search radius ($\theta$) and the distance cutoff ($D$) of \msecond\ to vary, based on the clustering visible in position/proper motion space. Additionally, the parallax error criteria is slightly relaxed as follows:

    \begin{enumerate}
        \item $(\alpha - \alpha_0)^2\cos^2\delta_0 + (\delta-\delta_0)^2 \leq \theta^2$
        \item $|d-d_0| \leq D$
        \item $\parallax / \sigma_{\parallax} \geq 3.$
    \end{enumerate}

    \item \mfour: Finally, we adopt a procedure in which we first retrieve stars from a relatively large volume using \mthird. Then we visually estimate the cluster's angular size $\theta$ in position space. We then select all stars lying within this angular region and having parallax consistent within $1\sigma$ error to the same region. The criteria are:
    
    \begin{enumerate}
        \item $(\alpha - \alpha_0)^2\cos^2\delta_0 + (\delta-\delta_0)^2 \leq \theta^2$
        \item $\parallax + \sigma_{\parallax} \geq \frac{1}{d_0(1+\theta)}$ and $\parallax - \sigma_{\parallax} \leq \frac{1}{d_0(1-\theta)}$
        \item $\parallax / \sigma_{\parallax} \geq 3.$
    \end{enumerate}
    
\end{itemize}

For each FU Ori star in our sample, the selection criteria used for identifying neighbors is finalized based on observed clustering in position and proper motion space, as well as in parallax. In general, if a stellar clustering appears following the application of one of the selection methods above, we then use cuts on position, parallax, and proper motion wherever possible to obtain a cleaner sample of candidate cluster stars. As would be seen in the following section on a case-by-case basis, despite our selection criteria using position space coordinates, clustering is typically more apparent and easier to separate in the \pmra-\pmdec\ space i.e. on a vector point diagram (VPD), than in position space.

We then consider three methods for isolating cluster stars in the VPD from the field star contamination, so as to define a cleaner cluster sample. In decreasing order of computational sophistication, these are: (1) Extreme deconvolution based Gaussian mixture modelling (XDGMM), that accounts for measurement errors and fits the distribution to two Gaussian components, (2) Simple Gaussian mixture modelling (GMM) with two components, and (3) Visually inspecting the concentrations on the VPD, then fitting a unimodal bivariate Gaussian and removing the more likely non-members by $3\sigma$ clipping. We observe that both XDGMM (implemented using the \texttt{astroML} Python library) and GMM (implemented using \texttt{scikit-learn}) fail to reliably identify the cluster stars in most cases. The bivariate Gaussian fitting followed by $3\sigma$-clipping, by contrast, works well, and is thus used for the final analysis in each case. While this method does not ensure a highly pure sample of the cluster stars, it works well and any residual field star contamination is small enough to not alter our final results significantly.

With a defined cluster sample, we then compare the FU Ori star kinematics with those of the cluster, and quantify the likelihood of cluster membership in terms of the number of standard deviations between the FU Ori kinematic measurements and the cluster distributions.  
We also examine the visual appearance of selected sources in position space, overlaid on images at 22, 12 and 3.4 $\mu m$ (mapped to R, G, B respectively) from the \textit{WISE} mission \citep{wise_overview}. 
This provides context for the position of the FU Ori star relative to the local stellar density distribution and any cloud/nebula structure.

\subsection{Assessing Physical Association}

Our starting diagnostic in identifying potential stellar associations with which our FU Ori sample stars may be associated, is clustering in position and proper motion spaces. If a potential concentration stands out, either spatially or in the vector point diagram (VPD, plot of \pmdec\ vs \pmra), 
we check for consistency of the FU Ori candidate with that concentration in two dimensions. To do so, we use the cluster star sample obtained above and calculate the distance of the FU Ori from the center (mean) of the sample in position and in proper motion, by fitting bivariate Gaussians to the star distributions in [RA, Dec] and in [\pmra\ , \pmdec], and measuring the separation in terms of number of standard deviations ($n-\sigma$) in each case. Results from these measurements are reported in Table~\ref{tab:n-sigma}.

We then also look for consistency of the line-of-sight components of parallax and RV.  The parallax data have higher fractional errors compared to the positions and proper motions. 
We considered both the direct inverse-parallax distances and the photogeometric distances of \cite{Bailer_Jones_2021}; for our FU Ori targets these are consistent within uncertainty, in all cases.
RV measurements are both much sparser and high in error, so are not able to be considered in our analysis. Thus our diagnostic power is more limited perpendicular to the plane of the sky compared to the in-plane position and proper motion measurements. 

We check for consistency in individual phase space components using two approaches:
\begin{enumerate}
    \item Using the full set of stars retrieved in the \textit{Gaia} search (i.e. not only candidate members separated by $3\sigma$-clipping from the kinematic concentration),  we create individual histograms for both proper motion components and for parallaxes.  We then fit a bimodal Gaussian that accounts for both the cluster stars and the general field population. While this is likely to have a high degree of field contamination, the bimodal Gaussian adequately takes care of accounting for both cluster and field.
    \item Using only the cleaner sample of stars retrieved by $3\sigma$-clipping on the kinematic concentration (and an additional cut in position space in the case of North America Nebula, Sec~\ref{sec:application}), we fit a unimodal Gaussian to fit the cluster sample alone.
\end{enumerate} 
In both cases a simple histogram does not take measurement errors ($\Delta$\pmra, $\Delta$\pmdec\ and $\sigma_{\parallax}$) into account. So we create histograms where each measurement, instead of adding a weight of unity to a single bin, adds a total weight of unity distributed over multiple bins in a Gaussian manner with mean equal to the measured value (e.g \pmra) and variance equal to the reported error (e.g. $\Delta$\pmra$^2$). This approach is similar to creating kernel density estimators, but instead of a uniform broadening width as used in KDEs, we set the width equal to the measurement error. These distributions are henceforth referred to as Gaussian-smoothed histograms and is the same presentation as in our previous analysis of the Orion sources \citep{TR_LAH_2024}. 

In principle, this procedure could also be applied in the 2D spaces for position (RA, Dec) and proper motion (\pmra, \pmdec).  However, the number of free parameters for a bimodal bivariate Gaussian is then twelve, and for nearly all of our fields, gave visually poor fits. This is also pointed to by the fact that XDGMM and GMM are unable to reliably isolate the clusters in the VPD. We thus base our analysis on the histograms of individual components, which produced credible fits in all cases.

In the next section, we apply these techniques to each of the FU Ori stars in Table~\ref{tab:object-list} with results appearing in Table~\ref{tab:n-sigma}. 

\begin{table}[h]
\begin{center}
\begin{tabular}{cccccccc} \hline
FU Ori Star & $\alpha$, $\delta$ & Sky Position & \pmra\  & \pmdec\  & Proper Motion& $\parallax$ & Parallax \\ 
& [deg] &  $n-\sigma$  & [mas yr$^{-1}$] & [mas yr$^{-1}$] & $n-\sigma$ & [mas] & $n-\sigma$ \\ \hline
HBC 722 & 314.571, 43.8954 & 2.34 & -0.45 $\pm$ 0.01 & -3.06 $\pm$ 0.01 & 1.09 & 1.31 $\pm$ 0.01 & 0.54\\
V1057 Cygni & 314.7239, 44.2579 & 2.24 & -2.70 $\pm$ 0.03 & -4.09 $\pm$ 0.03 & 1.86 & 1.10 $\pm$ 0.02 & 0.95\\
V1515 Cygni & 305.9501, 42.2072 & 1.01 & -2.61 $\pm$ 0.02 & -5.63 $\pm$ 0.02 & 1.23 & 1.11 $\pm$ 0.02 & 0.01 \\
V1735 Cygni & 326.8361, 47.5344 & non-Gaussian\footnote{Associated with the IC 5146 streamer, which is an asymmetric structure without a Gaussian density profile.} & -4.20 $\pm$ 0.09 & -2.81 $\pm$ 0.08 & 1.07 & 1.45 $\pm$ 0.08 & 0.7 \\
V2494 Cygni & 314.5879, 52.4910 & non-Gaussian\footnote{Associated with Braid nebula, which is asymmetric with an extended structure that does not fit a Gaussian density profile.} & 0.10 $\pm$ 0.29 & -2.47 $\pm$ 0.29 & 2.11 & 1.88 $\pm$ 0.26 & 1.35 \\

\hline
\end{tabular}
\end{center}
\caption{Results on consistency between each FU Ori star and the position, proper motion, and parallax parameters of the clusters listed in Table ~\ref{tab:object-list}, presented in the table as $n-\sigma$ values.} 
\label{tab:n-sigma}
\end{table}

\section{Application} \label{sec:application}

\subsection{HBC 722, V1057 Cygni and the North America Nebula}

The two well-known FU Ori sources, HBC 722 and V1057 Cygni, 
are located in the same spatial region as the North America/Pelican Nebula complex (henceforth referred to as NAP) with an established mean distance of $795$ pc \citep{Kuhn_2019}.  Their individually 
measured parallaxes correspond to distances of $763 \pm 8$ and $907 \pm 19$ parsecs respectively \citep{Gaia_2023}.

\begin{figure*}[t!]
 \includegraphics[width=6.8cm]{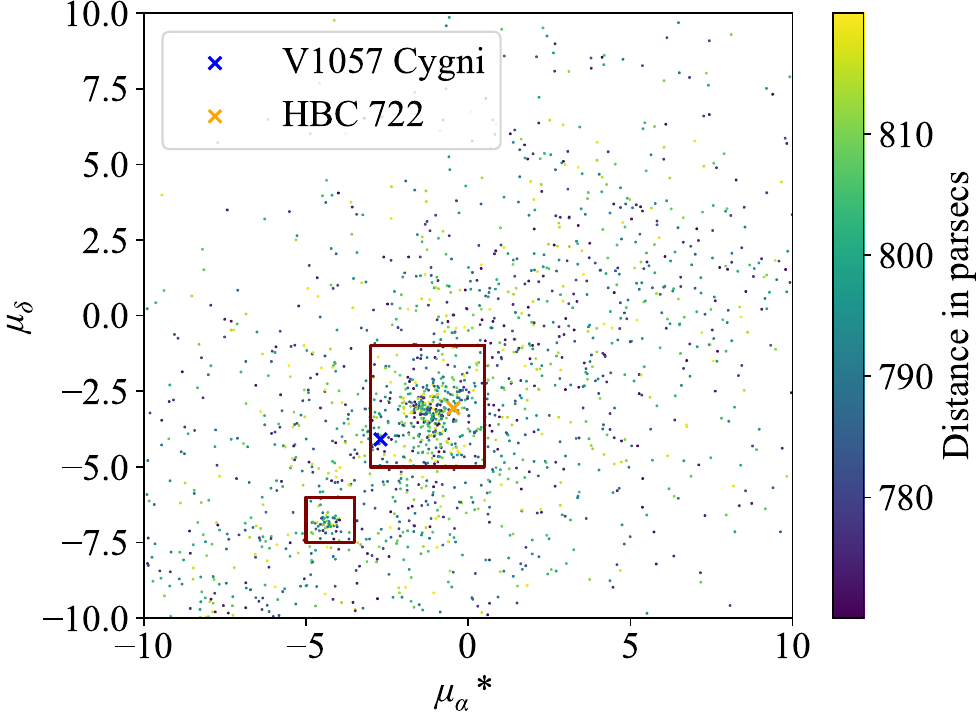}
 \includegraphics[width=6.3cm]{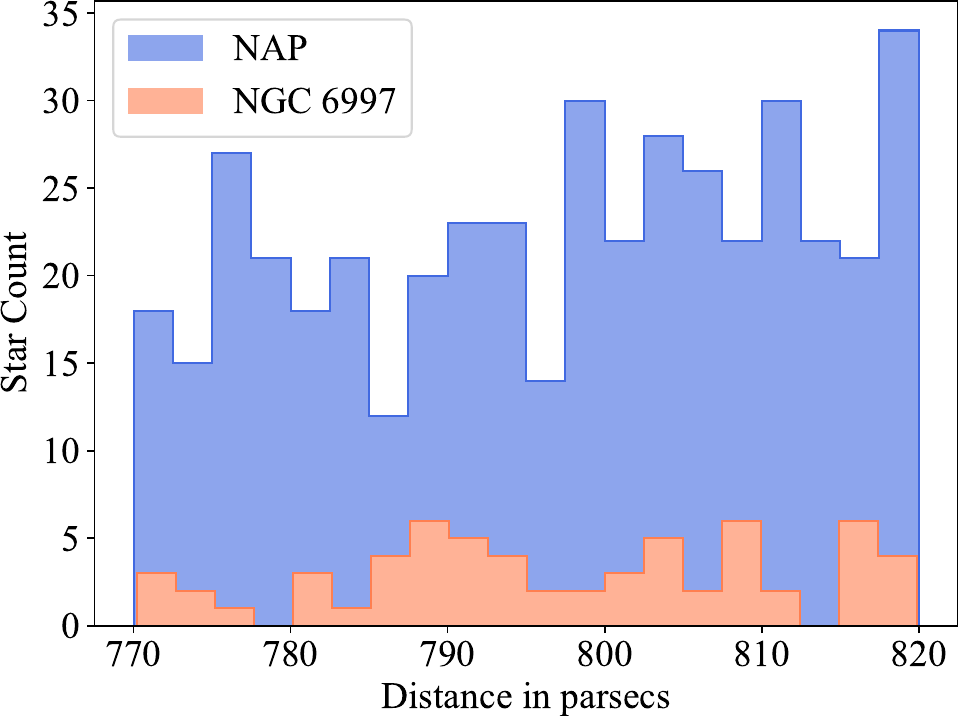}
 \includegraphics[width=5.2cm]{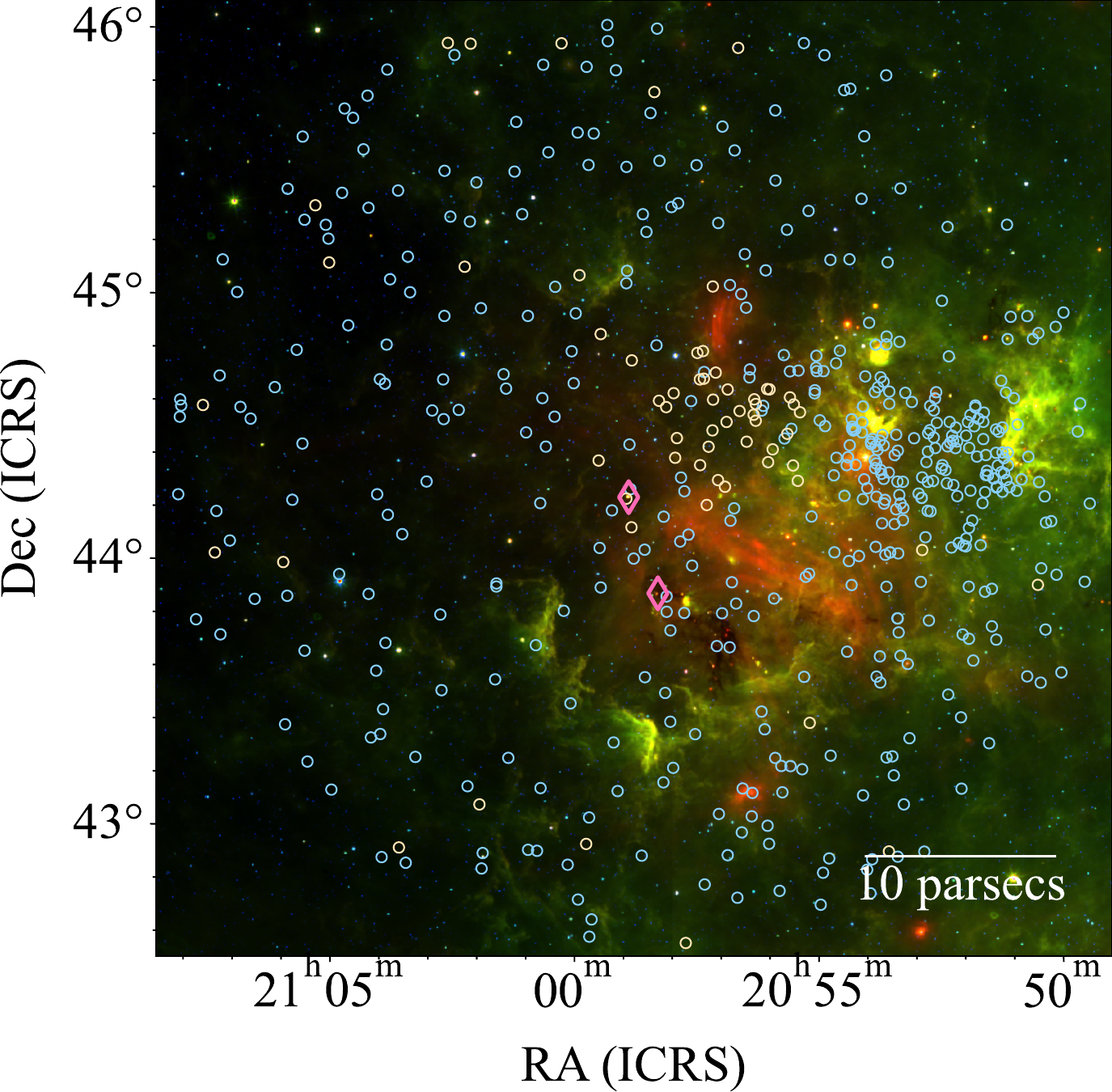}
    \caption{\textit{Left:} Vector point diagram for the sample of stars selected using method \mfirst\ (25 pc around centre of NAP) with colorbars indicating the corresponding distance in pc.  Two kinematic components are marked with boxes (the larger corresponds to NAP, and the smaller corresponds to NGC 6997). 
    The proper motion values of HBC 722 and V1057 Cygni are also marked, and both sources are consistent with the more populated kinematic component. \textit{Middle:} Distance histograms for stars within the two kinematic boxes; there are no clear peaks in the distribution.
    \textit{Right:} Image of the NAP region in infrared \textit{WISE} bands 22, 12 and 3.4 $\mu$m mapped to R,G,B. Stars belonging to the two kinematic components are marked in blue (larger) and orange (smaller), with V1057 Cygni and HBC 722 indicated by pink diamonds. 
    Some marginal spatial clustering can be seen for both components, which appear to have different spatial peaks.
    }
    \label{fig:25pc_snr5_vpd}
\end{figure*}

\begin{figure*}[hb!]
	\includegraphics[width=7cm]{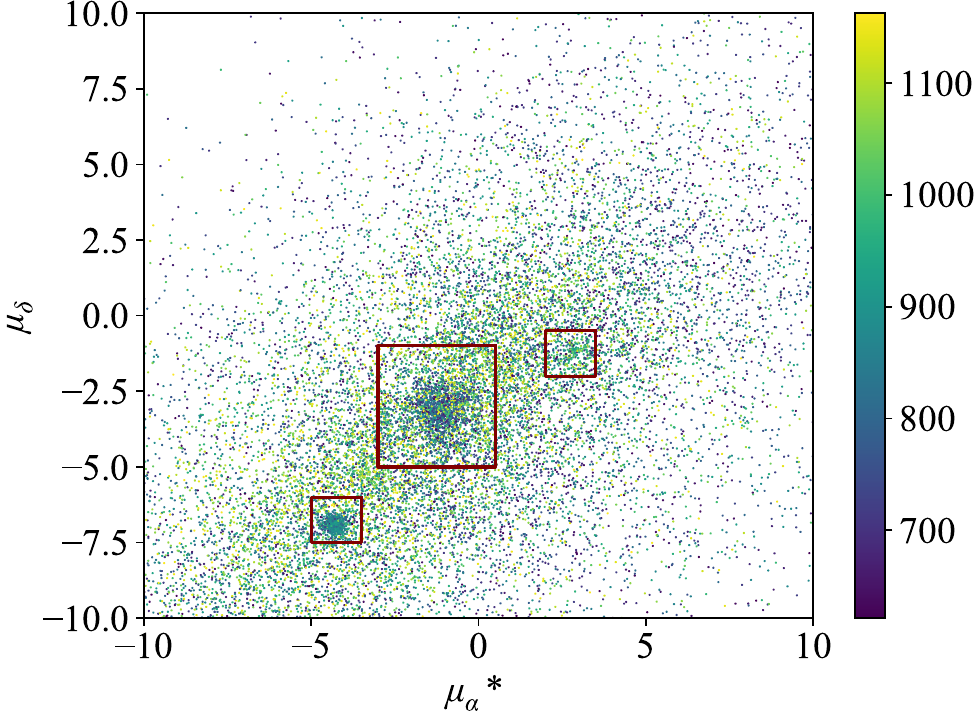}\
 \includegraphics[width=7cm]{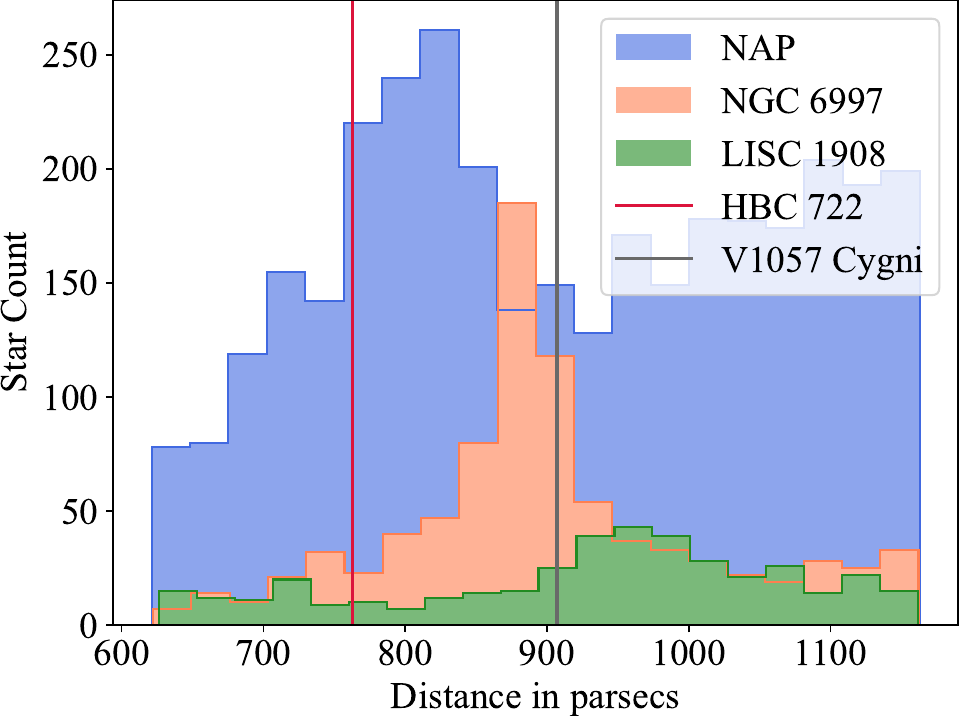} \\
 \includegraphics[width=5cm]{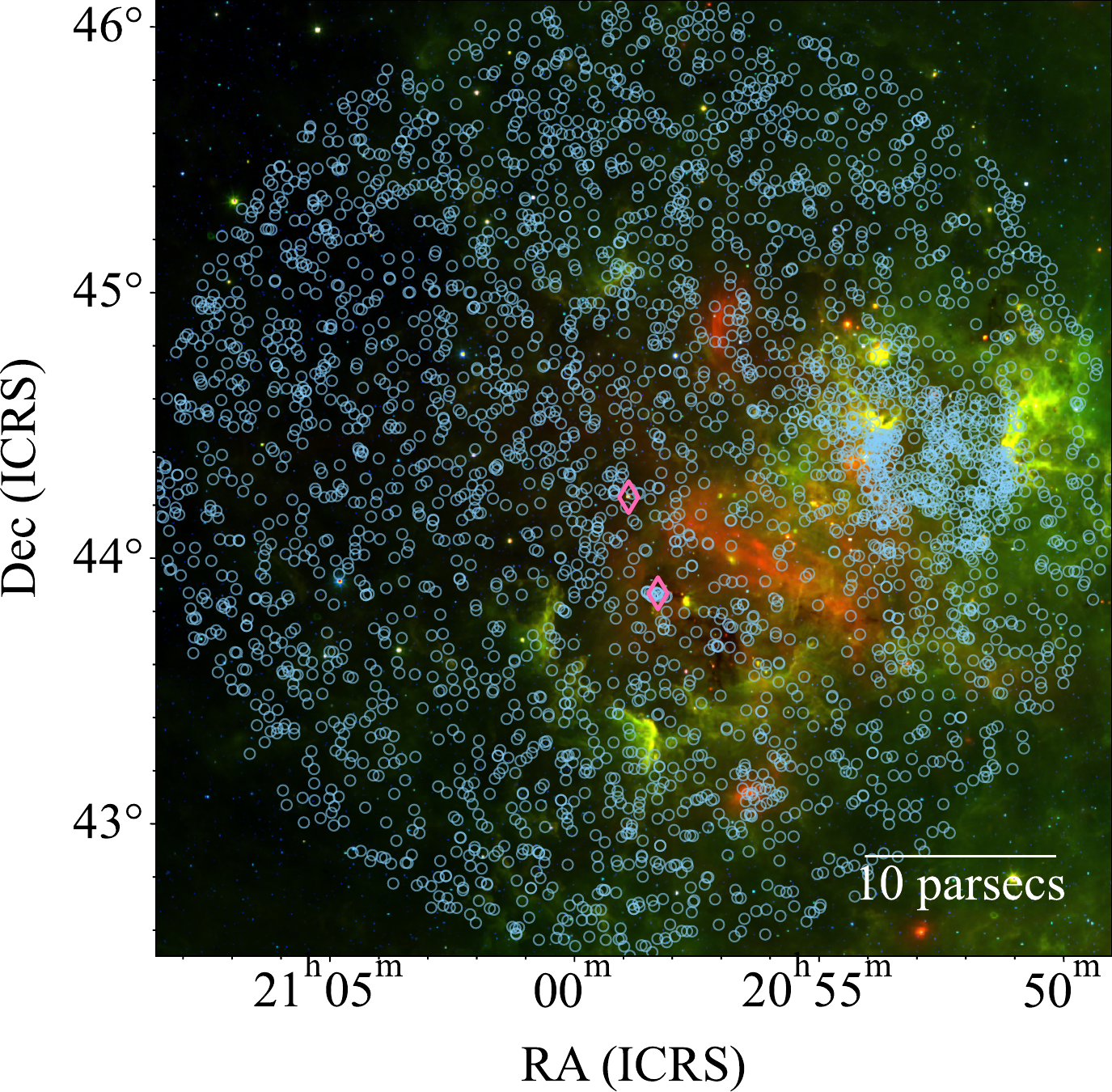}\
 \includegraphics[width=5cm]{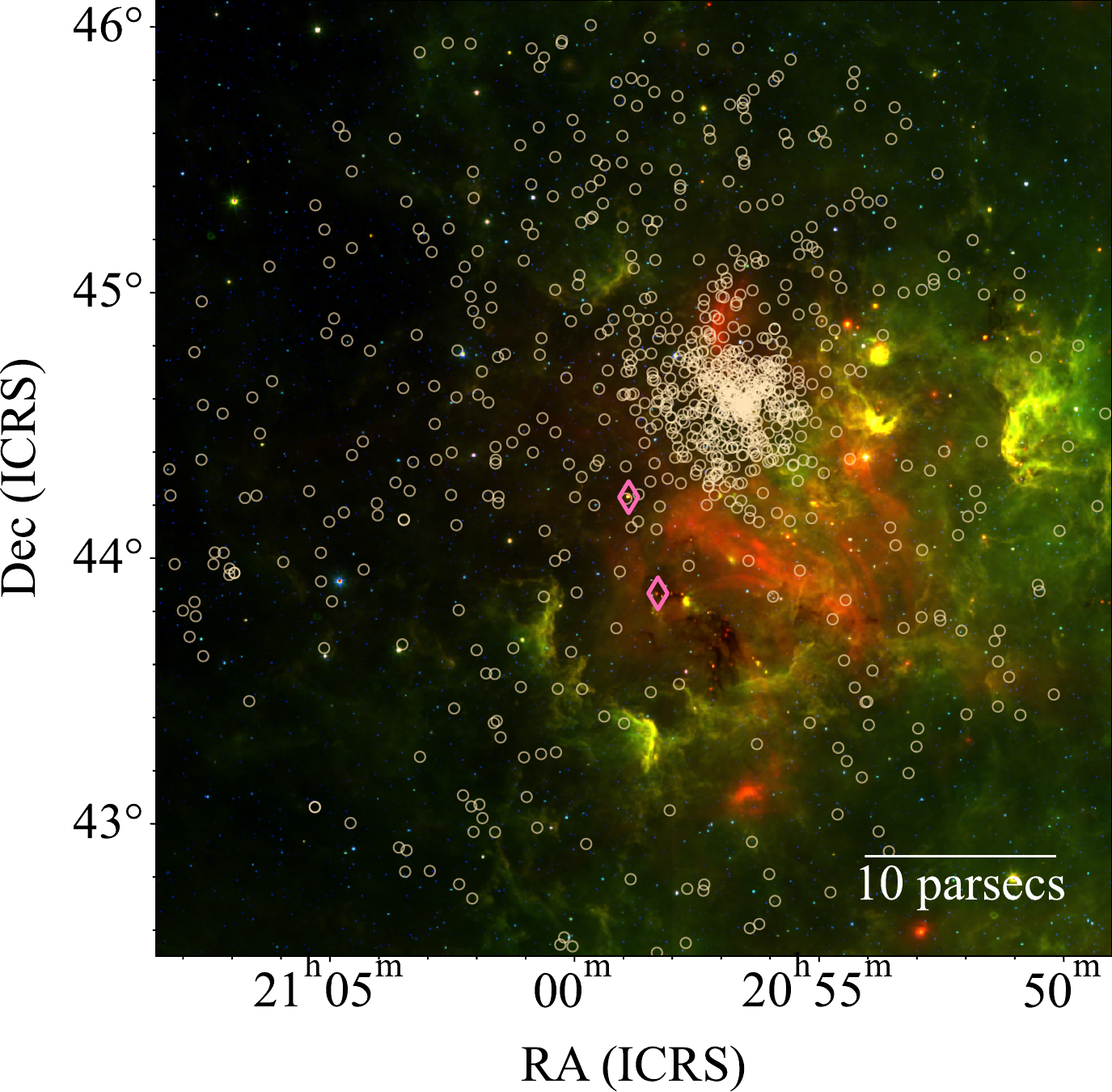}\ 
 \includegraphics[width=5cm]{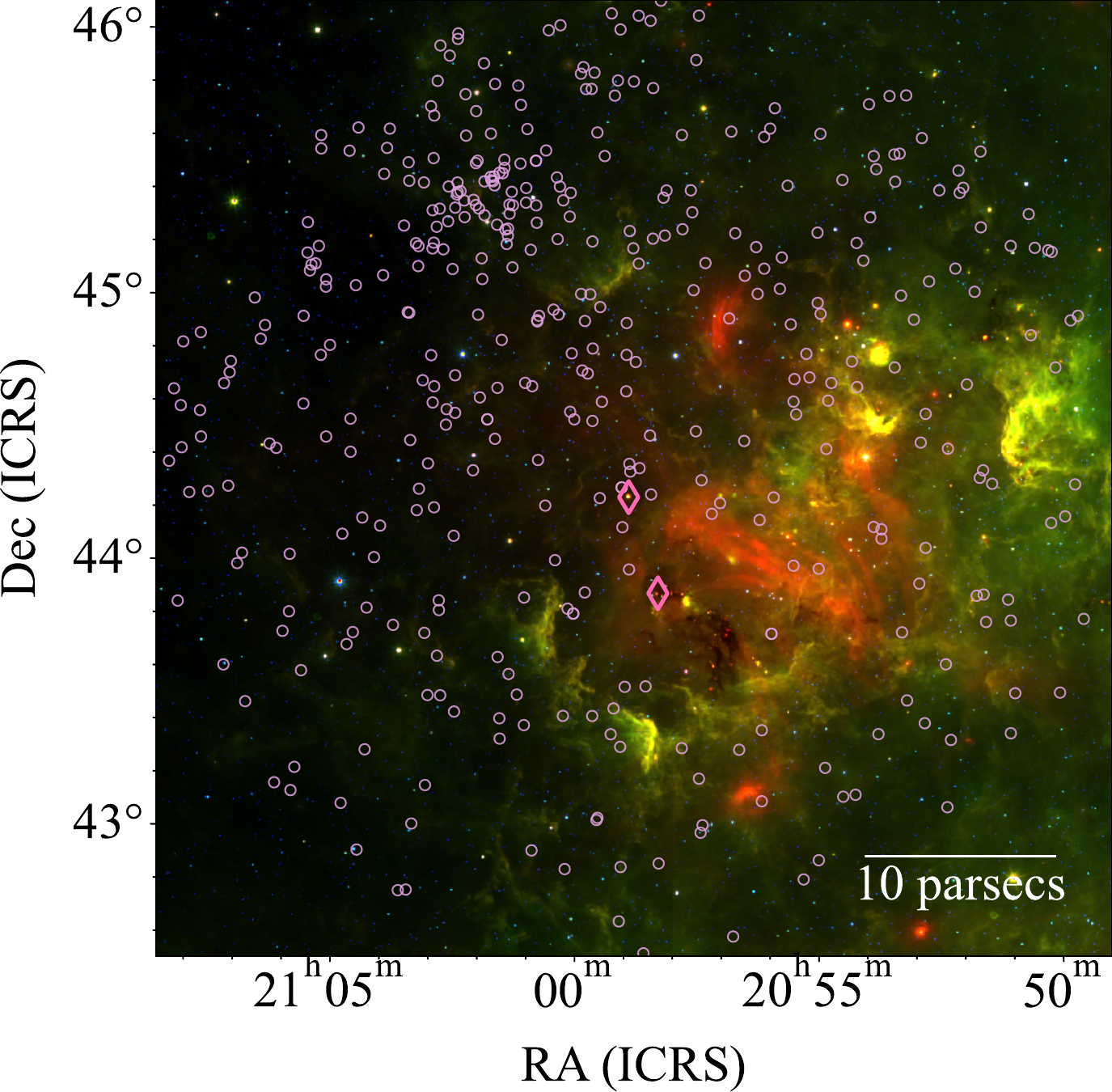}
    \caption{\textit{Top panels}: VPD and histogram of distances for the sample obtained by \msecond\ with a parallax range of 0.86--1.61 (taken from \citealt{Kuhn_2019}). Three kinematic components are clearly seen as overdensities on the VPD (marked in boxes, from left to right: NGC 6997, NAP, LISC 1908); each of these has a distinct peak in their distance histogram. \textit{Bottom panels}: \textit{WISE} images of the three kinematic groups of NAP, NGC 6997 and LISC 1908 in blue, light orange and purple respectively (HBC 722 and V1057 Cygni are marked with pink diamonds). Spatial clustering is evident for all three groups, which can be identified with NAP, NGC 6997 and LISC 1908. 
    }
    \label{fig:nap-3components}
\end{figure*}

We first applied method \mfirst\ with a 25 pc physical range projected at 795 pc. The VPD 
(Fig~\ref{fig:25pc_snr5_vpd}) showed two clear concentrations, as do the individual components in the Gaussian-smoothed histograms. 
The larger, more extended component can be identified with the North America/Pelican (NAP) nebula, while the smaller one is NGC 6997. Clustering in sky position (Fig~\ref{fig:25pc_snr5_vpd}) was also visually apparent. However, a histogram in parallax (Fig~\ref{fig:25pc_snr5_vpd}) did not show any clear peaks for either of the kinematic components of the VPD. 
The 25 pc range around the NAP adopted distance does not encapsulate the individually measured distances to the target FU Ori stars themselves, but we increase the distance range below in application of method \msecond, to try to find and isolate any related kinematic components.

 We use \msecond\ first with a simply expanded search box to 50 pc range on either side for parallaxes. This increased the number of retrieved stars but still did not produce any peaks in the parallax histograms. Next, the parallax search was even further relaxed to the 0.86 to 1.61 range of \cite{Kuhn_2019}. With this larger volume, the two previously seen kinematic components still stand out both in the VPD (Fig~\ref{fig:nap-3components}) and in the Gaussian-smoothed histograms. Now, however, clear peaks for both kinematic components are also clearly discernible in parallax histograms (Fig~\ref{fig:nap-3components}), with the NAP peak much broader than the NGC 6997 peak. Additionally, a third smaller kinematic component appears in the VPD, which is also clustered in position space and shows a peak in the parallax histogram as well (Fig~\ref{fig:nap-3components}). 
 This stellar group can be identified with open cluster LISC 1908 \citep{Cantat-Gaudin2020}.

 \begin{figure*}[b!]
	\includegraphics[width=6.4cm]{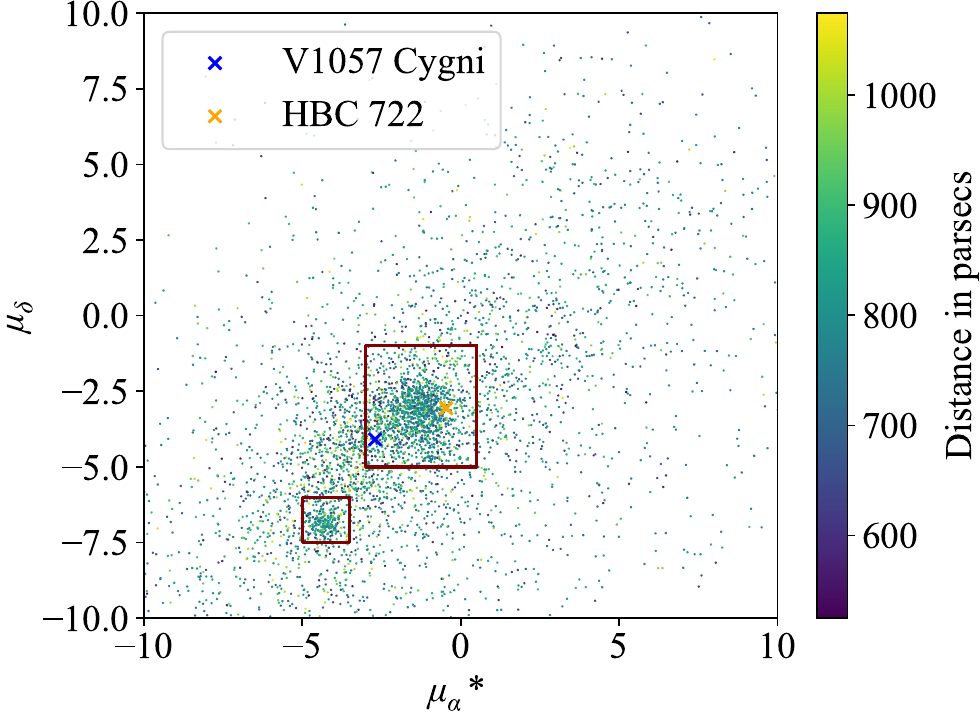}
 \includegraphics[width=6.4cm]{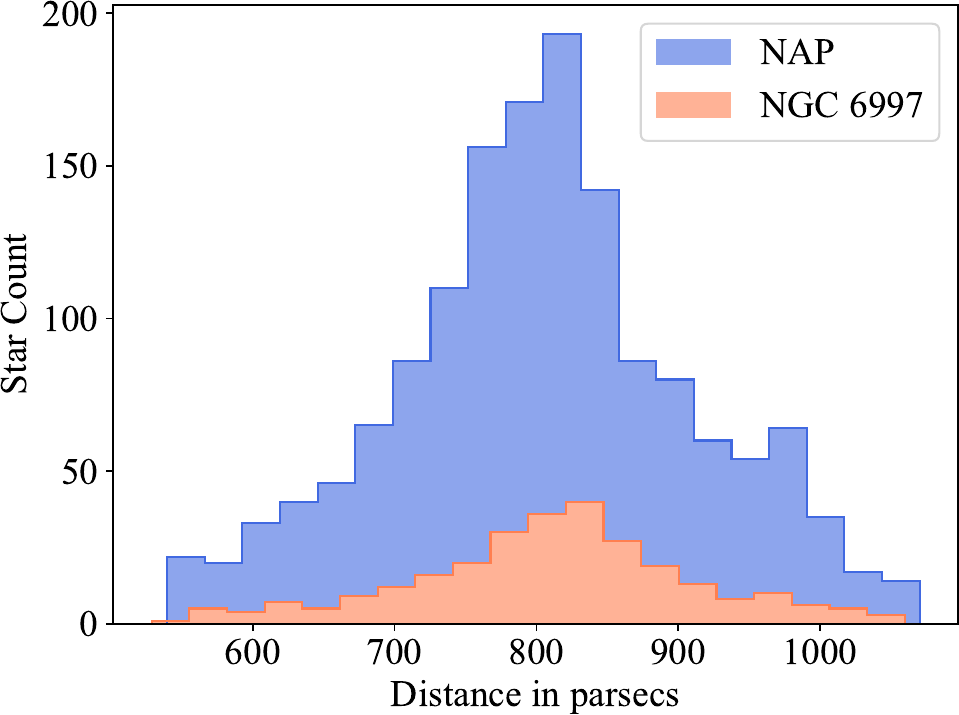}
 \includegraphics[width=5.6cm]{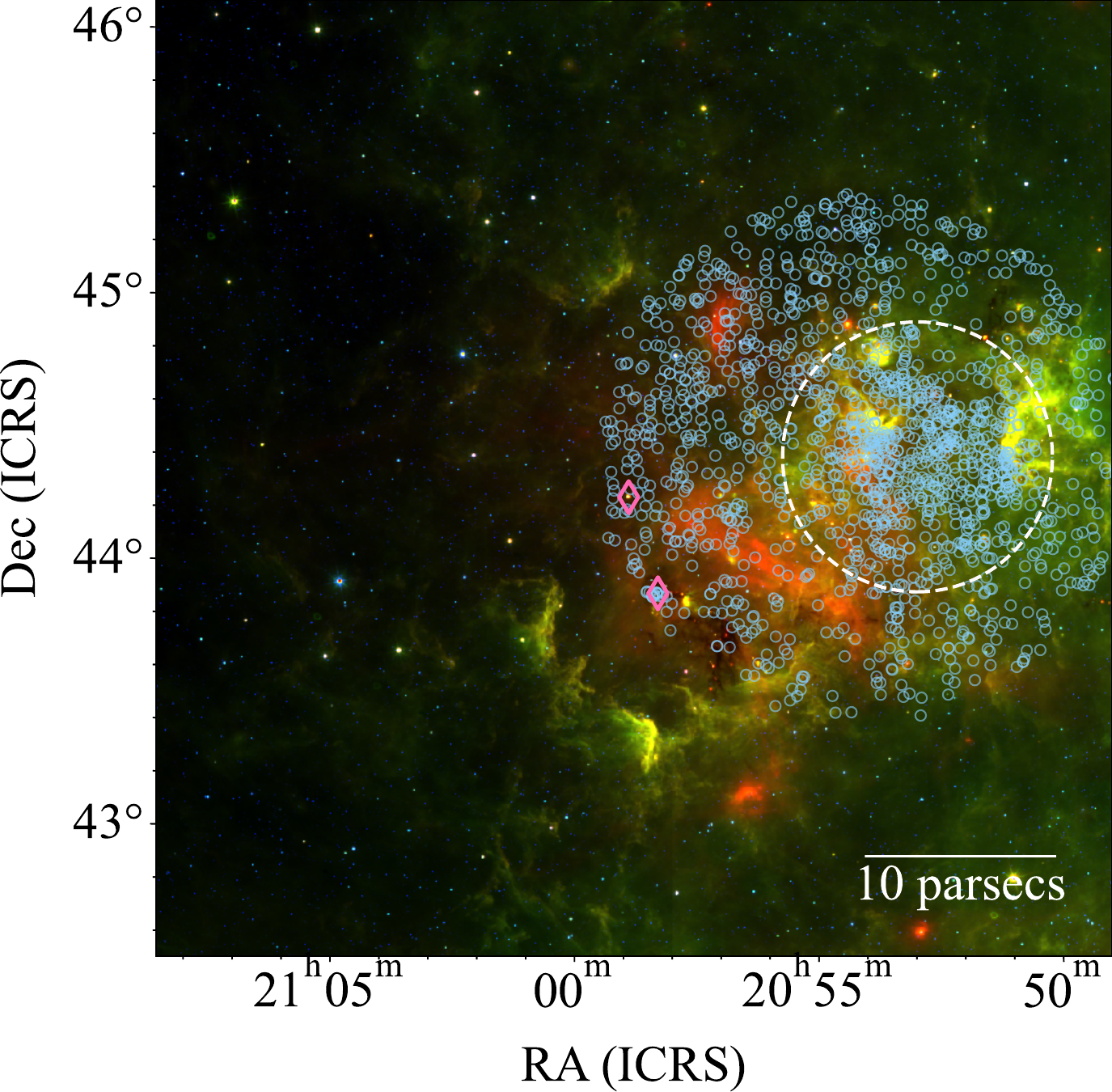} \\ 
 \includegraphics[width=12.5cm]{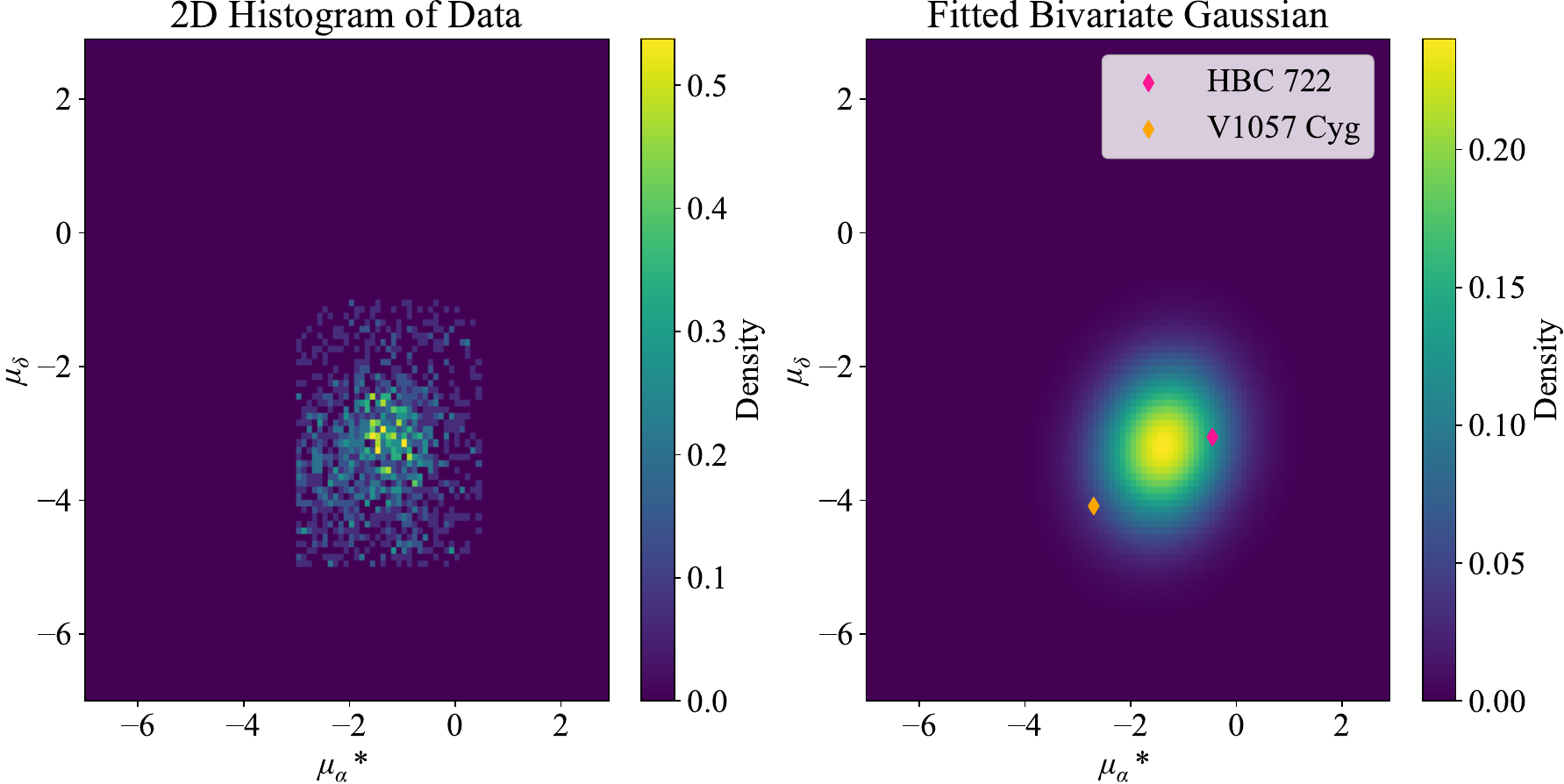}
 \includegraphics[width=5.6cm]{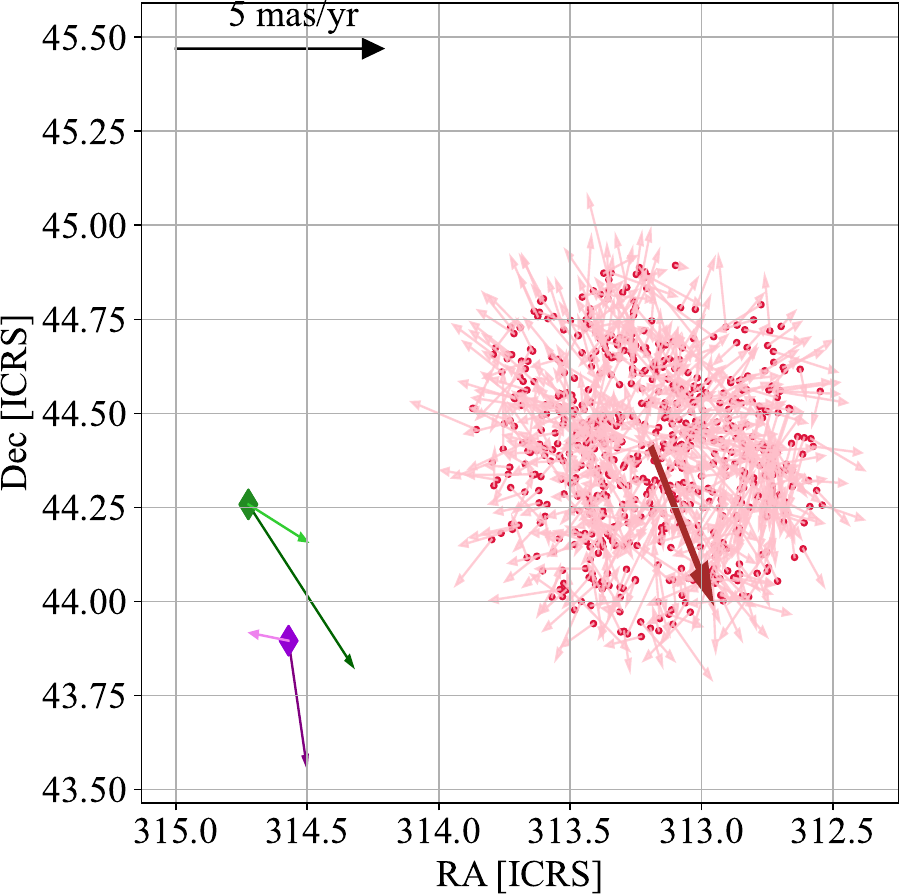}
    \caption{\textit{Top panels:} VPD of stars obtained using \mfour\ ($1^\circ$ search around NAP and NGC 6997, including $1\sigma$ distance consistency) with the kinematic overdensities corresponding to NAP and NGC 6997 marked in the larger and smaller boxes respectively. $1^\circ$ translates to $13.88$ pc at NAP distance. Distance histograms for both components show clear peaks. The right panel shows \textit{WISE} 22, 12 and 3.4 $\mu$m images, mapped to R,G,B, with blue circles marking the members of the kinematic concentrations of NAP, with the white dashed boundary showing our visual cut on position to separate the cluster. \textit{Bottom panels:} 2D histogram and fitted bivariate Gaussian to the NAP kinematic component in the left and middle, and the proper motions shown for the same on the right. The thick brown arrow shows the mean proper motion of the cleaned sample. Mean-subtracted proper motions of individual stars are shown in pink for cleaned sample. The diamonds show V1057 Cygni (green) and HBC 722 (purple). The lighter arrows show their mean-subtracted proper motion, and the darker ones show the overall proper motion which can be seen to be broadly consistent with the mean proper motion of the cluster itself.
    }
    \label{fig:14pc-nap}
\end{figure*}

Finally, we applied method \mthird\ and then \mfour\  of star selection by estimating the cluster radius on the sky to use as the search radius in both position and distance or parallax, including parallax errors. Due to the projected sky proximity of the NGC 6997 cluster and the NAP, stars from the former would likely contaminate the sample of stars from NAP, irrespective of the search radius chosen. We hence decide to include both in the search radius, taking a $1^\circ$ circle centred at $(\alpha, \delta) = (313.5, 44.4)$. This corresponds to a physical search box of $13.88$ pc. 

The resulting VPD has a better isolated peak for the NAP stars. We use a starting search box of $-3<$\pmra$<-0.5$ and $-5<$\pmdec$<-1$, and then perform a bivariate Gaussian fitting + $3\sigma$-clipping as described in Sec~\ref{sec:methods}. The resultant distribution in position space is now devoid of stars from the NGC 6997 component but still has significant field contamination. Fitting a bivariate Gaussian in position space returns a very wide Gaussian profile with the $3\sigma$ boundary enclosing the entire sample. So we visually isolate the cluster overdensity to a central circle $(\alpha-313.2)^2\cos^2(\delta_0)+(\delta-44.4)^2\leq 0.5^2$. The $0.5^\circ$ radius (6.94 pc at NAP distance), apart from being visually motivated, also corresponds to 3--5 times the characteristic size (median distance of star from centre) of the stellar groups in the region \citep{Kuhn_NAP_2020}. The final sample consists of all the stars passing both these cuts for which the VPD and spatial distribution (with proper motions marked) is shown in Fig~\ref{fig:14pc-nap}. We note that V1057 Cygni and HBC 722 both appear to be spatially displaced to outside the boundary of the cluster that we are probing, but this is solely due to the $0.5^\circ$ cut we applied for a cleaner sample. Fig 4 in \cite{Kuhn_NAP_2020} and Fig 19 of \cite{Fang_2020} demonstrate the larger spatial extent of the NAP and its associated stars, that well covers the position of both our FU Ori targets.

\subsection{V1515 Cygni and the Cygnus-X region}
V1515 Cygni is spatially located to the southwest of the NAP region, in a part of Cygnus known as Cygnus-X that is
undergoing more rigorous star formation and traditionally thought to be further away than the NAP.  
V1515 Cygni has a measured parallax corresponding to a distance of $902 \pm 18$ pc \citep{Gaia_2023}. 
Cygnus-X spans a large volume nominally centered on the young cluster Cyg OB2 at $1.6\pm 0.1$ kpc \citep{Lim_2019} but with many associated substructures in all directions, spanning several hundred pc.  \cite{rygl2012}, for example. found a mean distance of 1.4 kpc towards various clouds to the northeast and south based on maser parallaxes.

Instead of making assumptions about the Cygnus-X distance distribution, we first used the measured distance of V1515 Cygni for our search. 
We employed \mfirst\ to retrieve stars in a 25 pc box. The corresponding VPD had no clear peaks or overdensities. The number of stars retrieved was too sparse owing to the strict condition with \texttt{parallax\_over\_error} $>$ 5. 

\begin{figure}[ht!]
 \includegraphics[width=6.7cm]{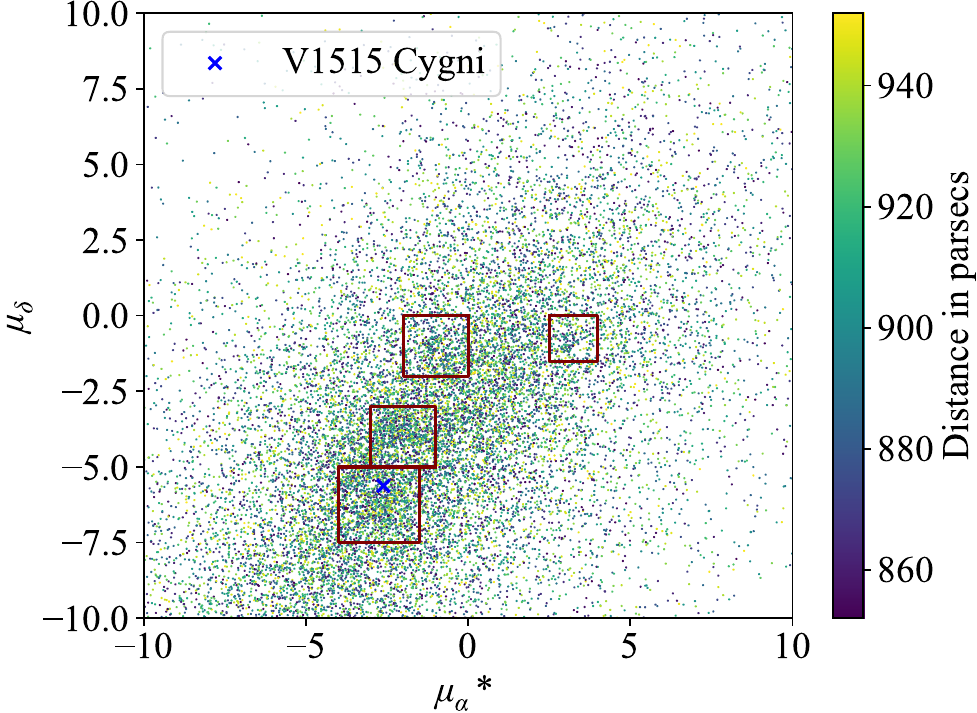}
 \includegraphics[width=6.4cm]{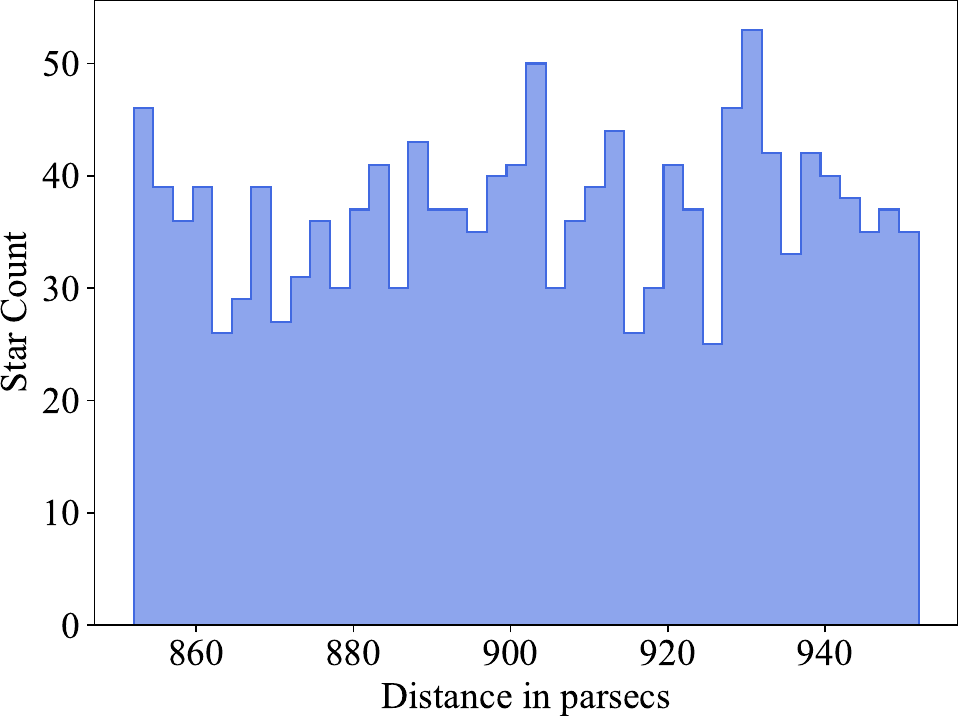}
 \includegraphics[width=5.3cm]{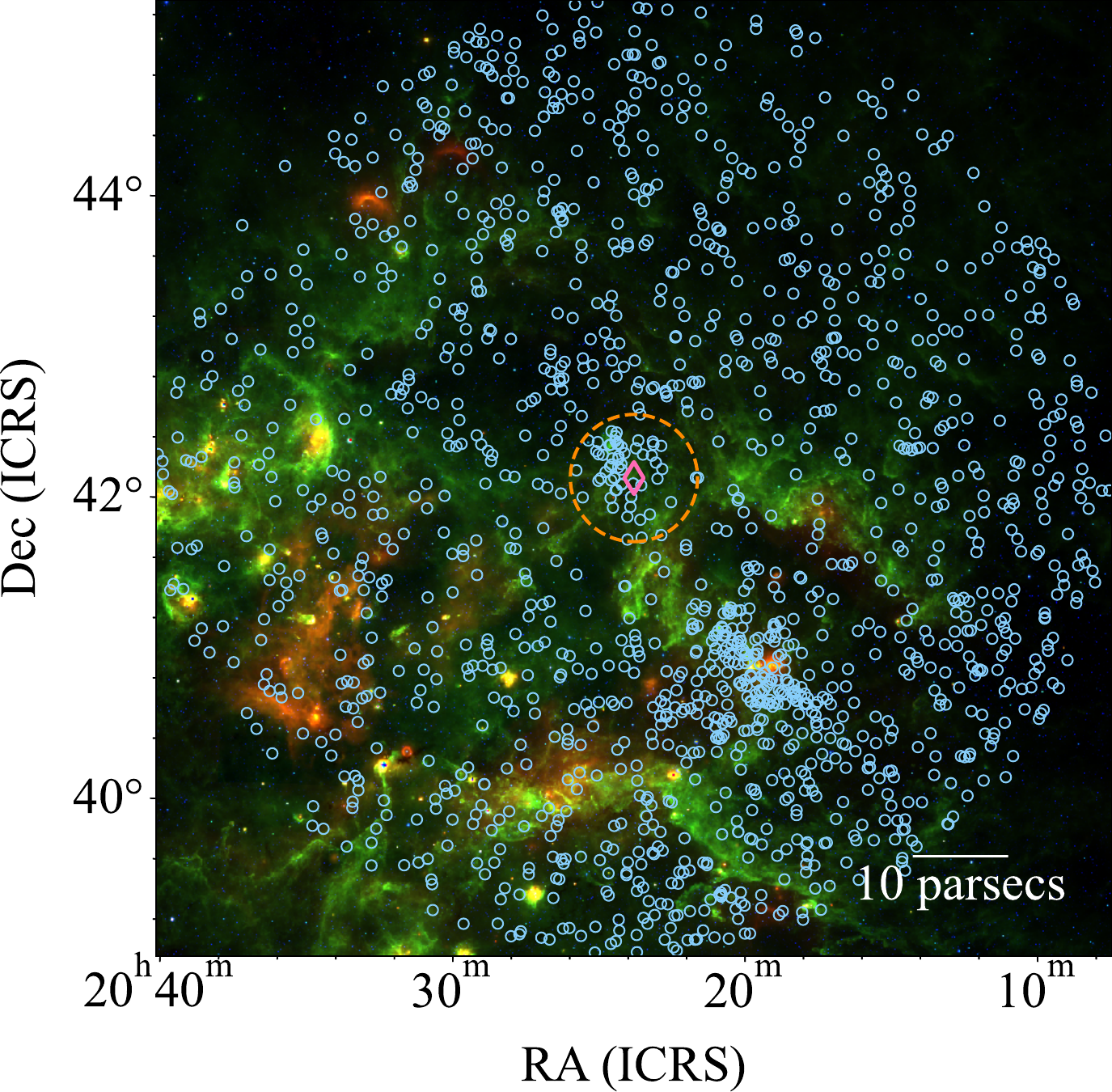}
    \caption{\textit{Left:} VPD of stars obtained using \mthird\ from the 50 parsec region around V1515 Cygni, with possible kinematic overdensities marked. One of these overdensities is consistent with the proper motions of V1515 Cygni \textit{Middle:} Distance histogram for stars belonging to the kinematic component containing V1515 Cygni. No clear peaks emerge. \textit{Right:} Stars in the kinematic concentration containing V1515 Cygni are marked in the \textit{WISE} image (22, 12 and 3.4 $\mu$m mapped to R,G,B), and can clearly be seen as clumping in two distinct groups. The three other overdensities marked in the left panel do not show any positional clustering and are thus likely not real clusters. V1515 Cygni is marked with a white diamond at the centre of the image. The NGC 6914 region is marked in an orange dashed circle.
    }
    \label{fig:50pc-cygx}
\end{figure}

Next, we used \mthird\ with both the angular size of the search range and the parallax range increased to $\pm 50$ pc, projected at 902 pc, and the error condition relaxed to \texttt{parallax\_over\_error > 3}. This gave some fuzzy overdensities in the VPD (Fig~\ref{fig:50pc-cygx}), but no clear peaks in the Gaussian-smoothed histograms. To check whether the marked overdensities in the VPD are credible physical structures, we look for clustering in position space. Three of the four marked overdensities in Fig~\ref{fig:50pc-cygx} show no spatial cluster and a roughly homogeneous distribution across the 50 pc (projected) search circle. The fourth component shows spatial clustering and the proper motions of V1515 Cygni are also consistent with those of this component. However, this group does not show any peak in the distance histogram (Fig~\ref{fig:50pc-cygx}), which can be attributed to the large number of field stars around the main clusters and the higher parallax errors.

\begin{figure}[h!]
	\includegraphics[width=6.7cm]{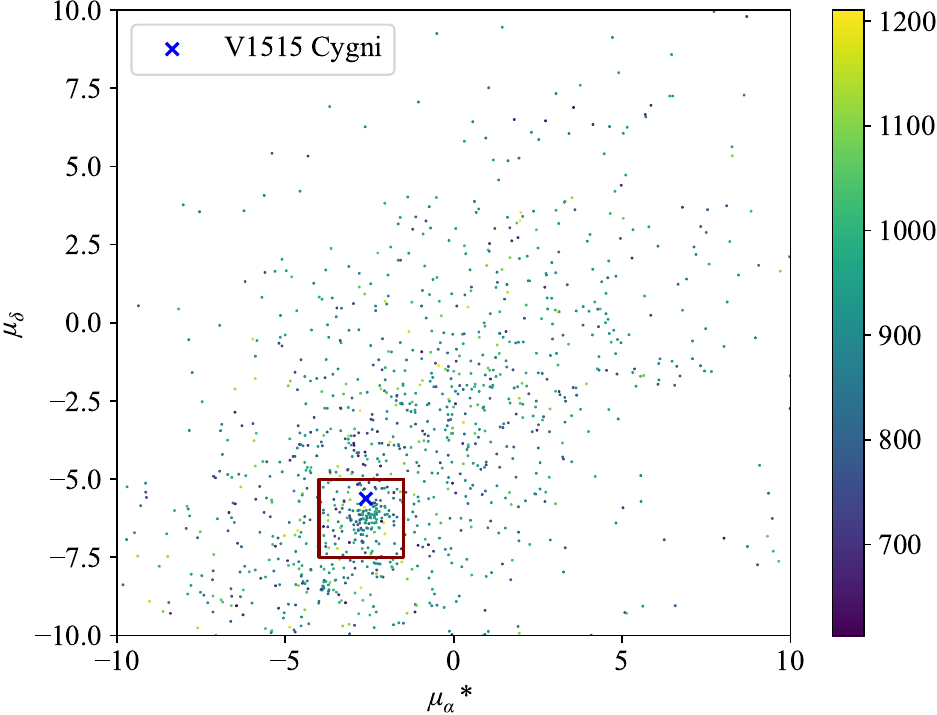}
 \includegraphics[width=6.4cm]{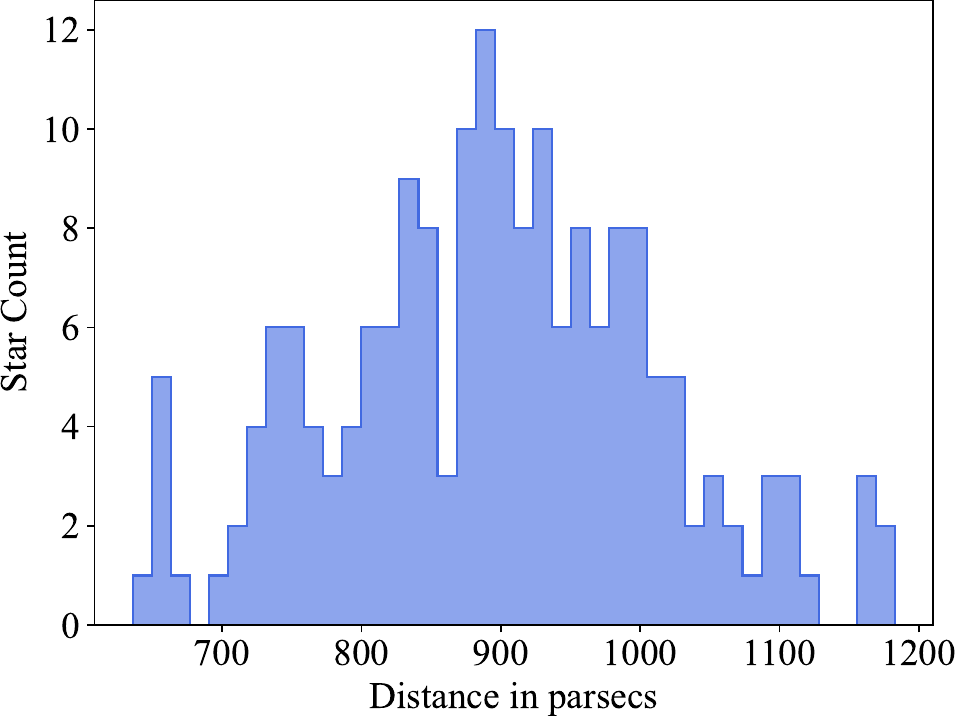}
 \includegraphics[width=5.3cm]{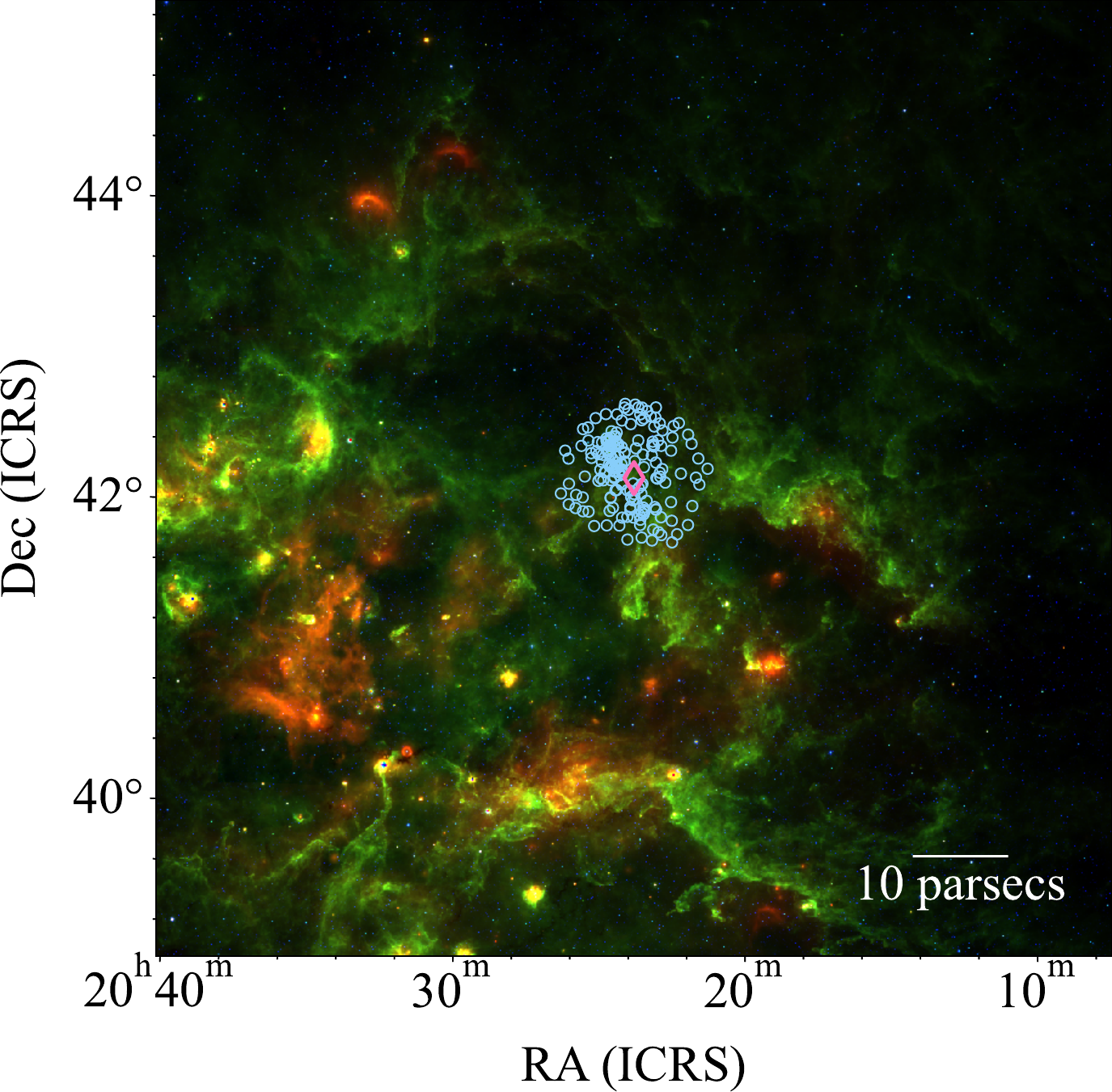}\\ 
 \includegraphics[width=12.5cm]{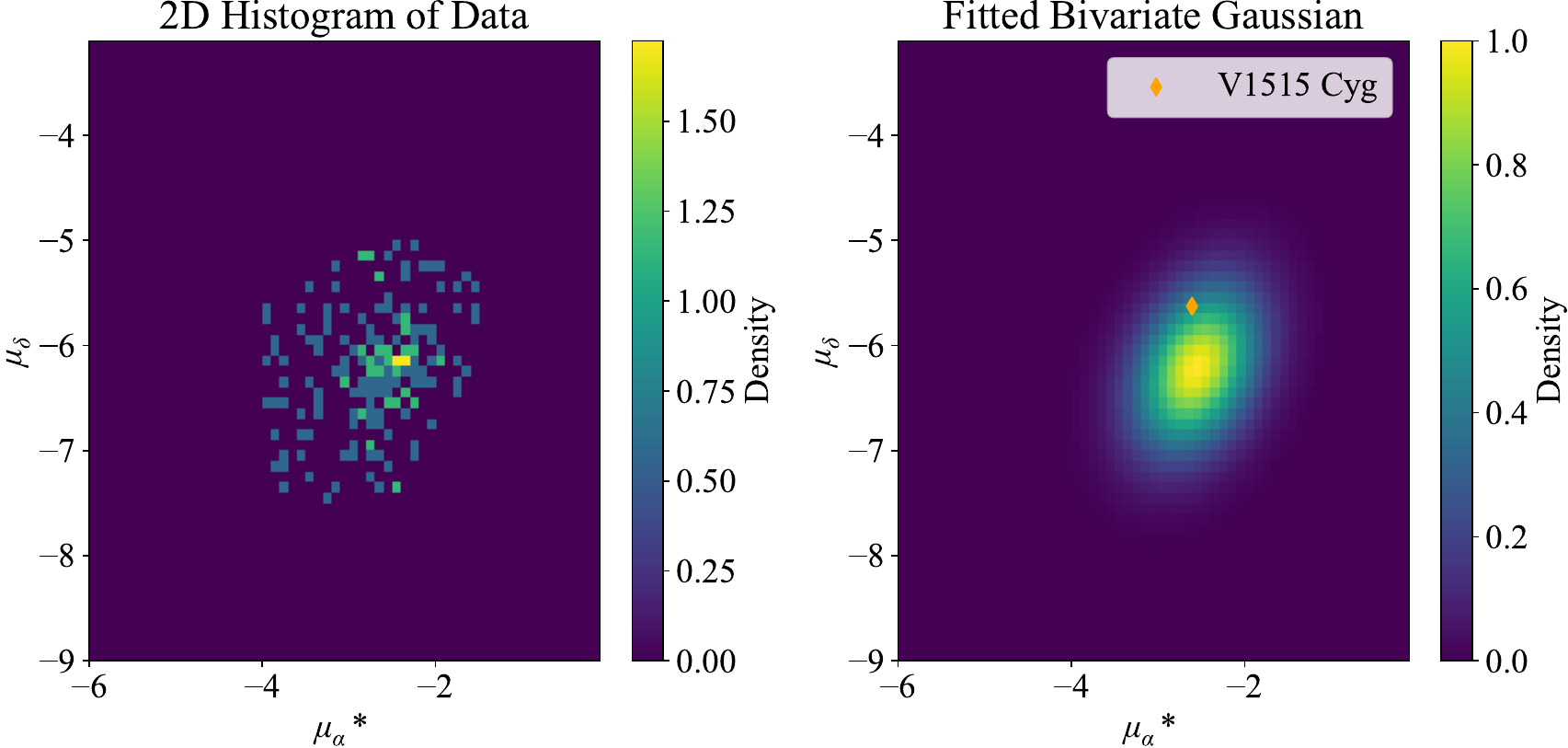}
 \includegraphics[width=5.6cm]{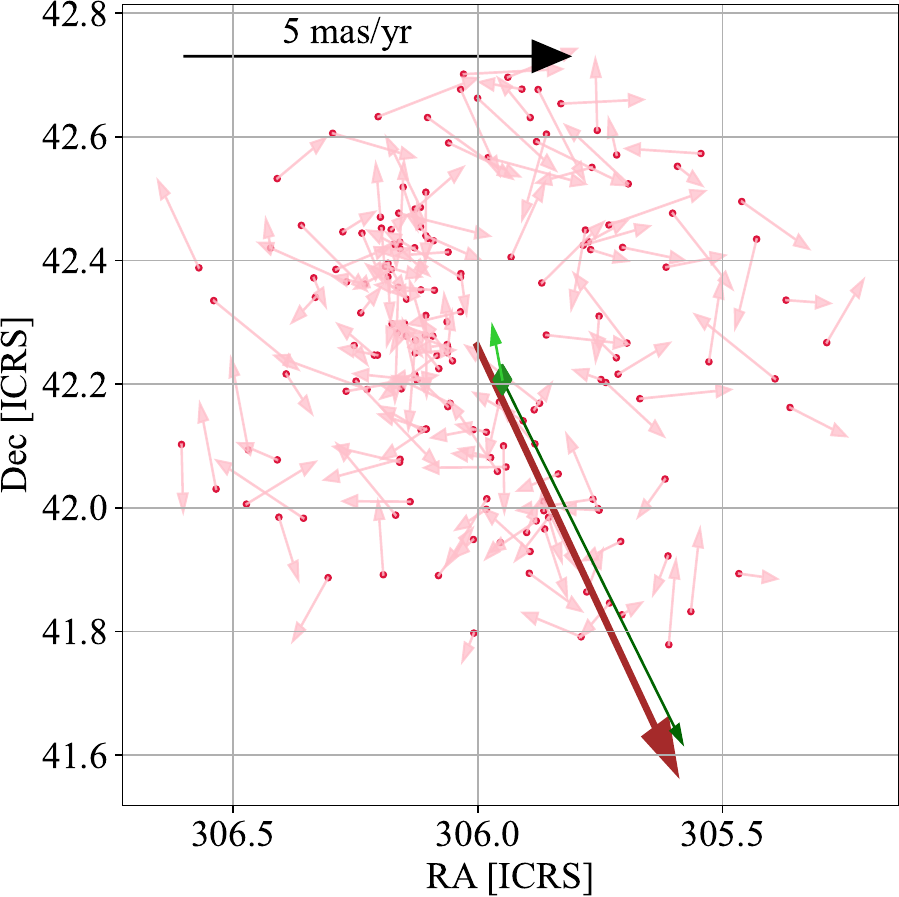}
    \caption{\textit{Top panels:} VPD of stars obtained using \mfour\ from the $0.5^\circ$ (7.88 parsec) region around V1515 Cygni. The kinematic overdensity consistent with V1515 Cygni and corresponding to NGC 6914 is marked, and can be seen more clearly relative to Fig~\ref{fig:50pc-cygx}. In the distance histogram for this component a peak is clearly discernible. The VPD clump stars (blue dots) and V1515 Cygni (white diamond) are overlayed on a \textit{WISE} color image (22, 12 and 3.4 $\mu$m mapped to R,G,B). \textit{Bottom panels:} The bivariate Gaussian fit to the kinematic concentration on the VPD, with the proper motion arrows in the right panel. Brown denotes the mean motion of the cleaned sample, the green diamond V1515 Cygni, with the darker arrow showing the full proper motion and the lighter showing the residual motion w.r.t. the cluster mean. The pink arrows show the residual proper motions of other member stars with respect to the mean.
    }
    \label{fig:8pc-cygx}
\end{figure}

The central cluster of points in this kinematic group corresponds to the cluster NGC 6914, one of the many components of Cygnus-X. Since it is a candidate associated cluster for V1515 Cygni as seen from the VPD, we attempt to isolate its component stars from the rest of the sample. 
For this, we use \mfour. The search radius is estimated using the approximate size of the NGC 6914 cluster on the sky, which is visually about $0.5^\circ$. This corresponds to a linear length of $7.88$ pc at 900 pc, and we used this value to define the parallax range. In the retrieved set of stars, NGC 6914 stands out clearly (see Fig~\ref{fig:8pc-cygx}), in position, in proper motion, as well as in the parallax histograms. It is also distinguished in the bimodal Gaussian-smoothed proper motion histograms. 

For cleaning the sample, we define an initial cut -$4<$\pmra$<-1.5$ and $7.5<$\pmdec$<-5$, and perform the bivariate Gaussian fit and $3\sigma$ clipping for the final sample. Since this is already well-localised in position, no further cuts are needed like NAP.

\subsection{V1735 Cygni and the IC 5146 streamer} 
V1735 Cygni is located in a region of Cygnus well to the northeast of the NAP region, and has a parallax corresponding to distance $690 \pm 36$ pc \citep{Gaia_2023}. A nearby star forming region, IC 5146, has projected angular distance on the sky of $1.08^\circ$, corresponding to about 13.1 pc at a distance of 690 pc.  However, along the line-of-sight, the IC 5146 cluster is thought to be somewhat further away, at $783 \pm 25$ pc.

We first consider stars in the vicinity of V1735 Cygni to see if it can be associated with a kinematic substructure that has not been previously recorded in the literature. 
We used the criteria of \mfour\ in our search, since it was effective in the preceding analysis in associating V1515 Cyg and NGC 6914 among the full background of the VPD. The search radius was set to $1^\circ$, corresponding to 12.2 pc at the distance of V1735 Cygni. However, no clear clustering emerged in the VPD, nor peaks in the Gaussian-smoothed histograms of \pmra\ and \pmdec, indicating the absence of any clearly detectable cluster or other substructure. 

\begin{figure}
	\includegraphics[width=6.4cm]{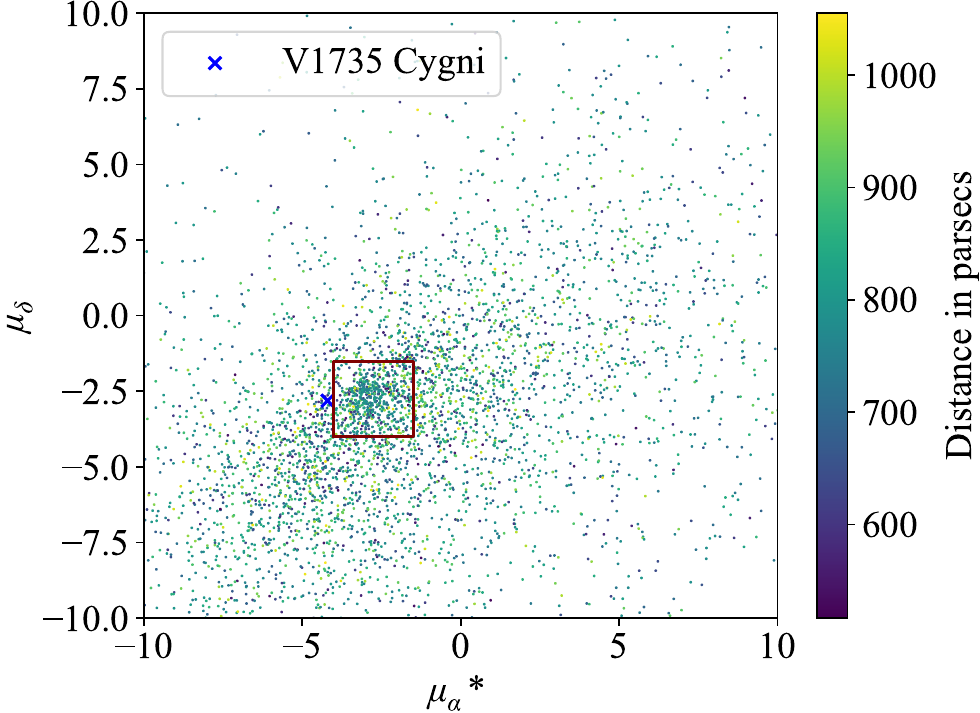}
 \includegraphics[width=6.4cm]{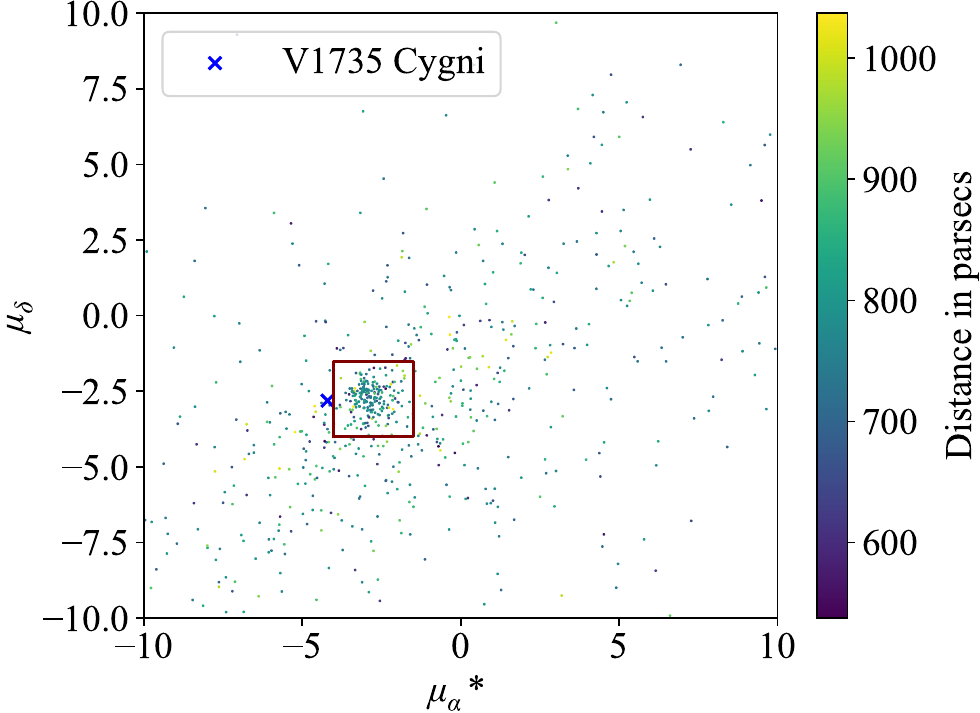} \\
 \includegraphics[width=6.4cm]{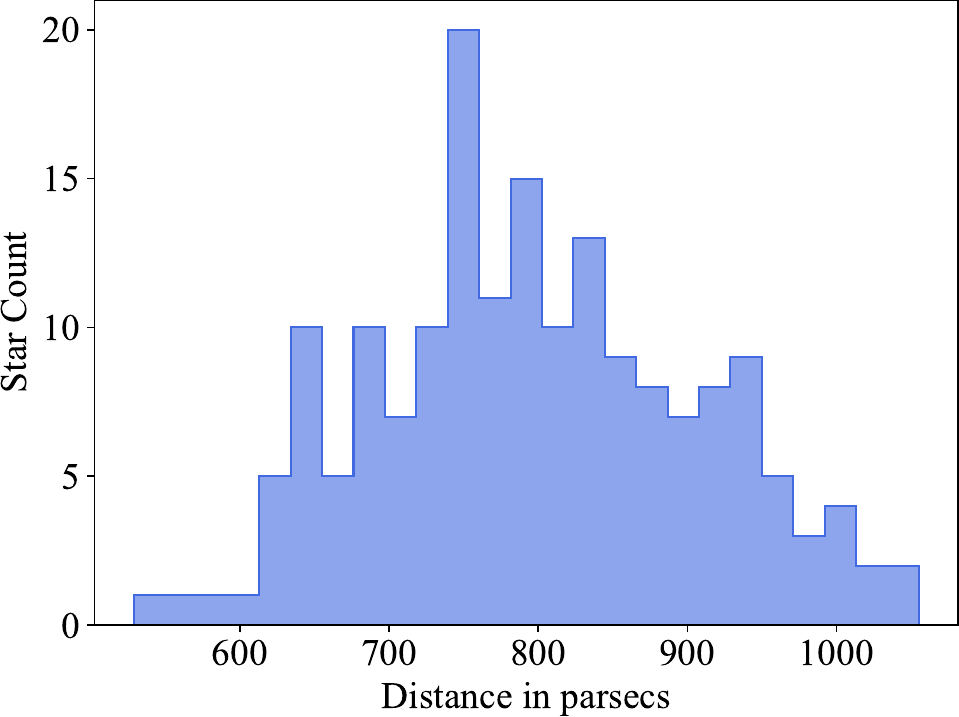}
 \includegraphics[width=5.6cm]{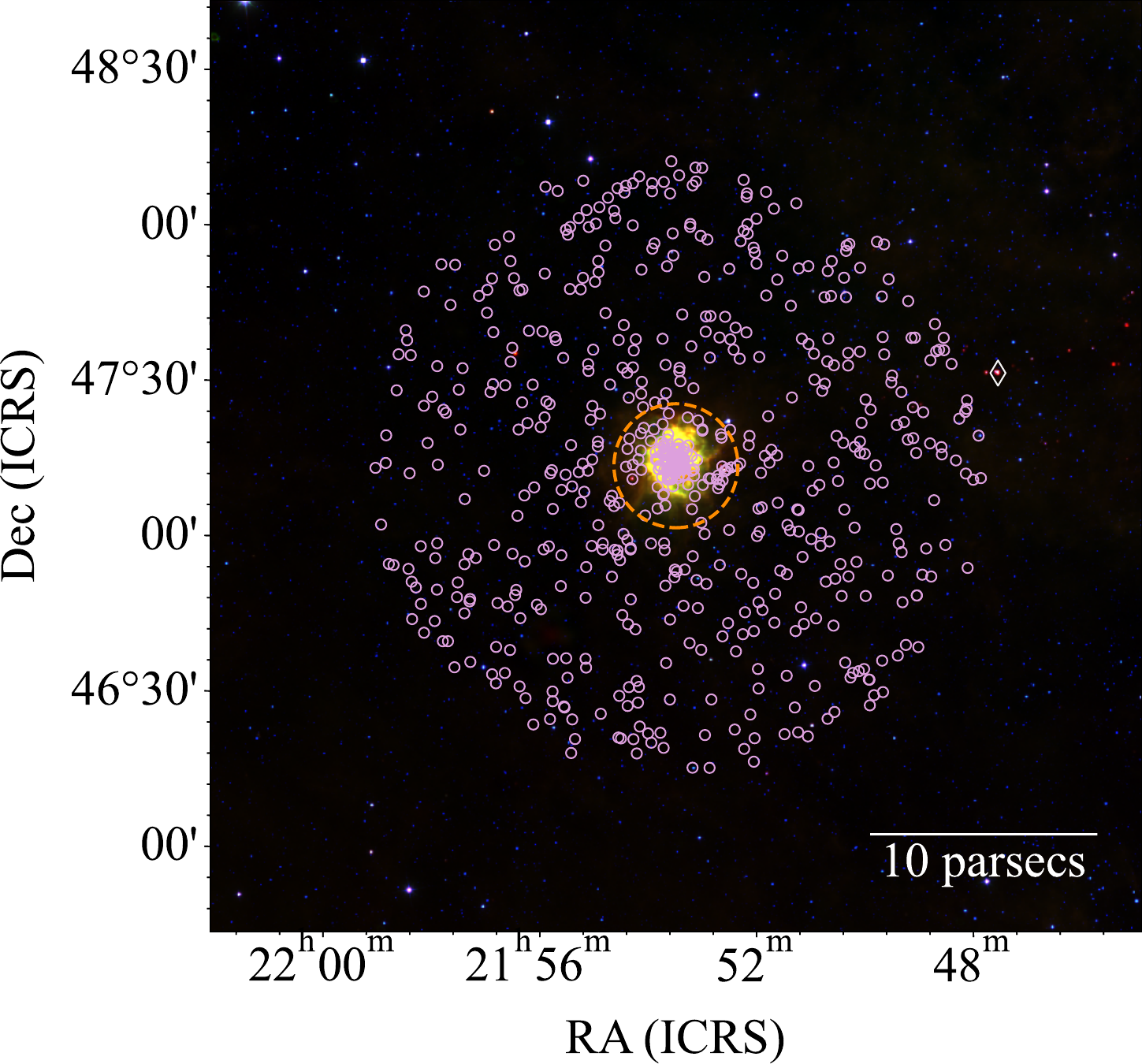}
    \caption{\textit{Top panels:} Vector point diagrams of the $1^\circ$ sample around the IC 5146 position on the left, and of the $0.4^\circ$ sample on right, obtained using \mfour. \textit{Bottom panels:} Distance distribution of the $1^\circ$ sample on the left. A clear peak can be seen at the literature distance for IC 5146 at $\sim 780$ pc. The corresponding \textit{WISE} image (22, 12 and 3.4 $\mu$m mapped to R,G,B) on the right shows the full $1^\circ$ sample in purple dots, and V1735 Cygni as a white diamond just outside the search radius at top right. The dotted orange circle shows the $0.4^\circ$ radius used in the final retrieved sample.
    }
    \label{fig:ic5146-vpd-WISE} 
\end{figure}

We then used the criteria of \mfour\ to retrieve stars based on the coordinates and distance of IC 5146, with the initial search radius again set to $1^\circ$. Although there is not a clear peak in the VPD, the retrieved stars do show clear clustering in RA-Dec at the position of IC 5146. Based on the clustering, the search radius was updated to $0.4^\circ$, corresponding to $5.5$ pc.  As shown in Fig~\ref{fig:ic5146-vpd-WISE} , the resultant set of stars shows a clear clustering in VPD as well as identifiable peaks in all of \pmra\, \pmdec\ and parallax.  This exercise demonstrates our ability to identify likely members of IC 5146.
However, the proper motion of V1735 Cygni is only marginally consistent with that of IC 5146 in RA ($\sim 3\sigma$ distance in \pmra, and $>3\sigma$ in full proper motion space), and its spatial position is inconsistent with being a member of the main cluster. This renders a kinematic association unlikely, and we do not attempt further cleaning of the sample. 

We note that star formation in the IC 5146 region comprises not only the identified cluster but also a `streamer' which appears as a filamentary extension to the North-West of the main cloud. \cite{Harvey_2008} found several YSOs both in the main cloud and the streamer based on \textit{Spitzer} observations. The positions of the streamer filament and associated YSOs are spatially coincident with the V1735 Cygni, and we thus analyze the kinematics of the streamer YSOs with respect to the IC 5146 cluster core stars.

\begin{figure}[h!]
	\includegraphics[width=6.0cm]{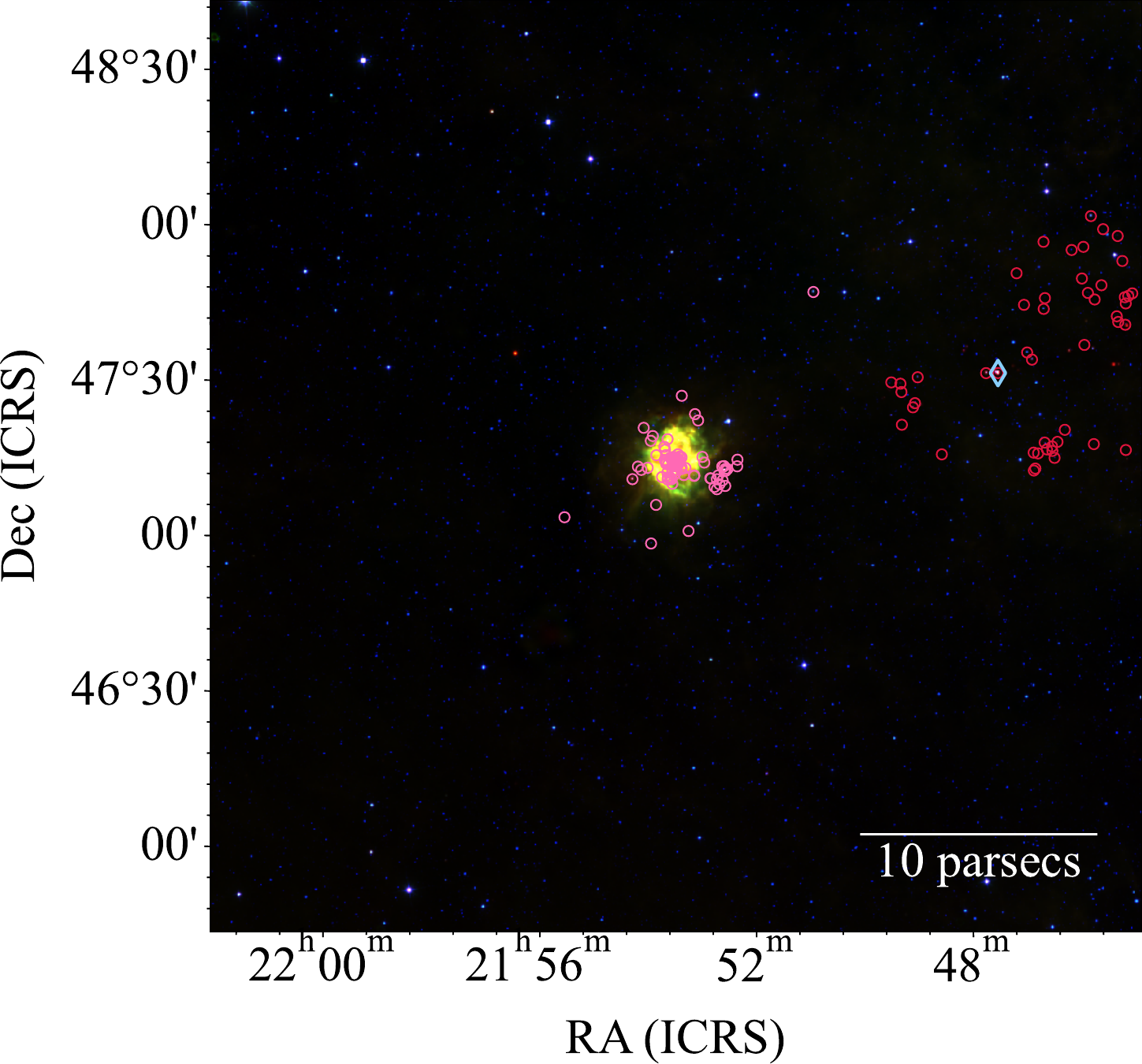}
 \includegraphics[width=7.4cm]{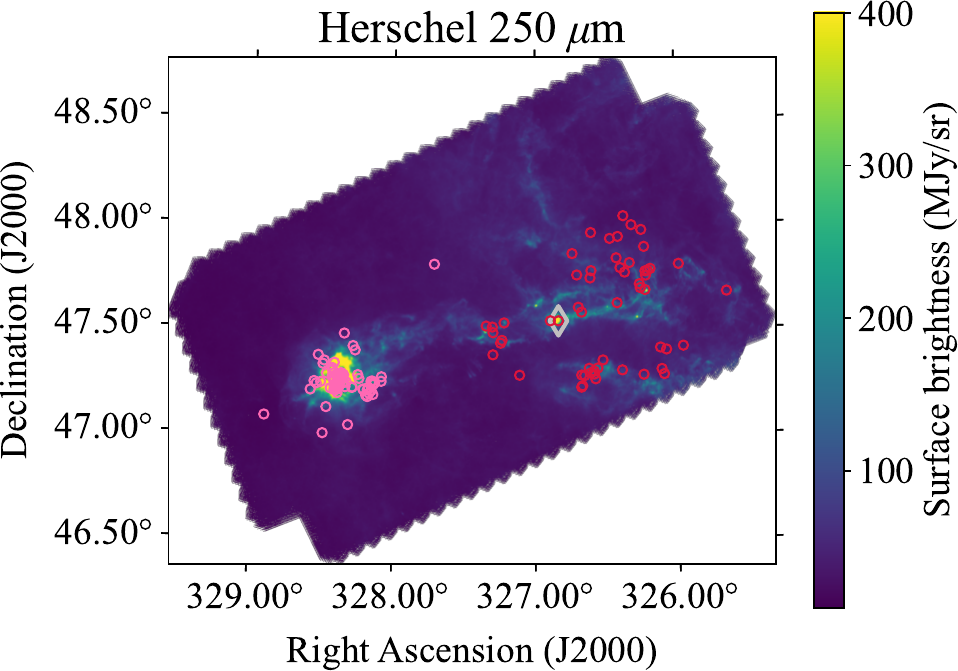}
    \caption{\textit{Left:} Distribution of \textit{Spitzer}-identified YSOs from \cite{Harvey_2008} study of the IC 5146 core region (pink circles) and streamer (dark circles), overlaid on a \textit{WISE} (22, 12 and 3.4 $\mu$m mapped to R,G,B) image. \textit{Right:} Same distributions overlaid on a \textit{Herschel} 250 $\mu$m image that illustrates the thermal emission from the cloud and streamer filament. V1735 Cygni is marked in both panels with a blue-grey diamond.
    }
    \label{fig:streamer-img}
\end{figure}

We retrieve \textit{Gaia} DR3 counterparts to the documented YSOs from \cite{Harvey_2008} by cross-matching their RA/Dec to within a radius of 1 arcsecond. Out of the 202 YSOs in their catalog, 148 had \textit{Gaia} counterparts (single stars within the crossmatching radius), of which 142 had proper motion values. Of these, 52 belonged to the streamer and 90 to the core. We do not impose any restrictions on the parallax or the measurement uncertainties of the retrieved stars. Their spatial distribution is shown in Fig~\ref{fig:streamer-img}.

\begin{figure}[h!]
	\includegraphics[width=6.1cm]{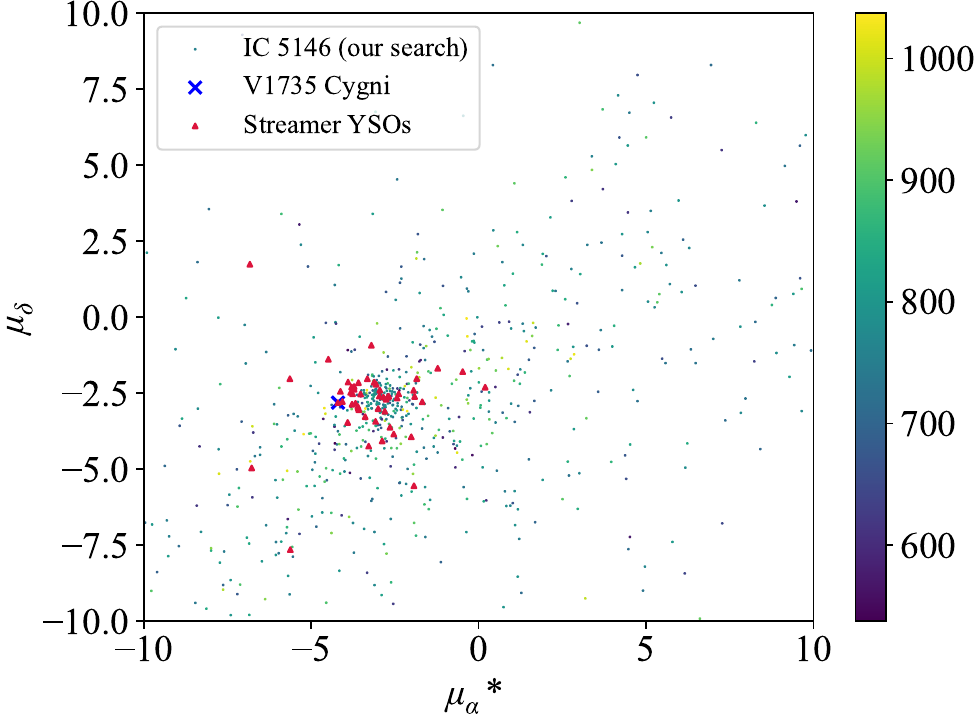}
 \includegraphics[width=6.1cm]{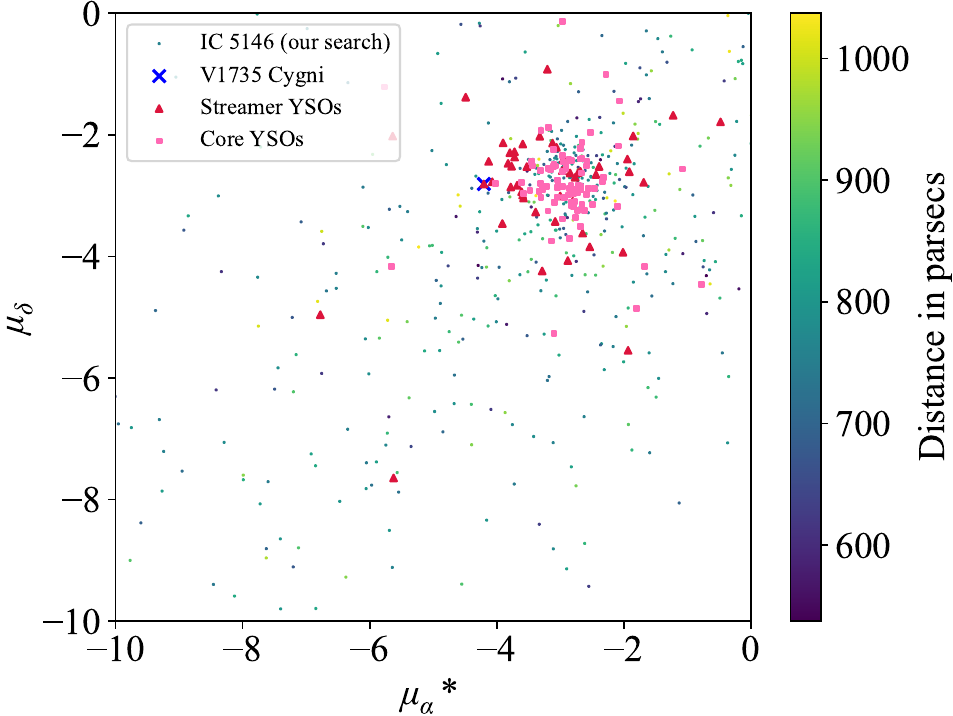}
 \includegraphics[width=6.1cm]{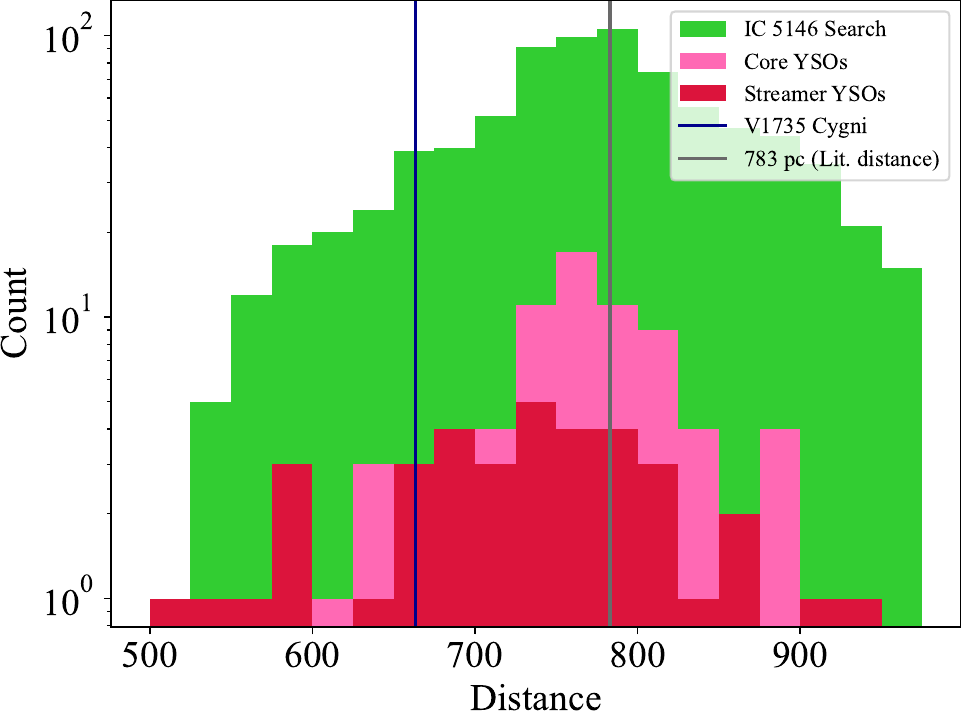} \\ 
 \includegraphics[width=9.1cm]{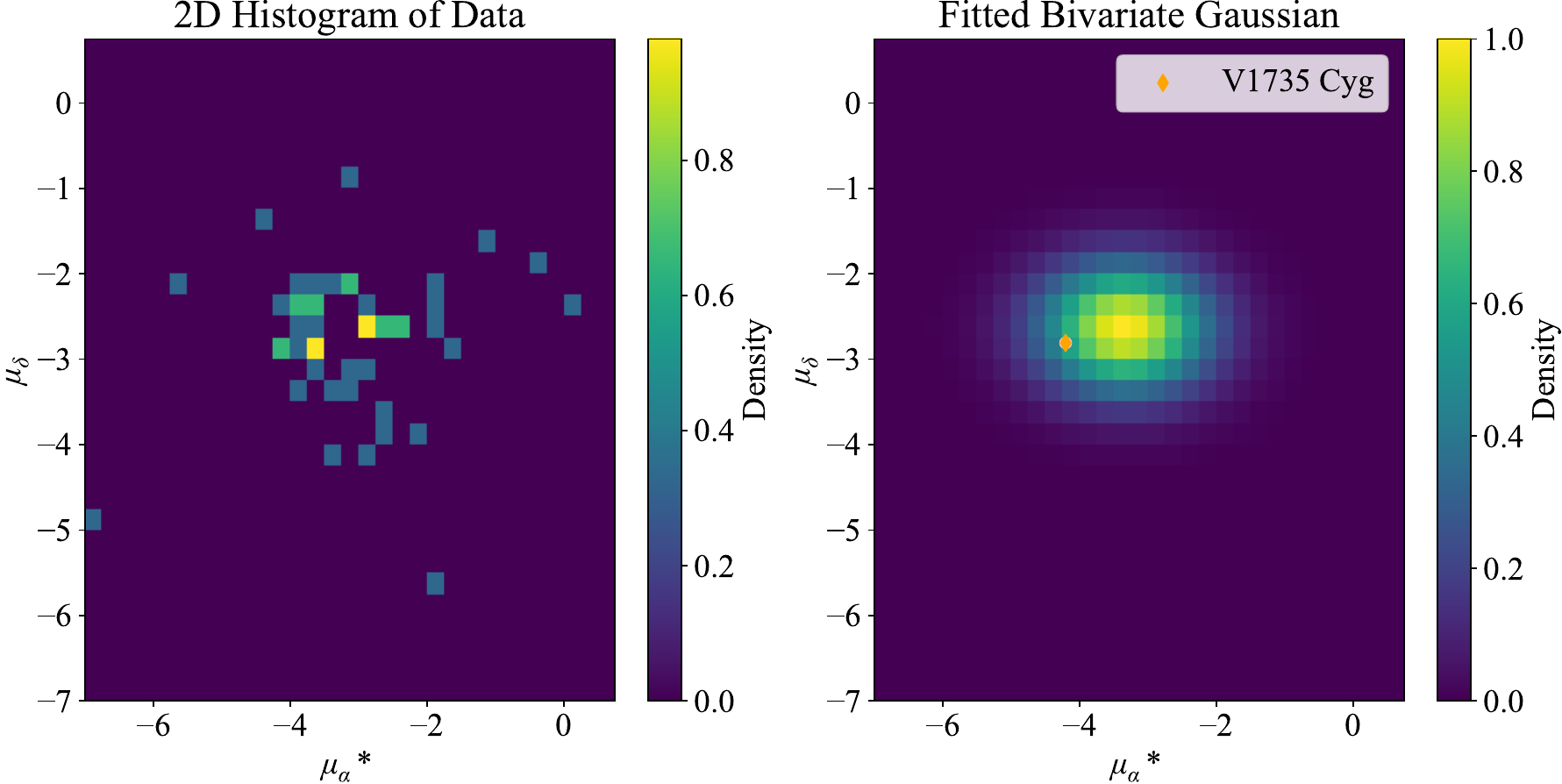} 
 \includegraphics[width=9.1cm]{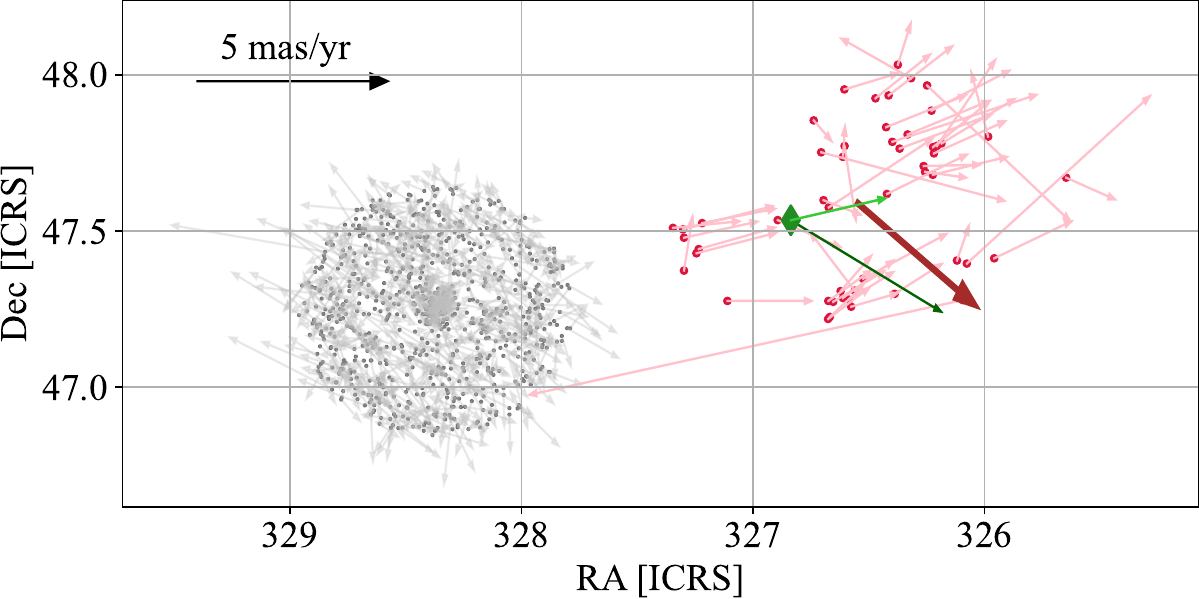} 
    \caption{\textit{Top panels:} Vector point diagram of the \textit{Spitzer}-identified streamer YSOs (in red) compared with our earlier retrieved IC 5146 sample. Zoomed-in version of the VPD, with the core YSOs (in pink) and streamer YSOs (red) is also shown. The core YSO proper motions are fully consistent with our retrieved IC 5146 sample, while the streamer YSOs have a slightly more negative \pmra\ than the core. V1735 Cygni is consistent with the streamer in the VPD. In the distance histograms of the full IC 5146 sample, core and streamer YSOs compared to V1735 Cygni, we see the streamer YSOs having a broader distribution and V1735 Cygni being consistent with that distribution.
    \textit{Bottom panels:} Bivariate Gaussian fit to the VPD of the streamer YSOs. We also show the mean and residual proper motions. The grey arrows indicate the core of IC 5146, and the pink/red dots show the streamer. To show the streamer structure moving away visually, the arrows for streamer stars show the residuals with respect to the mean motion of IC 5146. Mean proper motion of the streamer is in brown and V1735 marked with a green diamond, with the darker arrow showing its full proper motion and lighter arrow showing the resdiual w.r.t. the cluster
    }
    \label{fig:streamer-vpd}
\end{figure}

We construct VPDs for both core and streamer YSOs and compare them against the VPD of our original set of IC 5146 stars retrieved using \mfour\ in a $0.4^\circ$ radius. Further, we examine parallax histograms of our IC 5146 sample, the core YSOs and and the streamer YSOs separately, from \cite{Harvey_2008}.  
The resultant plots are shown in Fig~\ref{fig:streamer-vpd}.

Since this set is already reliably associated to be a coherent streamer structure i.e. free of field contamination, we do not perform further filtering in the VPD like NAP or NGC 6914. Visually we observe that the YSOs of both regions (streamer and core) form kinematic concentrations at approximately the same values of \pmra\ and \pmdec\ as our identified IC 5146 kinematic clustering. The concentration of streamer YSOs appears at a slightly more negative \pmra\ than the core YSOs. Since V1735 Cygni is spatially coincident with the streamer, we fit a bivariate Gaussian to the streamer YSOs in proper motion space to check for the separation of V1735 Cygni from this group. They are consistent within $1.1\sigma$.  Further, the distance histogram shows that the \textit{Gaia} distance to V1735 Cygni is consistent with the distance distribution of the streamer YSOs.

We conclude that the streamer YSOs identified by \cite{Harvey_2008} from \textit{Spitzer} color data form a distinct kinematic component. Furthermore, the streamer YSOs share nearly identical proper motions with the core YSOs identified by \cite{Harvey_2008}, as well as with our retrieved IC 5146 star sample, implying physical association.  The slightly more negative \pmra\ of the streamer YSOs relative to IC 5146 and their position to the west of the main IC 5146 cloud is consistent with the streamer members moving away from the core.  Finally, V1735 Cygni is a member of the streamer of IC 5146, consistent with both its proper motion and three-dimensional position.

\subsection{V2494 Cygni and Cyg OB7/Braid Nebula}
V2494 Cyg has a measured parallax indicating distance of $533 \pm 75$ parsecs \citep{Gaia_2023}.  A star forming region nearby in projection, Cyg OB7, of which Braid Nebula is a part, has a larger distance traditionally around 800 pc \citep{Dobashi_1996} but more recently assessed at about 594 pc \citep{Kuhn_2019B}. The region does not have a clear central cluster or nebula.

We based our initial search for associated stars around the position and nominal distance of V2494 Cygni itself.  Application of the \mfirst\ criteria did not reveal any definite peaks in the VPD. Further, in the absence of clear clustering in the VPD and without reliable distance measures in prior literature, we cannot use \msecond\, \mthird\ or \mfour\ effectively given that neither the search distances $D$ nor the hypothetical cluster's subtended angle on sky $\theta$ are known. Furthermore, the error in the distance of V2494 Cygni itself (75 pc) is larger than our typical search radii (10--50 pc).

We instead conducted a broad search centered at the Cyg OB7 position but over the distance range 500--900 pc in order to identify potential kinematic groups within the VPD.   
Two concentrations emerged: a weak one in the 500--600 pc bin, also faintly visible in the 600--700 pc bin, and a stronger one in the 800--900 pc bin.  The former appears nearest to the proper motion of V2494 Cyg while the latter corresponds to the general field population towards Cygnus. The measured Gaia parallax of V2494 Cyg is also coincident with the closer distance.  

We then used method \mthird\ to retrieve all stars within $1^\circ$ of V2494 Cygni, having distance between 500 and 650 pc with parallax SNR $\geq 3$. The resultant VPD had a clear concentration, though the \pmra\ and \pmdec\ of stars in this concentration appeared strongly correlated (i.e. along a diagonal instead of being a typical circular concentration). Stars from the kinematic component also illustrated spatial clustering in a region identifiable with the dark cloud LDN 1003 \citep{Lynds_1962} in the Braid Nebula, a substructure in Cygnus OB7. 

\begin{figure}[h!]
	\includegraphics[width=6.6cm]{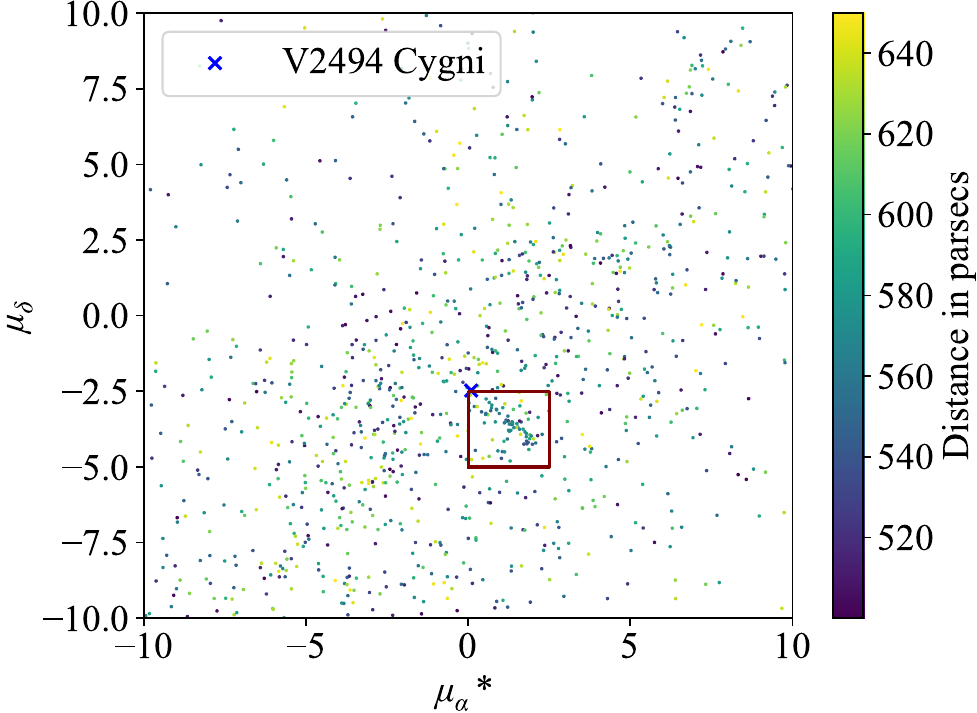}
 \includegraphics[width=6.3cm]{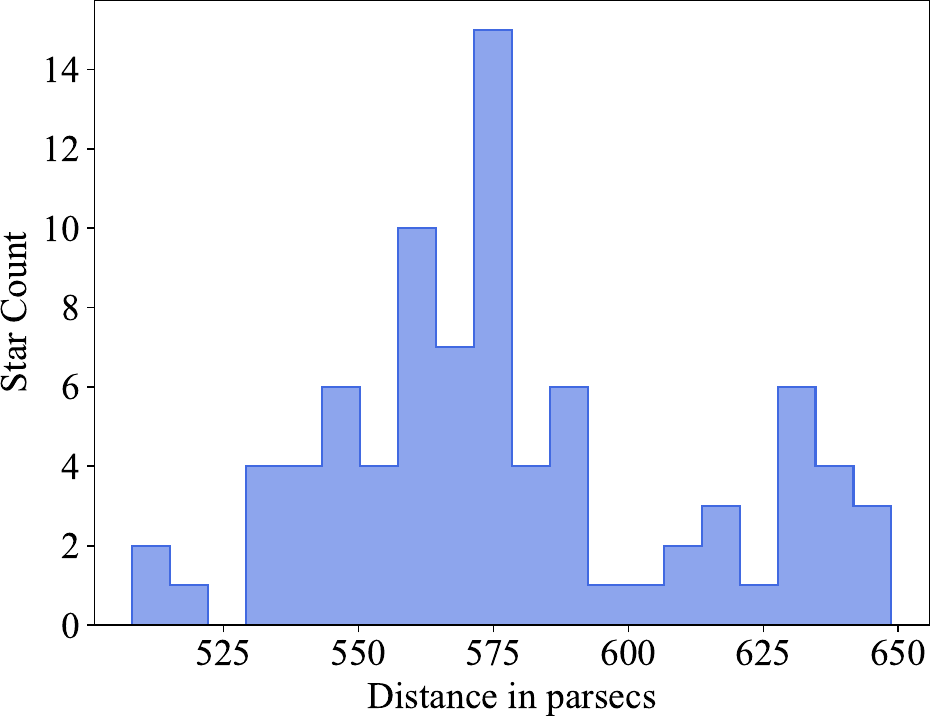}
 \includegraphics[width=5.3cm]{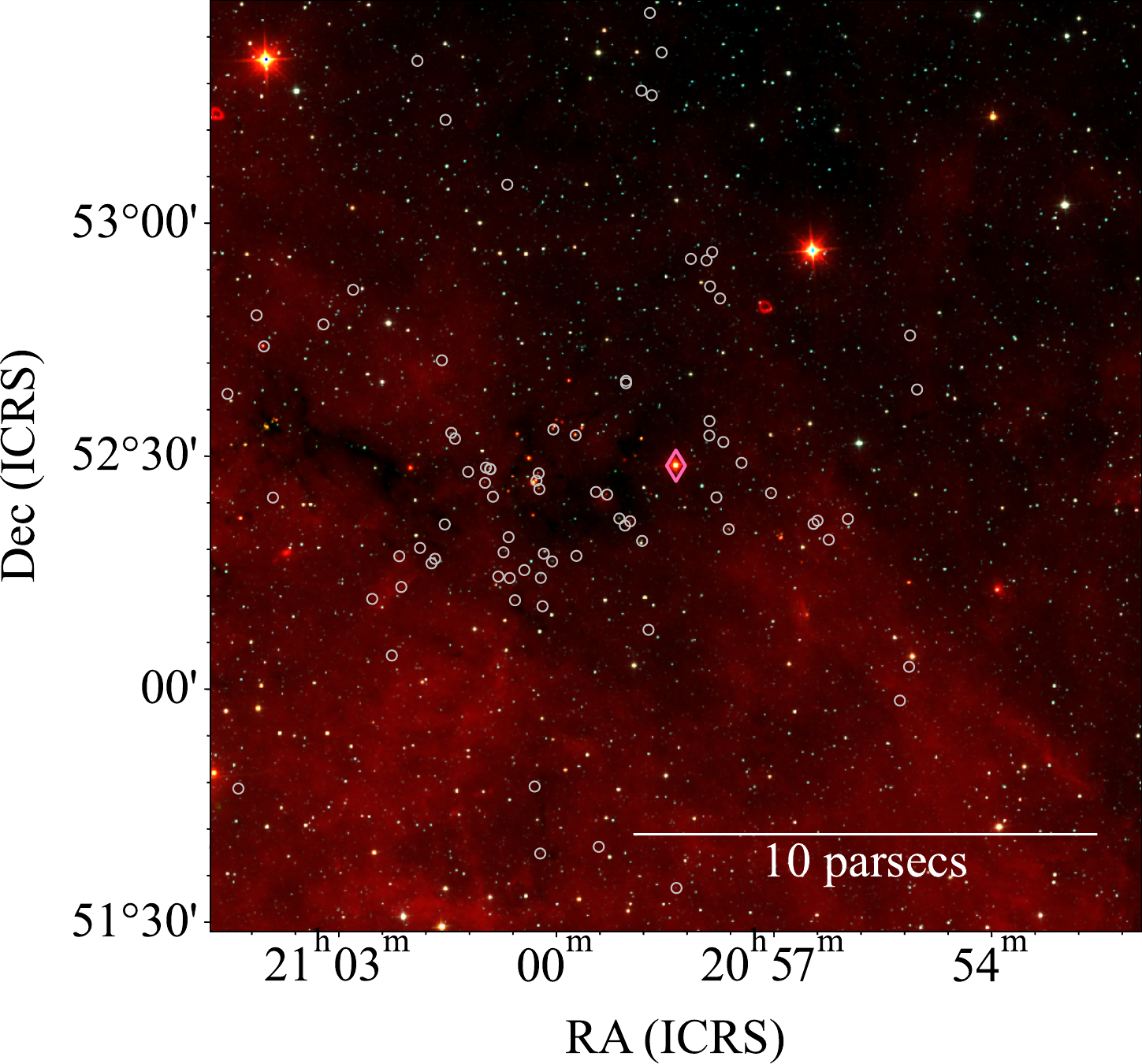}
    \caption{\textit{Left:} Vector point diagram of the sample selected using $M_4$ --- from a $1^\circ$ spatial around the Braid Nebula position, and constrained to 500--650 pc in distance. A kinematic concentration is marked, with which V2494 Cygni appears to be marginally consistent. \textit{Middle:} Distance distribution for stars in the kinematic concentration, with a clear peak around 560--580 pc. \textit{Right:} \textit{WISE} image (22, 12 and 3.4 $\mu$m mapped to R,G,B) of the region. White circles show the members of the kinematic component, and the diamond marks V2494 Cygni.
    }
    \label{fig:braid-vpd-WISE} 
\end{figure}

We thus repeat the same search between 500 and 650 pc but now centered on the coordinates of Braid Nebula. The results are illustrated in Figure~\ref{fig:braid-vpd-WISE}  where the VPD shows strong kinematic clustering and  the distance distribution of stars in this kinematic clump peaks at $\sim 575$ pc.   The spatial distribution is also shown, relative to a \textit{WISE} image.

\begin{figure}[h!]
	\includegraphics[width=6.4cm]{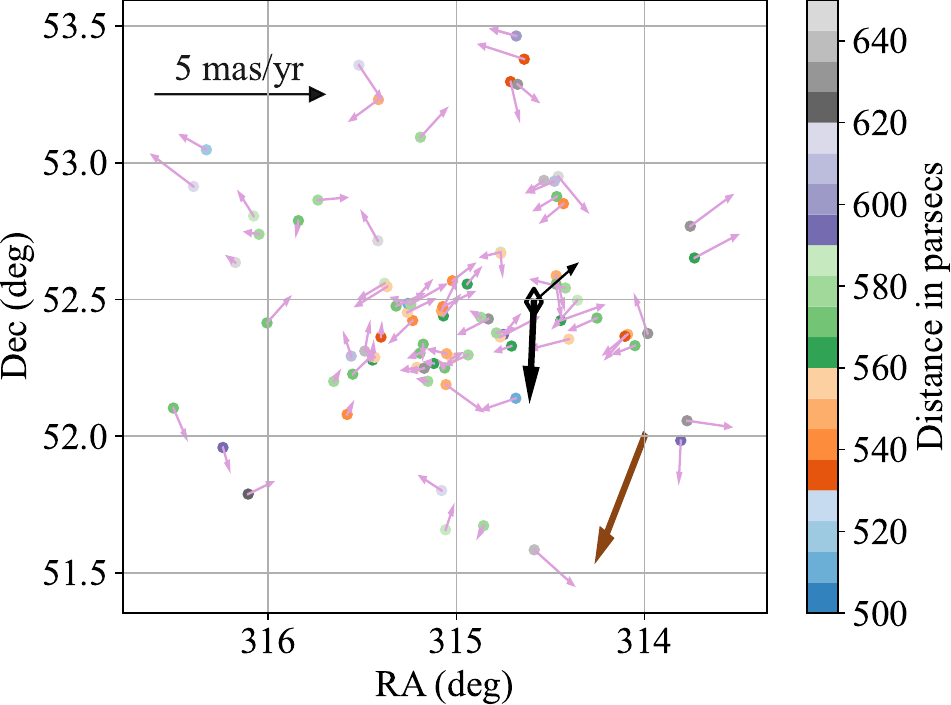}
 \includegraphics[width=6.4cm]{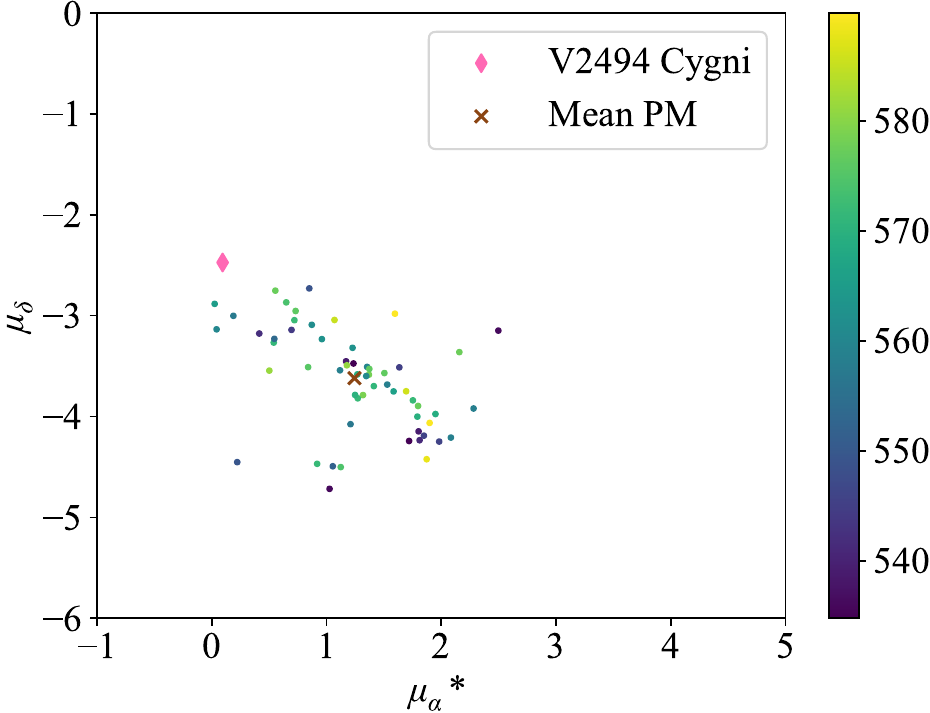}
 \includegraphics[width=5.6cm]{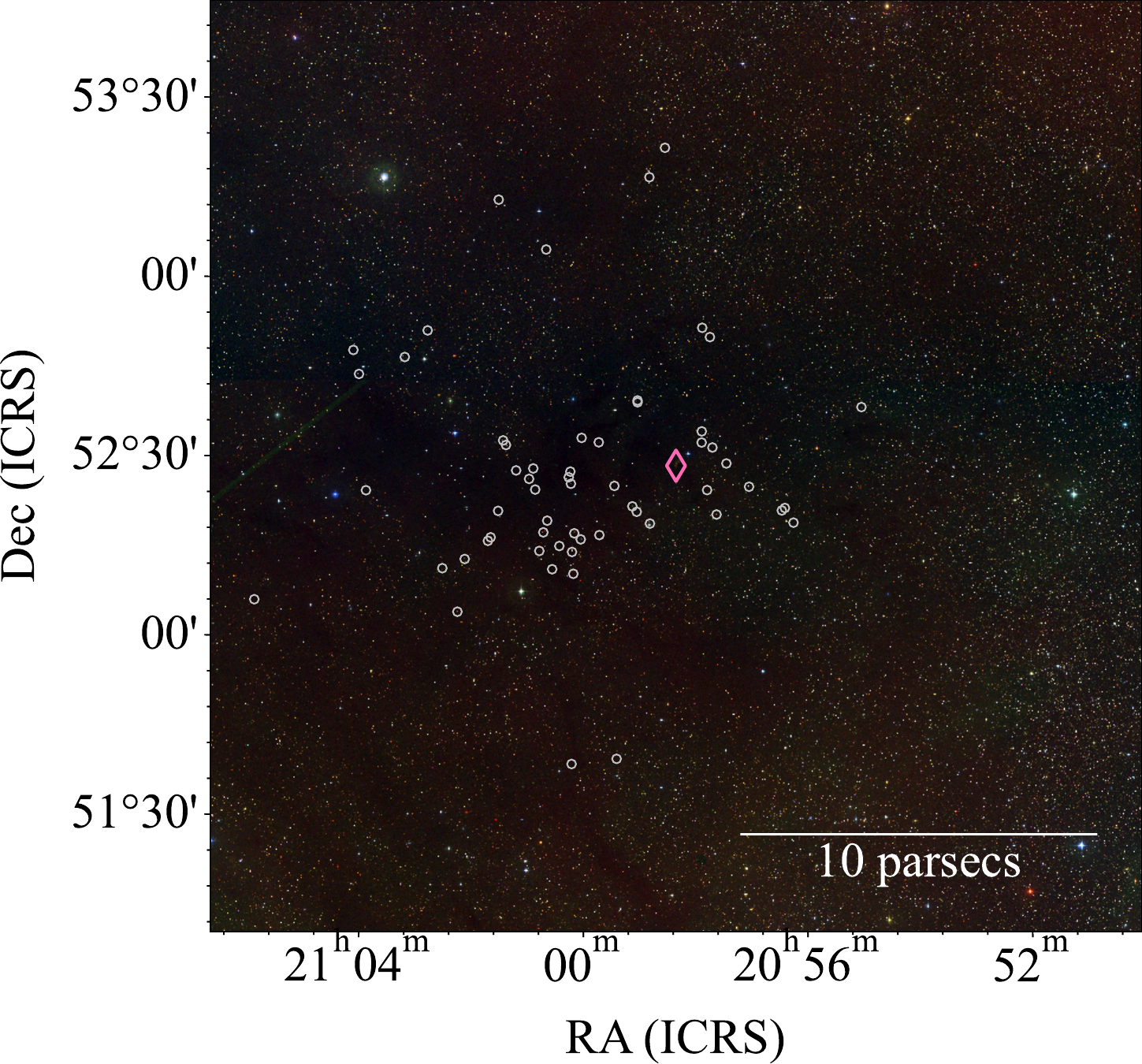} \\
 \includegraphics[width=12.5cm]{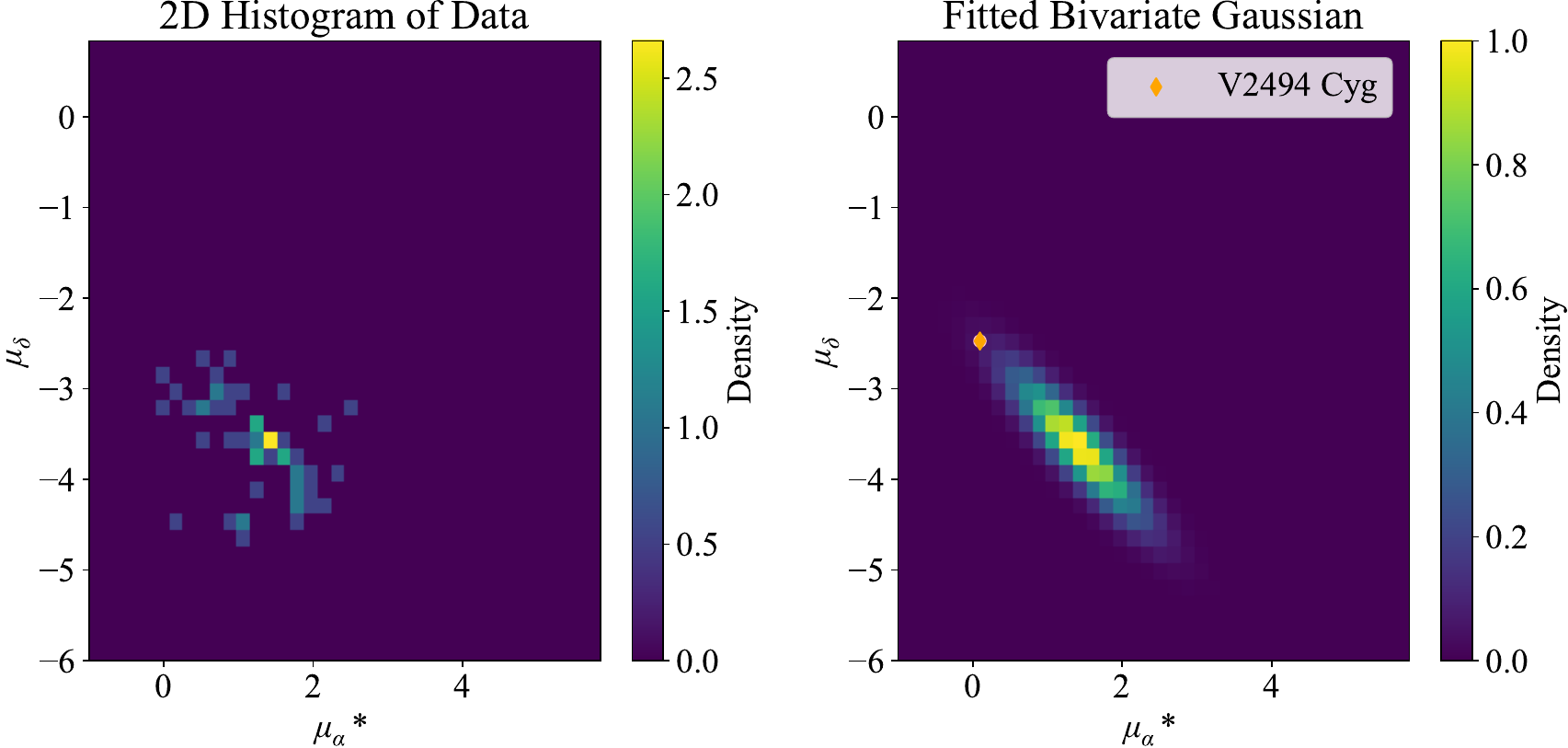}
    \caption{\textit{Top panels}: Proper motions and spatial$+$distance distribution of the stars within 500--650 pc which belong to the kinematic component marked in Fig~\ref{fig:braid-vpd-WISE}. Mean of the sample is marked with a brown arrow, and residual motions w.r.t. this mean are marked in light pink. V2494 Cygni is shown as a black diamond, with the thinner arrow showing its residual motion and the thicker arrow showing the full proper motion. We see stars clump roughly along a diagonal, corresponding spatially to LDN 1003. These have distances in the range 530--590 pc as indicated by the colour-coding. VPD of the stars from 530--590 pc range and an overlay of the truncated set on a DSS image of LDN 1003 is shown.
    \textit{Bottom panels:} Bivariate Gaussian fit to the 530--590 pc sample.
    }
    \label{fig:ldn1003-pm-dss}
\end{figure}

Since the spatial clustering is not so clear, we tried to isolate the stars associated with LDN 1003 by applying a cut on the distance range. Also, to further explore the correlation between the proper motion components, we calculate the residual proper motion of each star in the kinematic concentration by subtracting from it the mean \pmra\ and \pmdec\ of the concentration. 

We see in Figure~\ref{fig:ldn1003-pm-dss} that the the dark cloud and the corresponding distribution of stars is non-spherical, and the proper motion distribution is also clearly asymmetric. The highest variance axis of the proper motion distribution roughly coincides with the major axis of the positional distribution of the stars and the cloud shape. This could imply an expansion in the cloud's stellar members along the longer axis of the asymmetric structure. Such expansion has previously been observed in other asymmetric clusters such as $\lambda$ Orionis \citep{SS2024, Armstrong_2024} and Lagoon Nebula \citep{Wright_2019}.

This induces a visible correlation between the two components on the VPD. Additionally, we find that most stars that cluster in RA/Dec and have proper motions within the kinematic concentration lie in a distance range of 530--590 pc. We thus restrict our stars to this distance range, and also restrict the proper motions to satisfy $0<$\pmra$<2.5$ and $-5<$\pmdec$<-2.5$. We then apply the bivariate Gaussian fitting + $3\sigma$ clipping on this set for a cleaner sample. Like the IC 5146 streamer, the structure is not shaped like a typical cluster so it cannot be fit to a bivariate Gaussian profile in position space. The final set of stars is exactly spatially coincident with LDN 1003, as seen in Fig~\ref{fig:ldn1003-pm-dss}. V2494 Cygni is consistent with this set in both position (including parallax) and proper motions.

\section{Results and Discussion}
\label{sec:results}
\subsection{Gaussian-Smoothed Histograms}

\begin{figure}[!ht]
 \includegraphics[width=6.2cm]{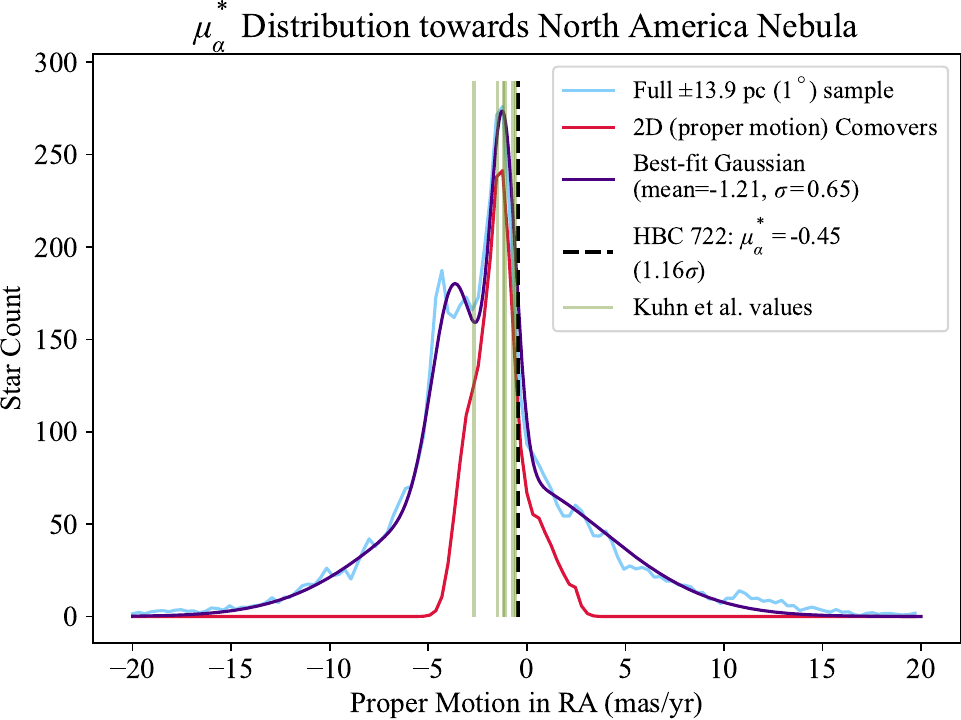}
 \includegraphics[width=6.2cm]{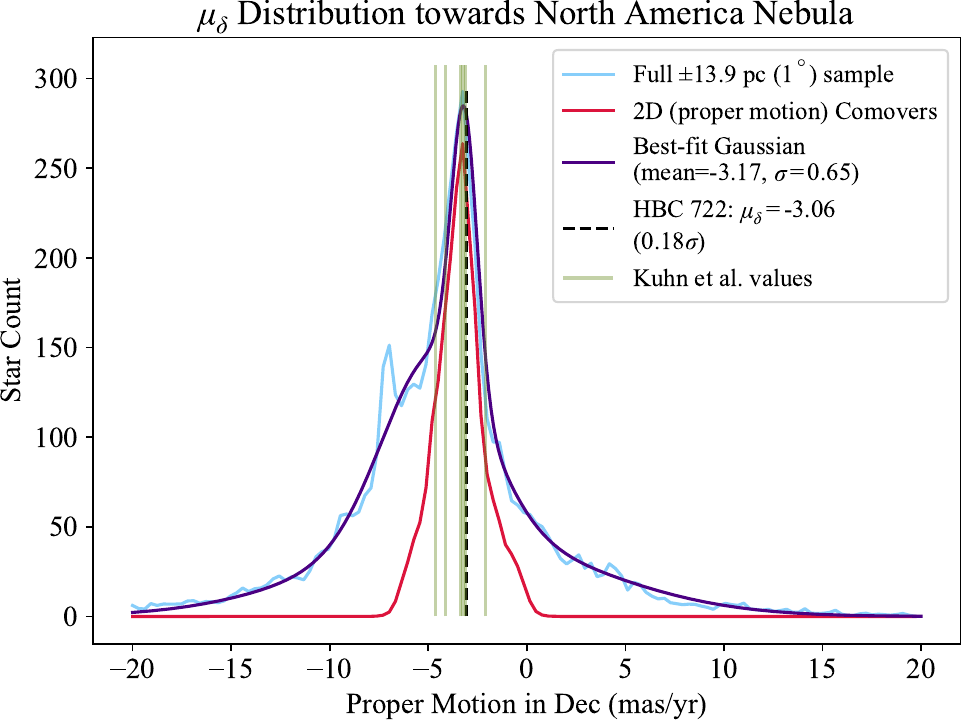}
 \includegraphics[width=6.2cm]{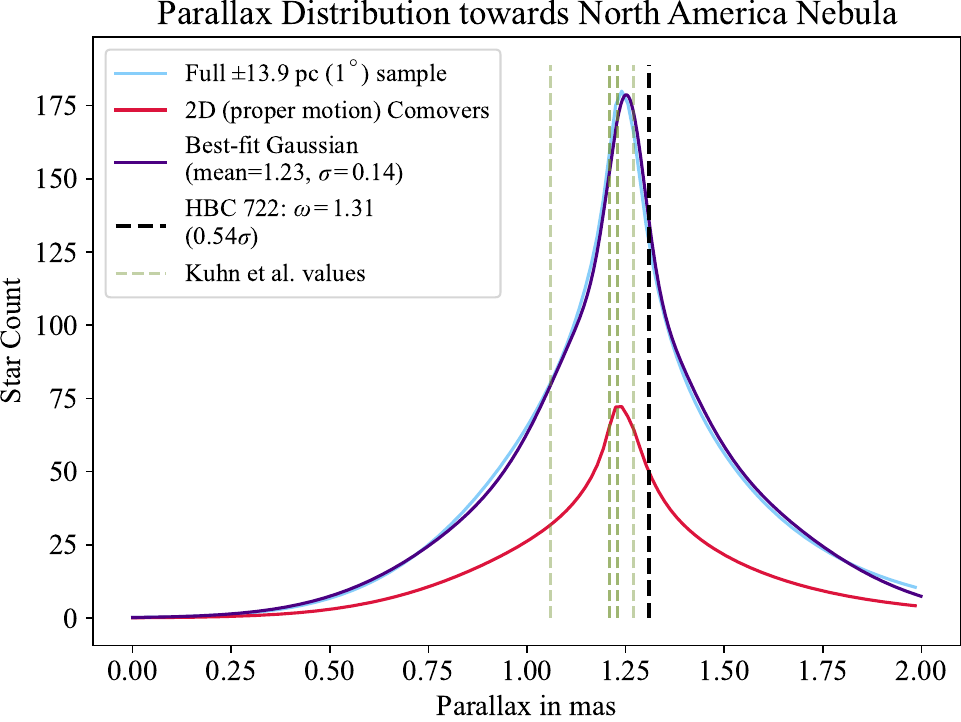} \\

 \includegraphics[width=6.2cm]{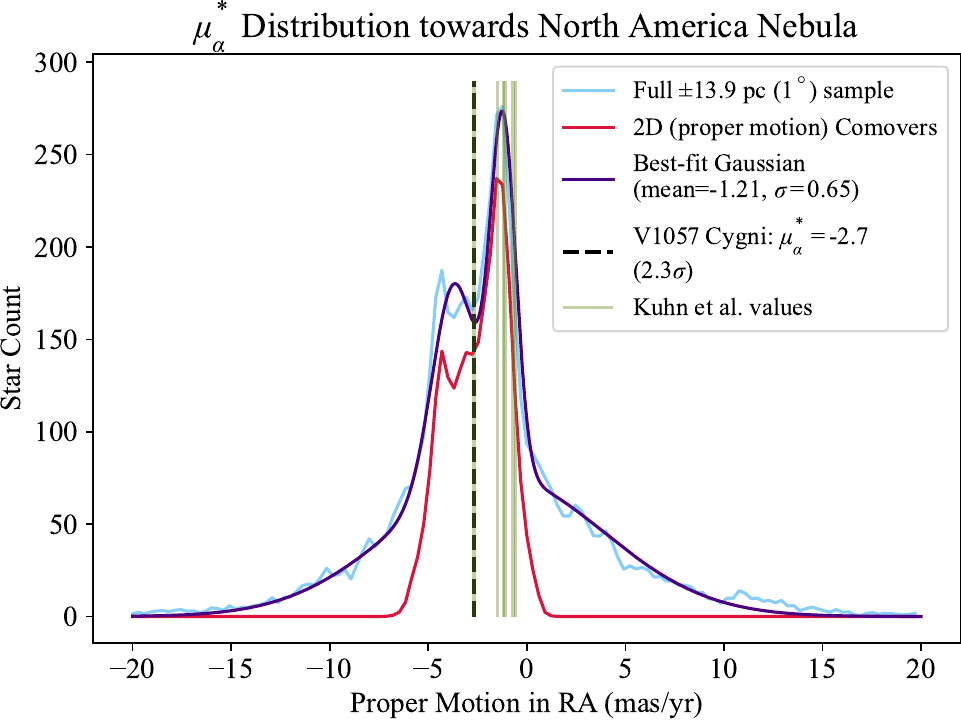}
 \includegraphics[width=6.2cm]{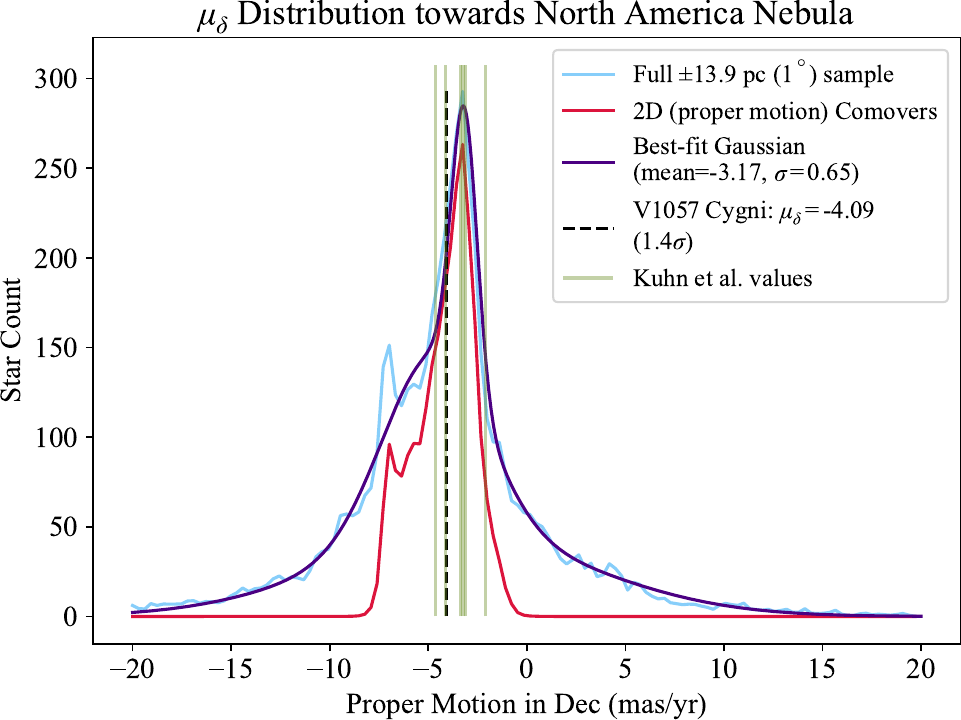}
 \includegraphics[width=6.2cm]{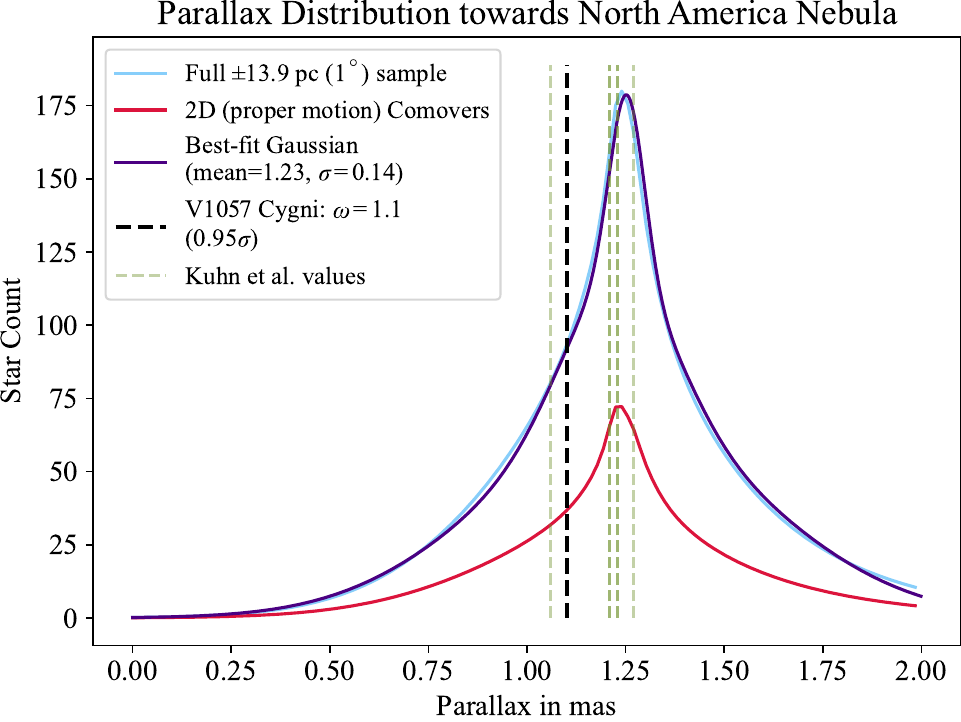} \\

 \includegraphics[width=6.2cm]{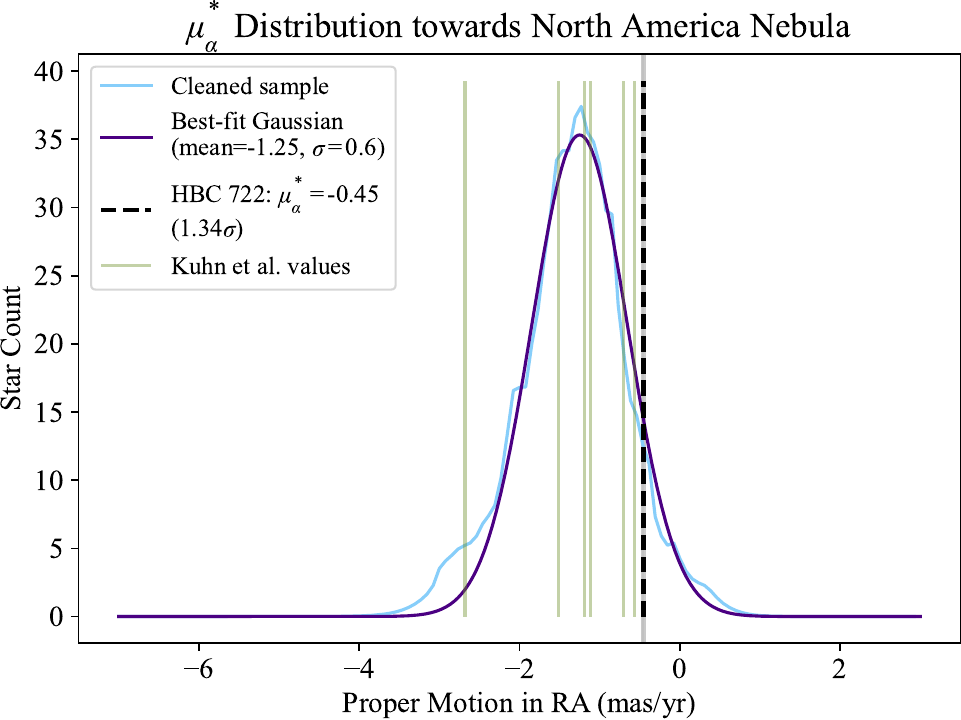}
 \includegraphics[width=6.2cm]{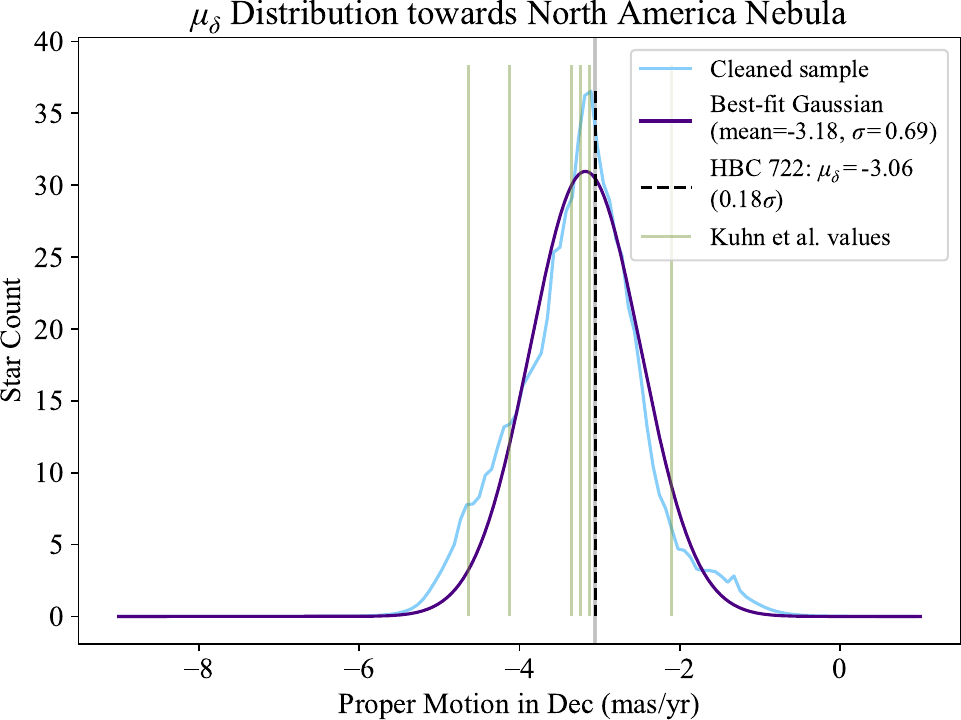}
 \includegraphics[width=6.2cm]{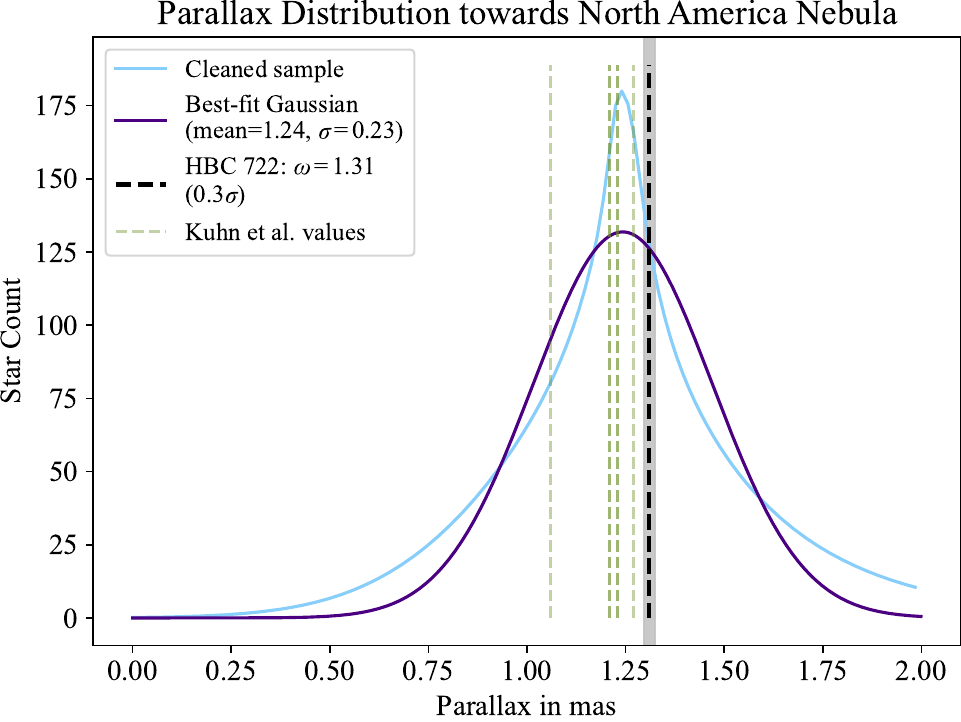} \\

 \includegraphics[width=6.2cm]{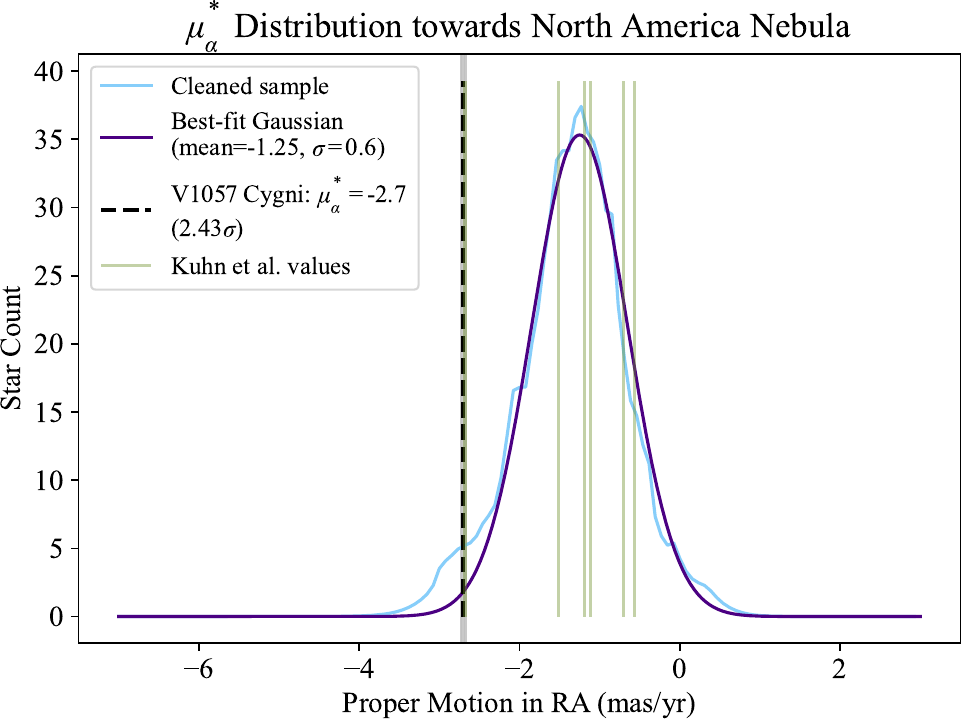}
 \includegraphics[width=6.2cm]{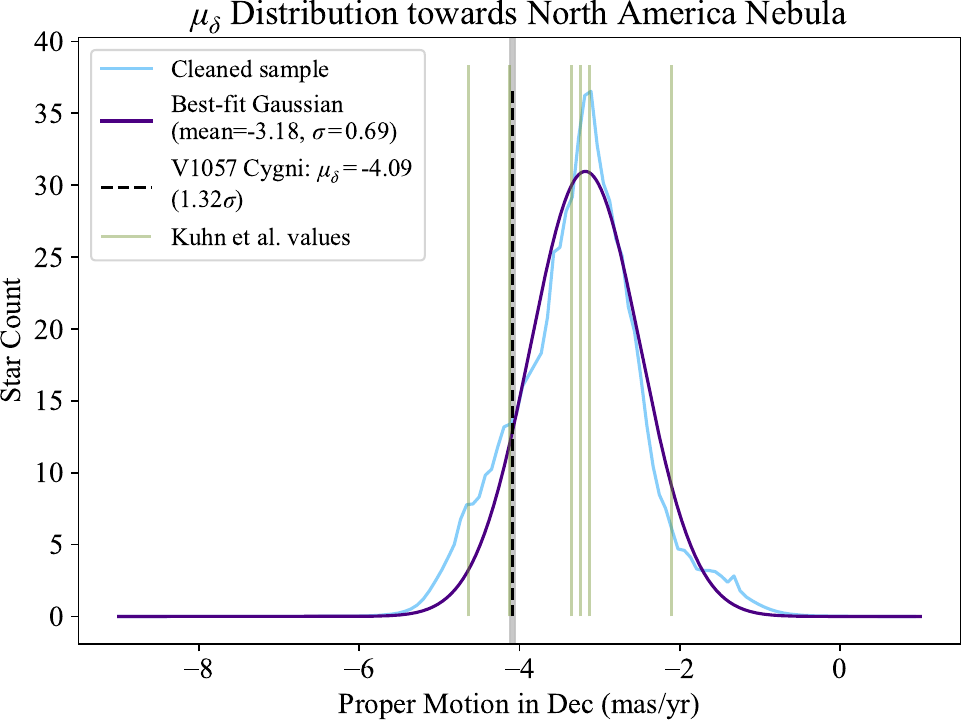}
 \includegraphics[width=6.2cm]{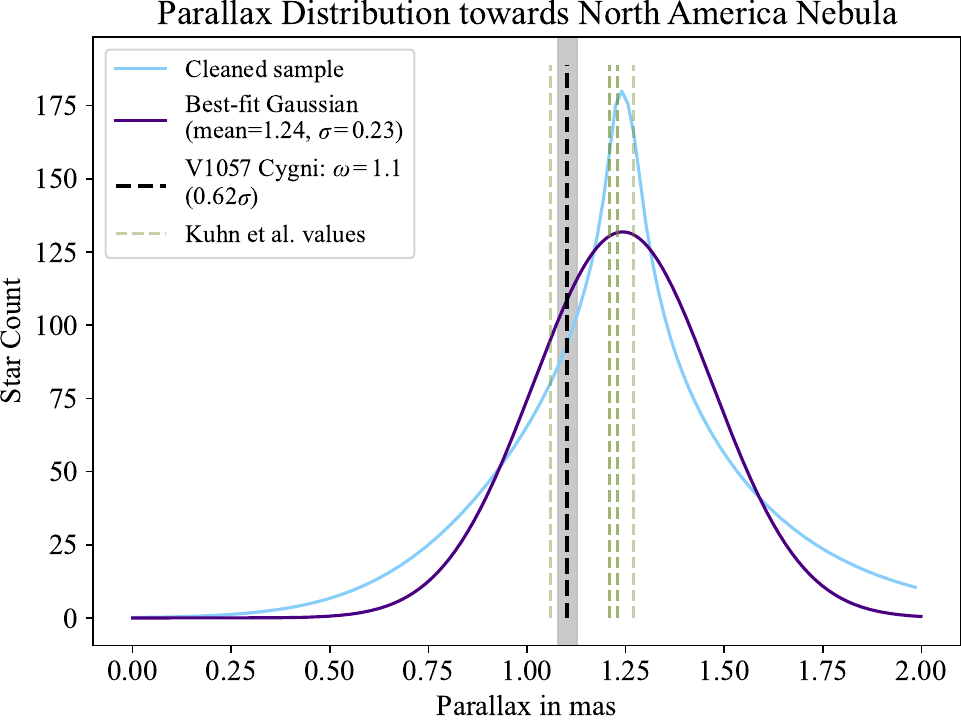} 
 
 \caption{Gaussian-weighted histograms of \pmra, \pmdec\ and parallax (across rows) for NAP with the full retrieved sample in the top two rows,  and the cleaned sample in the bottom two rows, 
 respectively compared to HBC 722 and V1057 Cyg. 
 For the full sample we fit a trimodal Gaussian (accounting for NAP, NGC 6997 and the field), and for the cleaned sample we fit a unimodal Gaussian. The tested FU Orionis star along with their $n-\sigma$ differences with respect to the cluster means, are indicated in each panel. For the cleaned samples we also indicate the $1\sigma$ measurement uncertainty band around each FU Ori.
    }
    \label{fig:histograms}
\end{figure}

\begin{figure}
 \includegraphics[width=6.2cm]{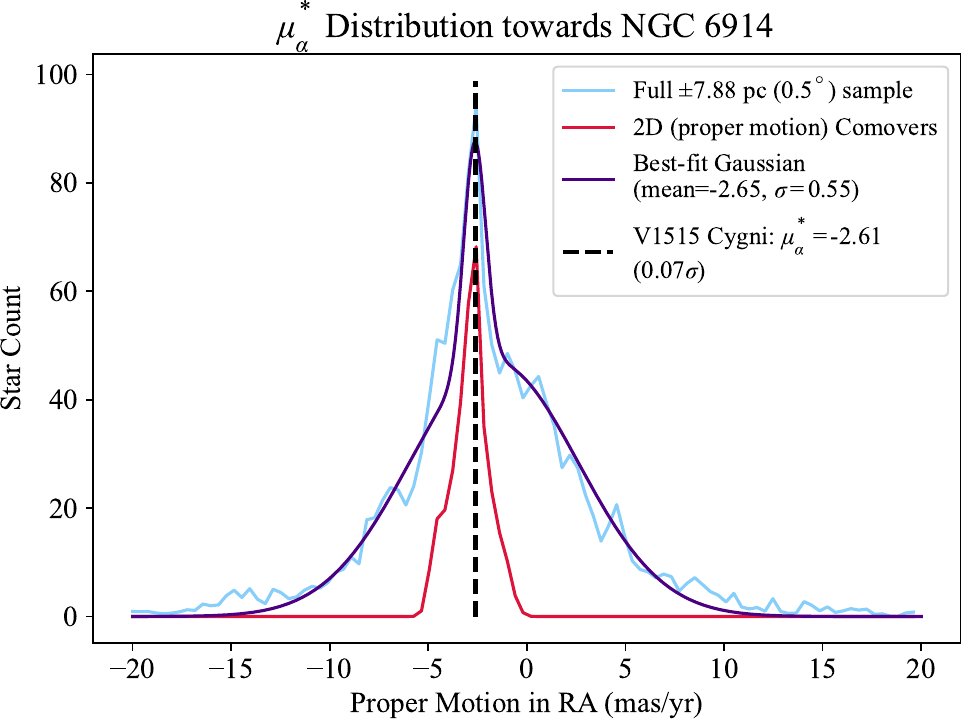}
 \includegraphics[width=6.2cm]{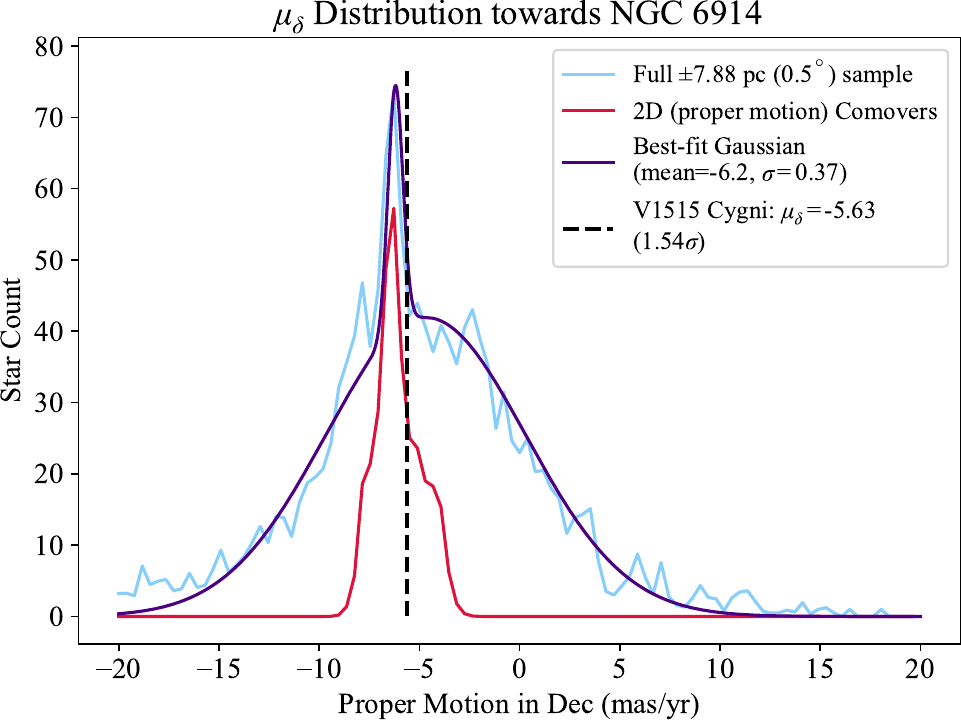}
 \includegraphics[width=6.2cm]{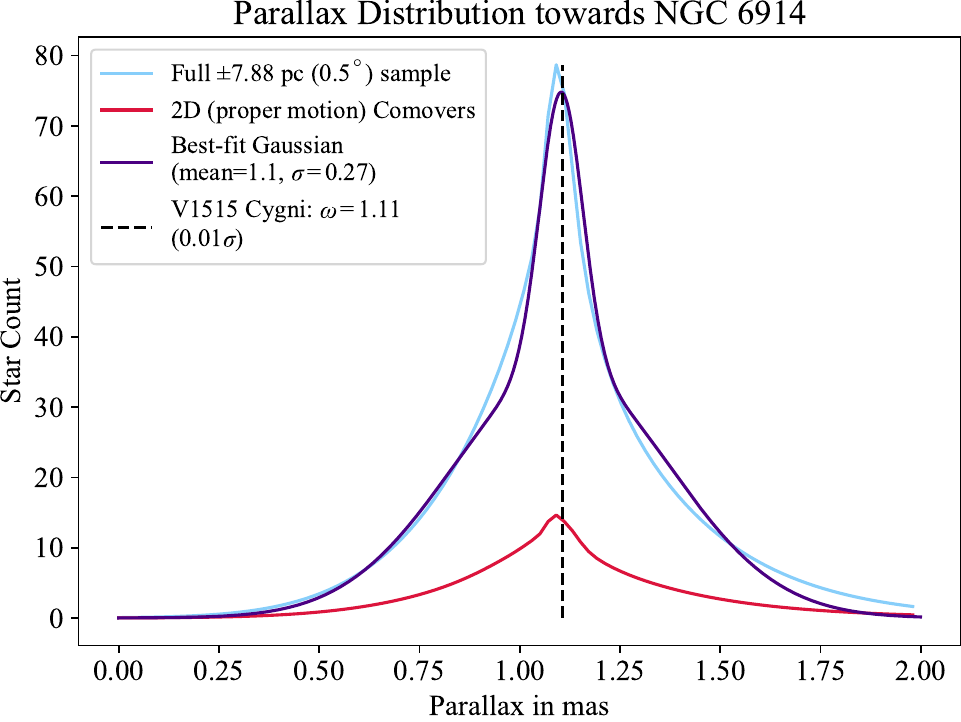} \\
 
 \includegraphics[width=6.2cm]{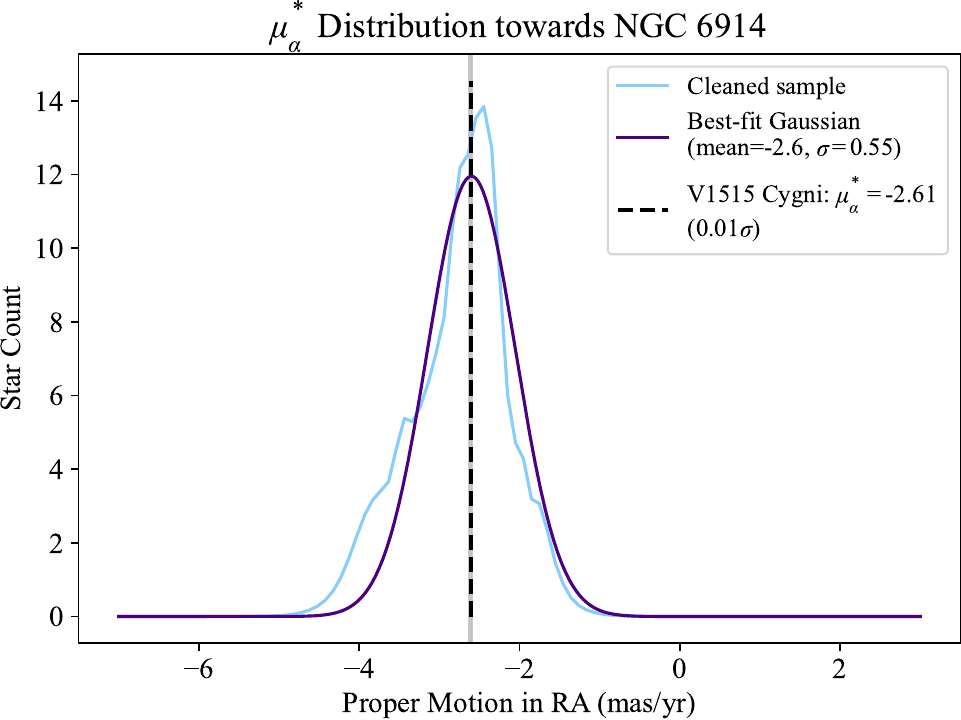}
 \includegraphics[width=6.2cm]{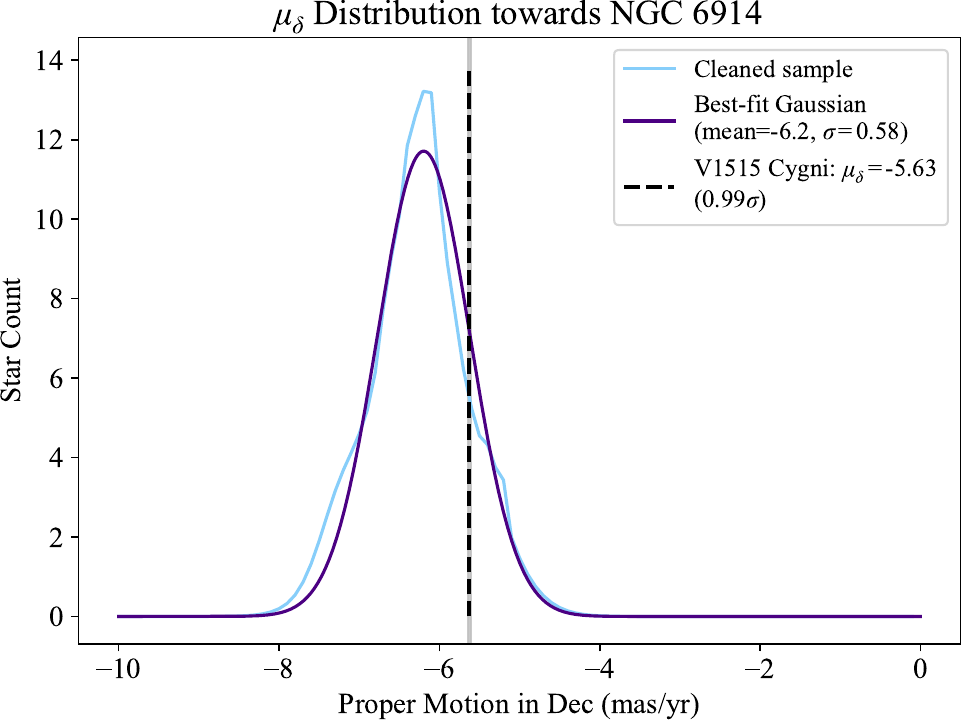}
 \includegraphics[width=6.2cm]{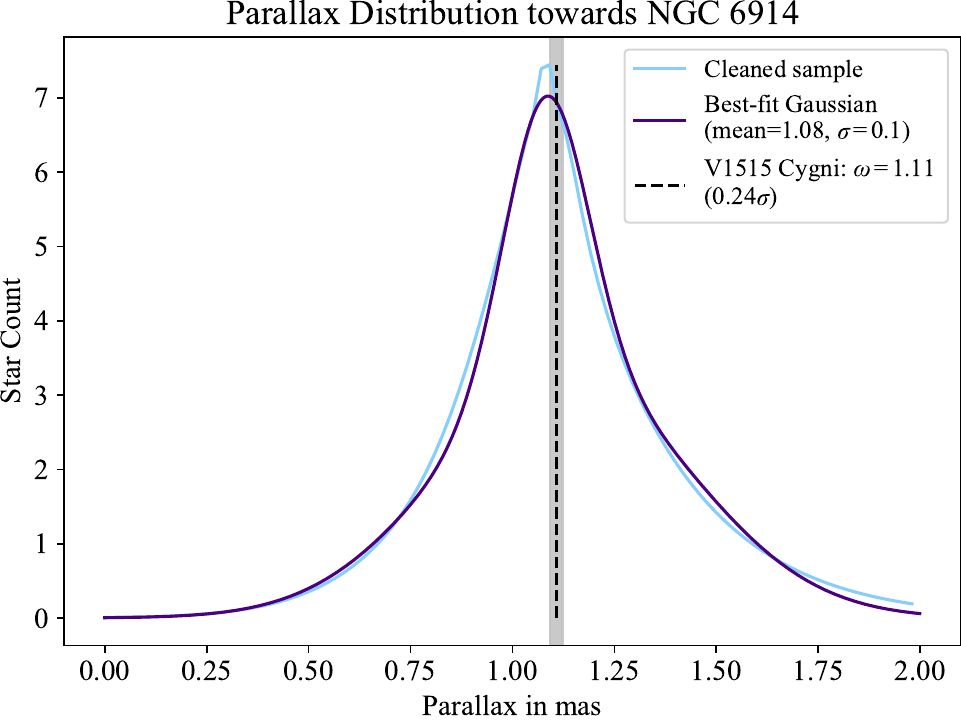} 

 \caption{Gaussian-weighted histograms of \pmra, \pmdec\ and parallax (across rows) for NGC 6914 showing the full retrieved sample (top row) and the cleaned sample (bottom row), 
 compared to V1515 Cyg.  For the full sample rows, we fit a bimodal Gaussian, and for the cleaned samples we fit a unimodal Gaussian. The tested FU Orionis star along with the $n-\sigma$ differences with respect to the cluster means, are indicated in each panel. The cleaned sample plots also indicate the $1\sigma$ measurement uncertainty band around each FU Ori.
    } 
    \label{fig:histograms2}
\end{figure}

\begin{figure}
\includegraphics[width=6.2cm]{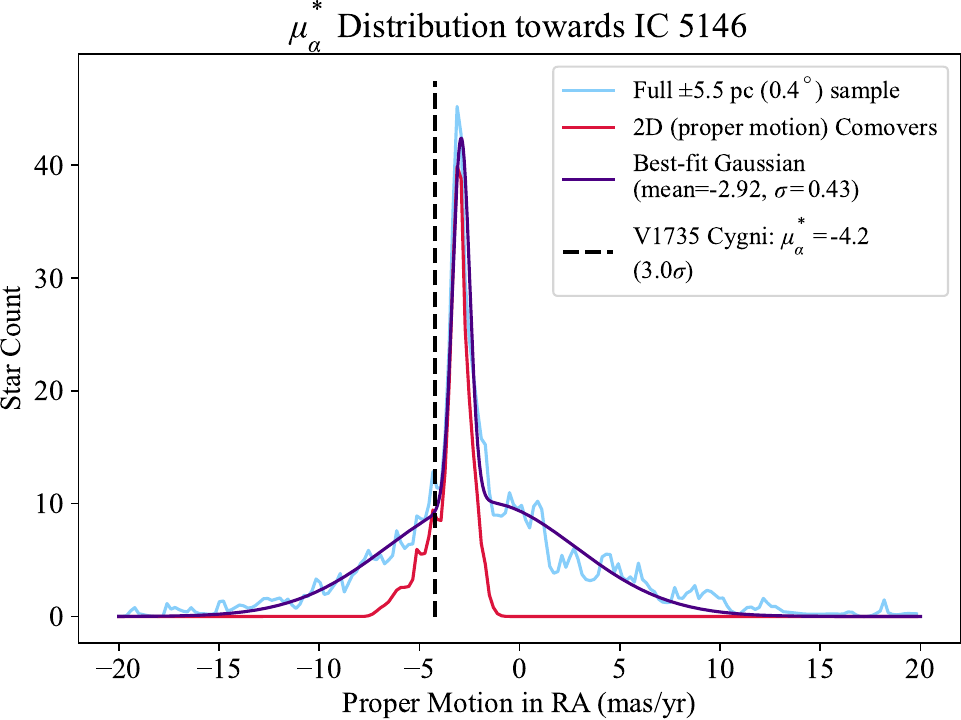}
 \includegraphics[width=6.2cm]{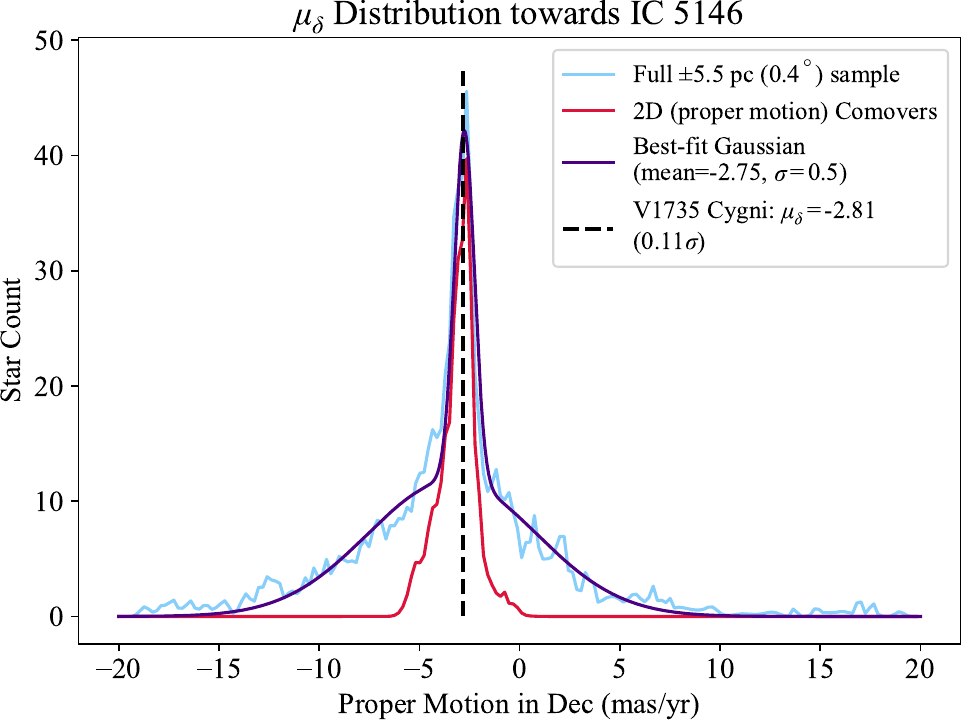}
 \includegraphics[width=6.2cm]{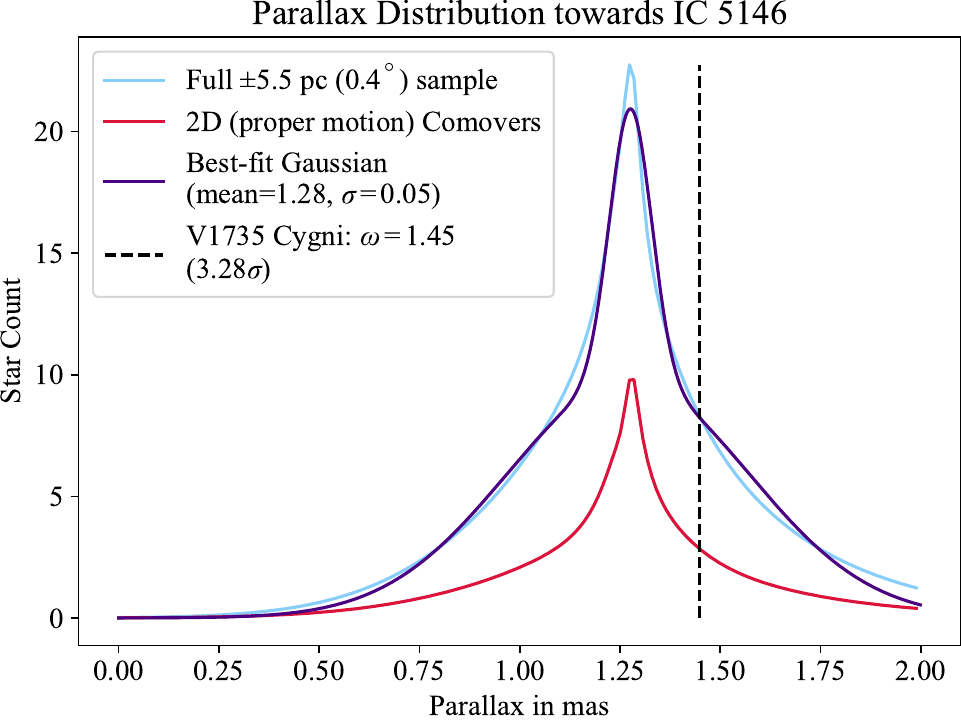} \\
 \includegraphics[width=6.2cm]{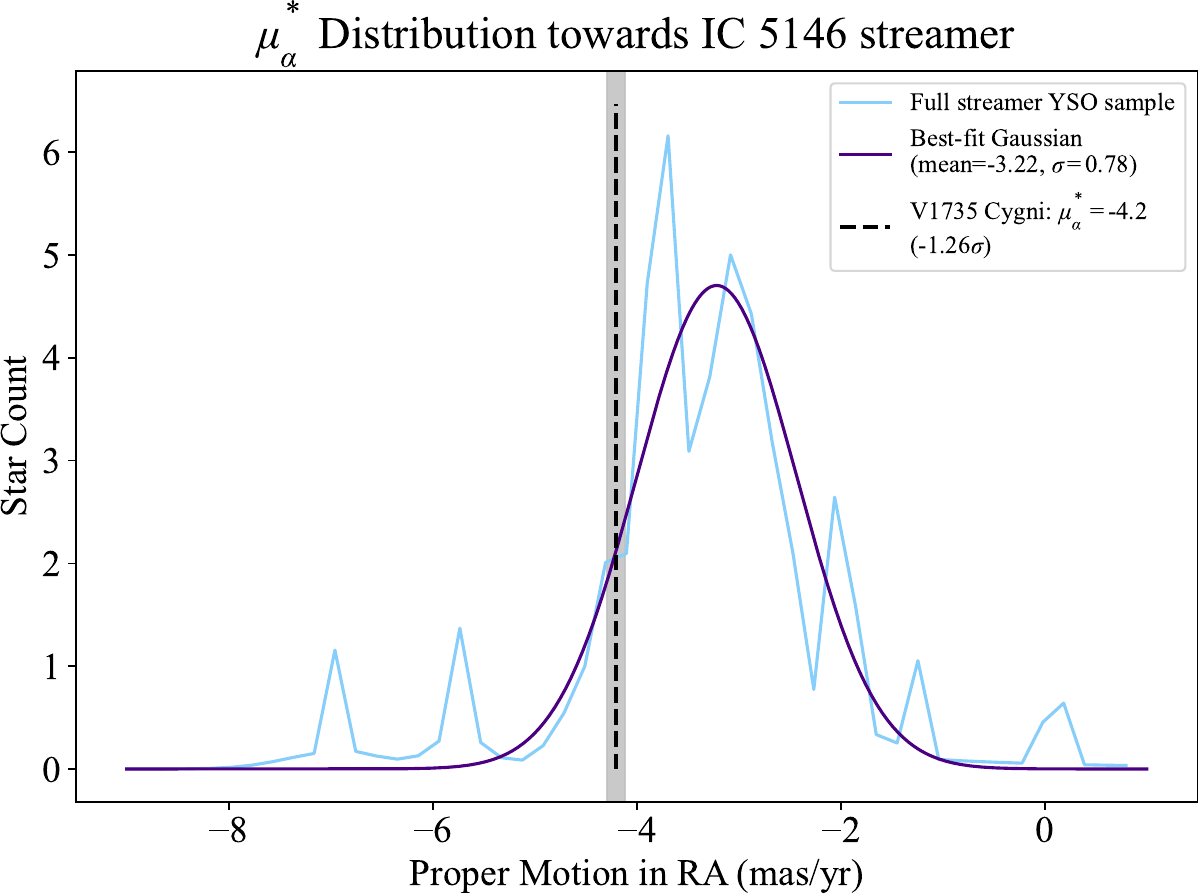}
 \includegraphics[width=6.2cm]{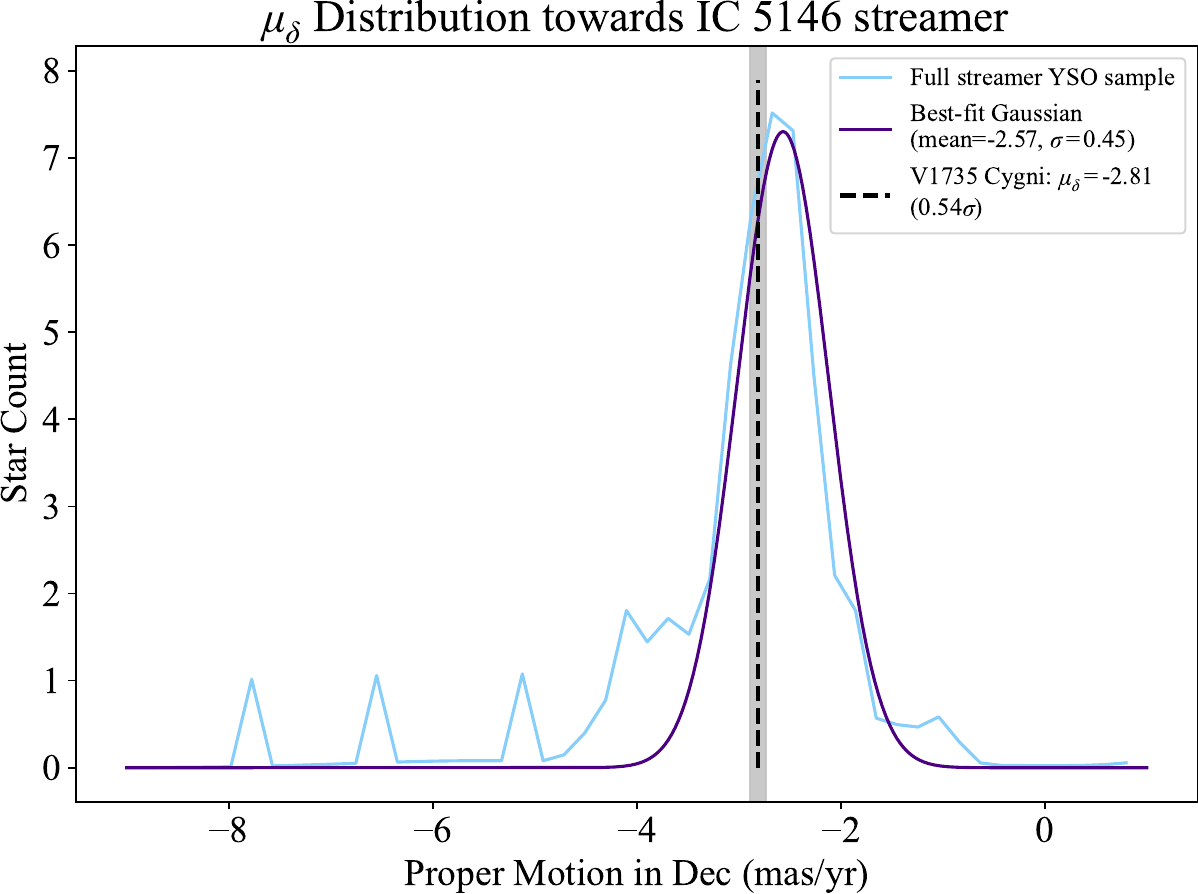}
 \includegraphics[width=6.2cm]{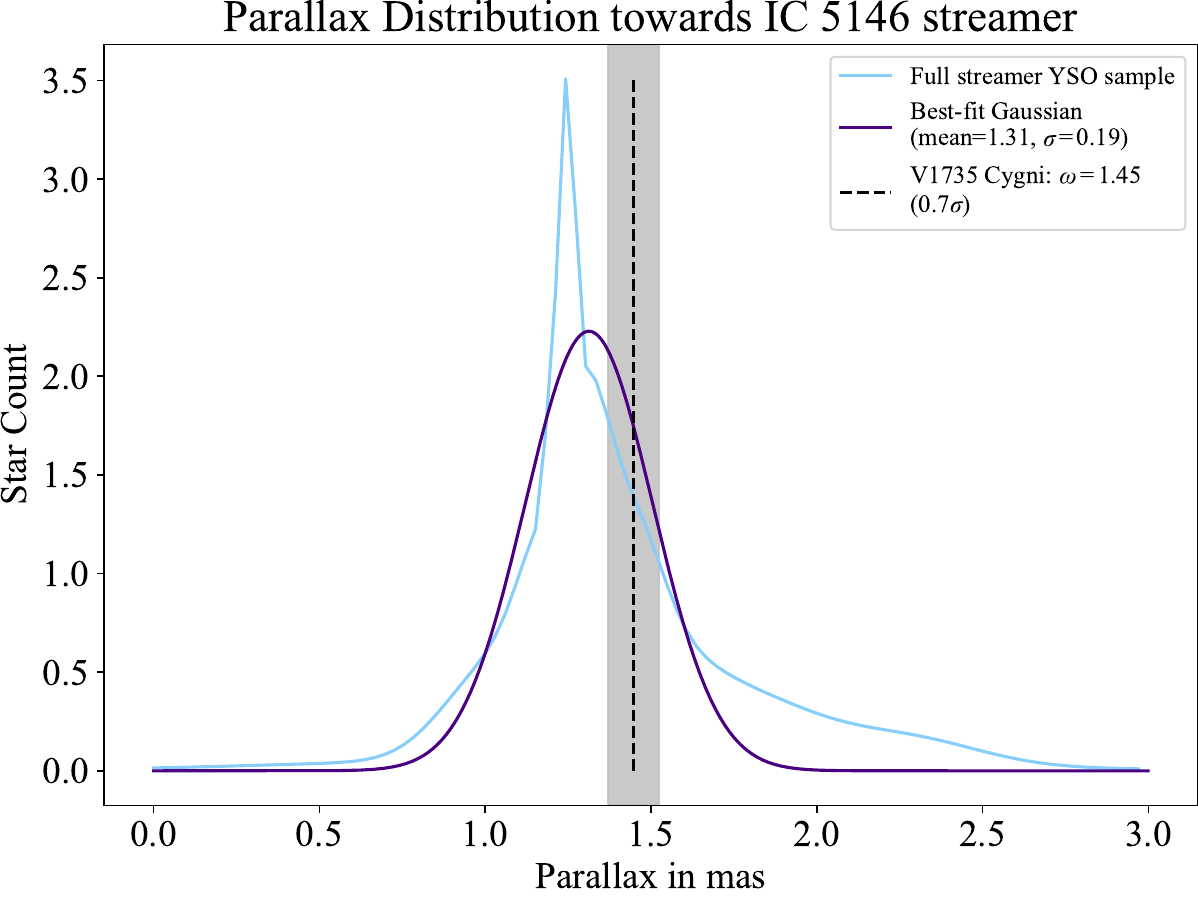} 
\caption{Same as \ref{fig:histograms} but
for the IC 5146 showing the full sample (top row) and the streamer region (bottom row), compared to V1735 Cyg. 
}
    \label{fig:histograms3}
\end{figure}

\begin{figure}
\includegraphics[width=6.2cm]{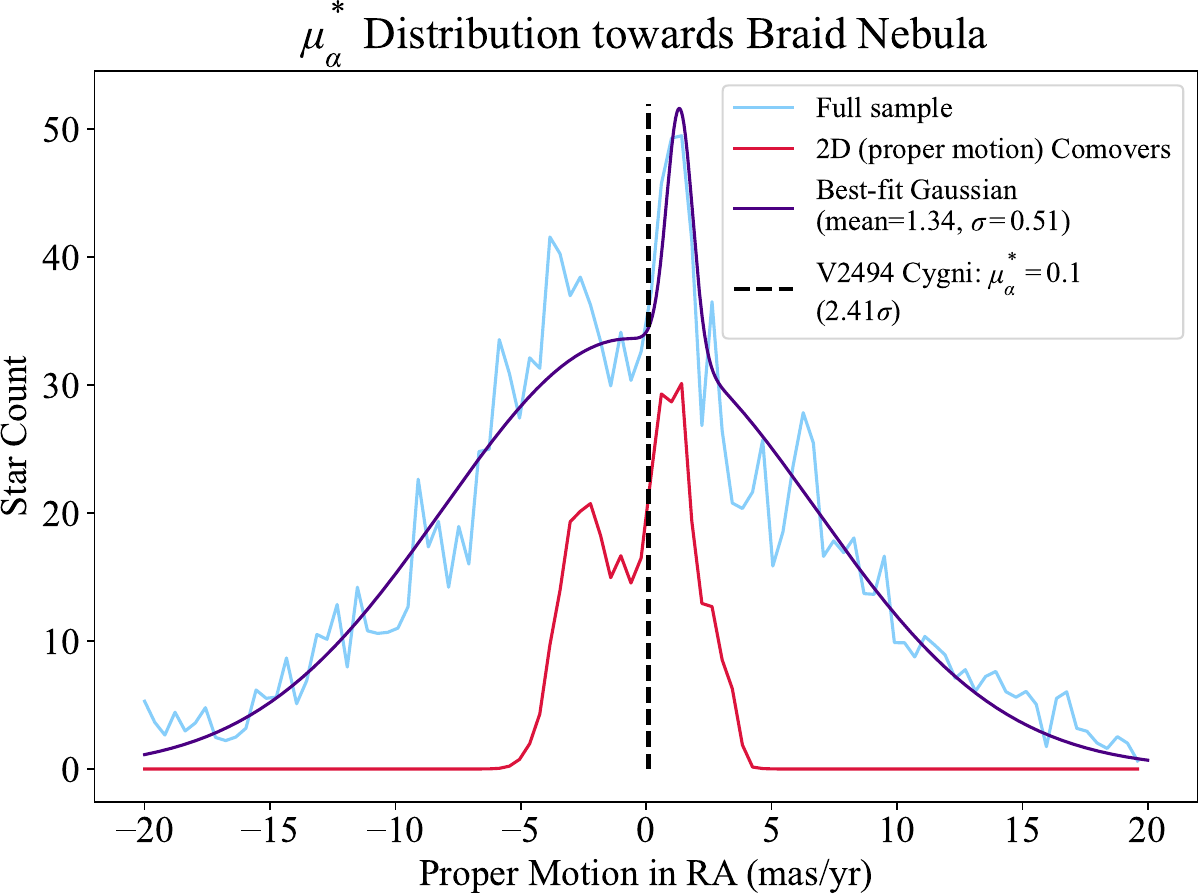}
 \includegraphics[width=6.2cm]{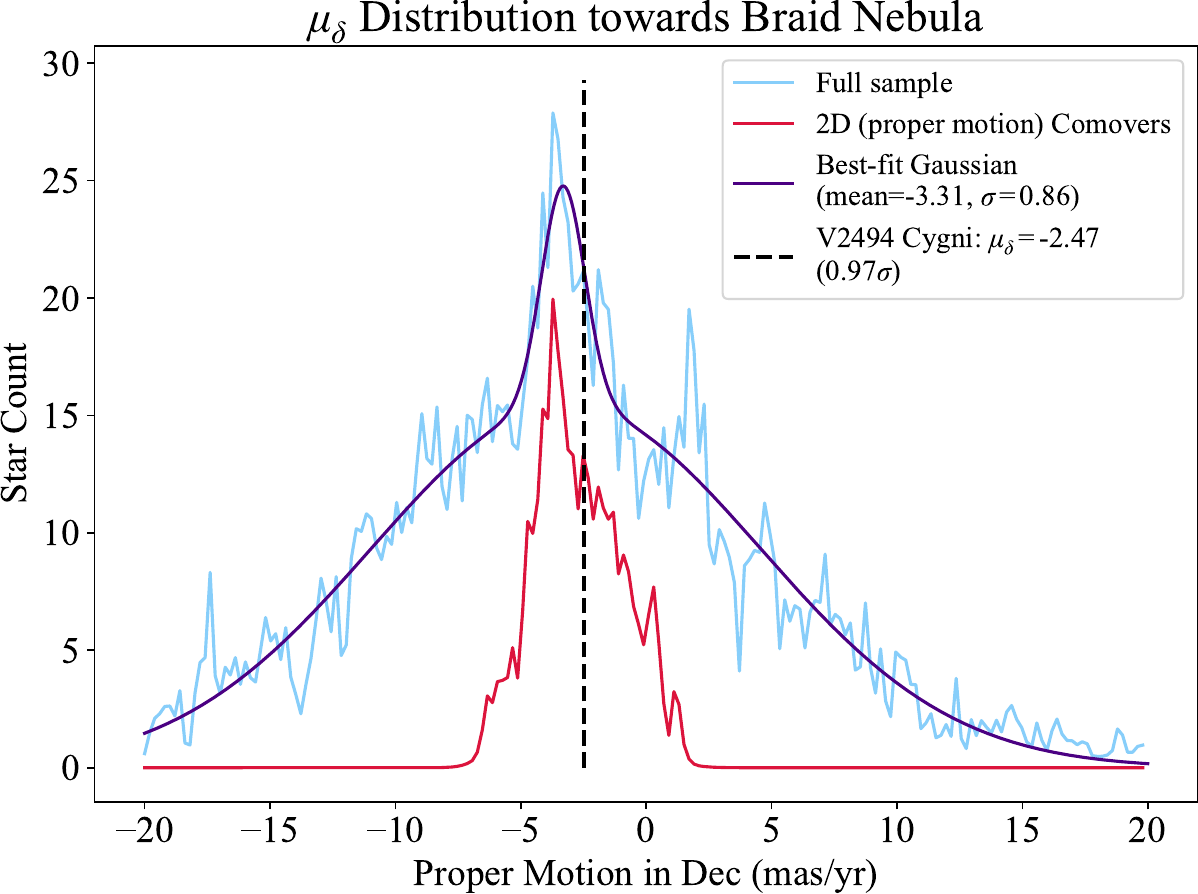}
 \includegraphics[width=6.2cm]{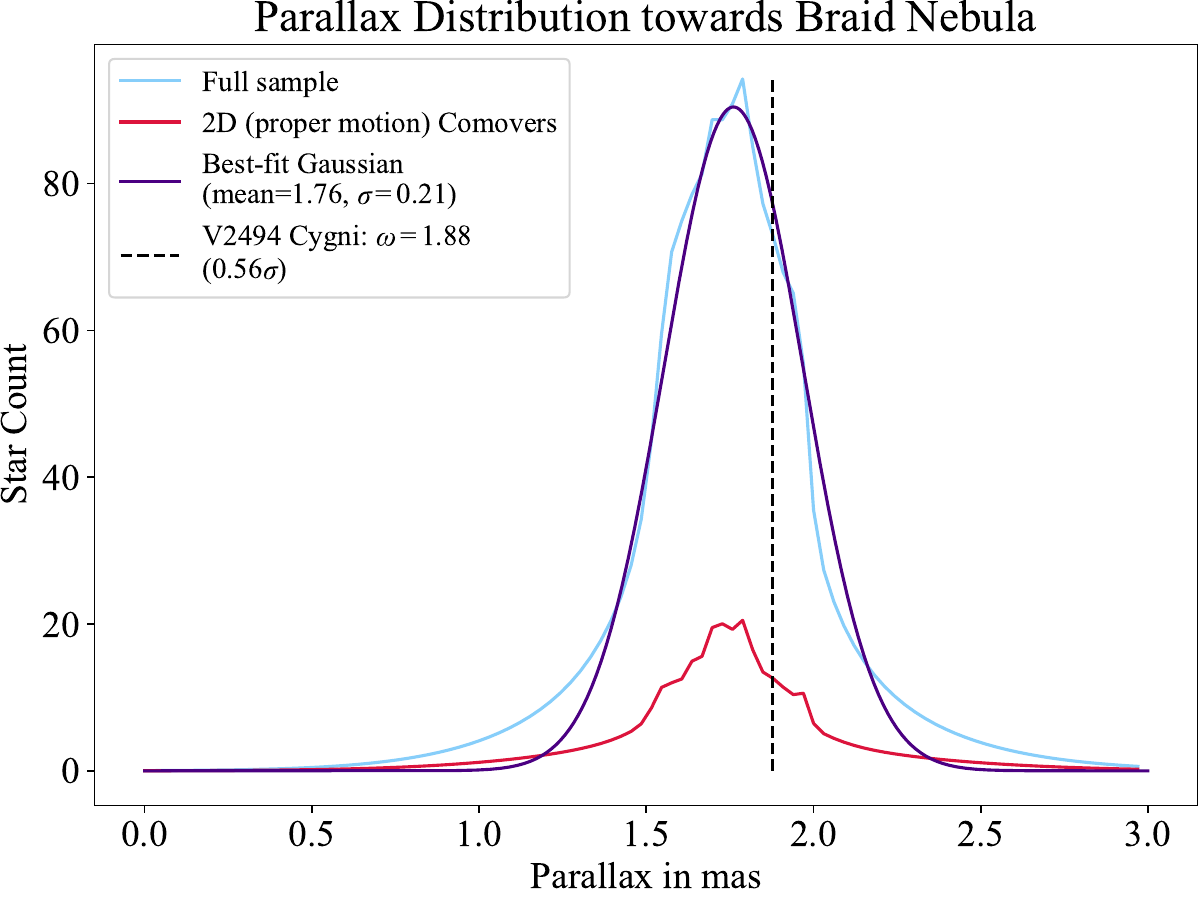} \\
 \includegraphics[width=6.2cm]{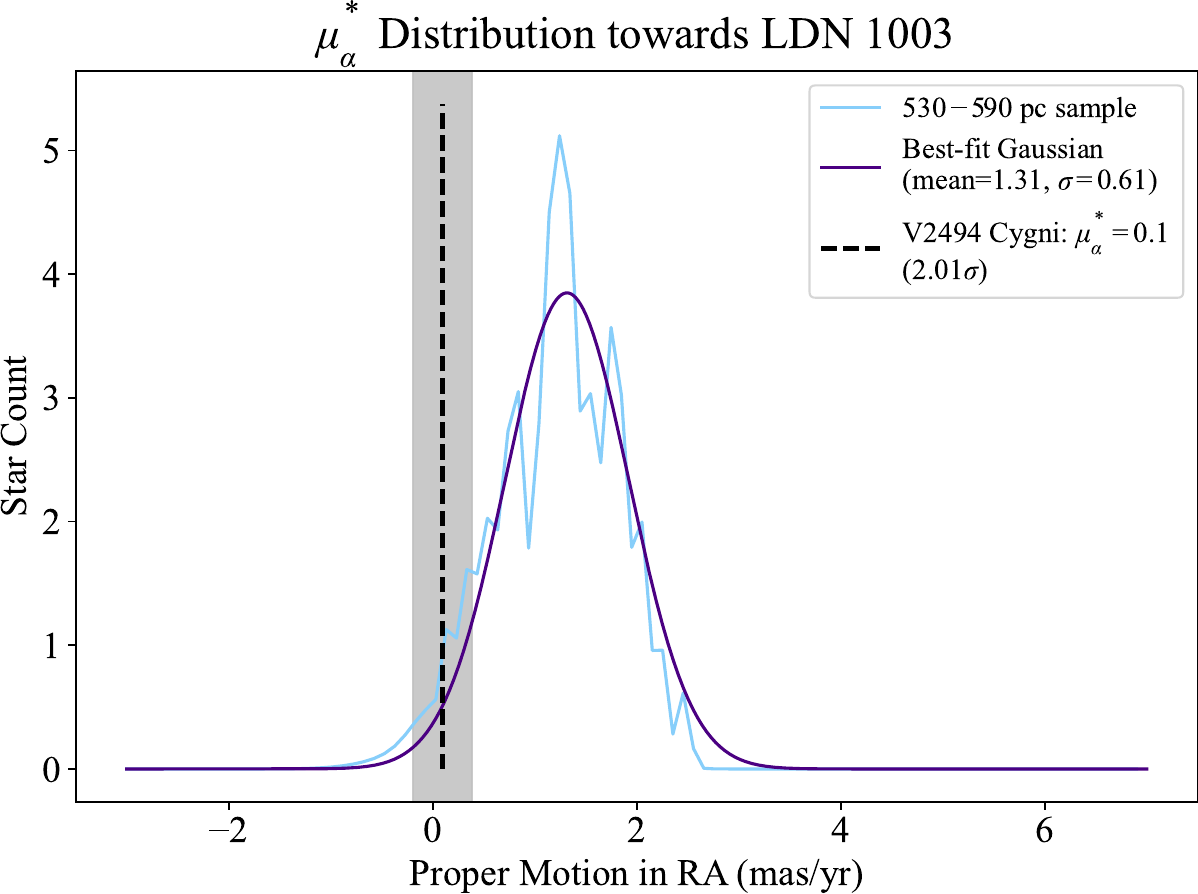}
 \includegraphics[width=6.2cm]{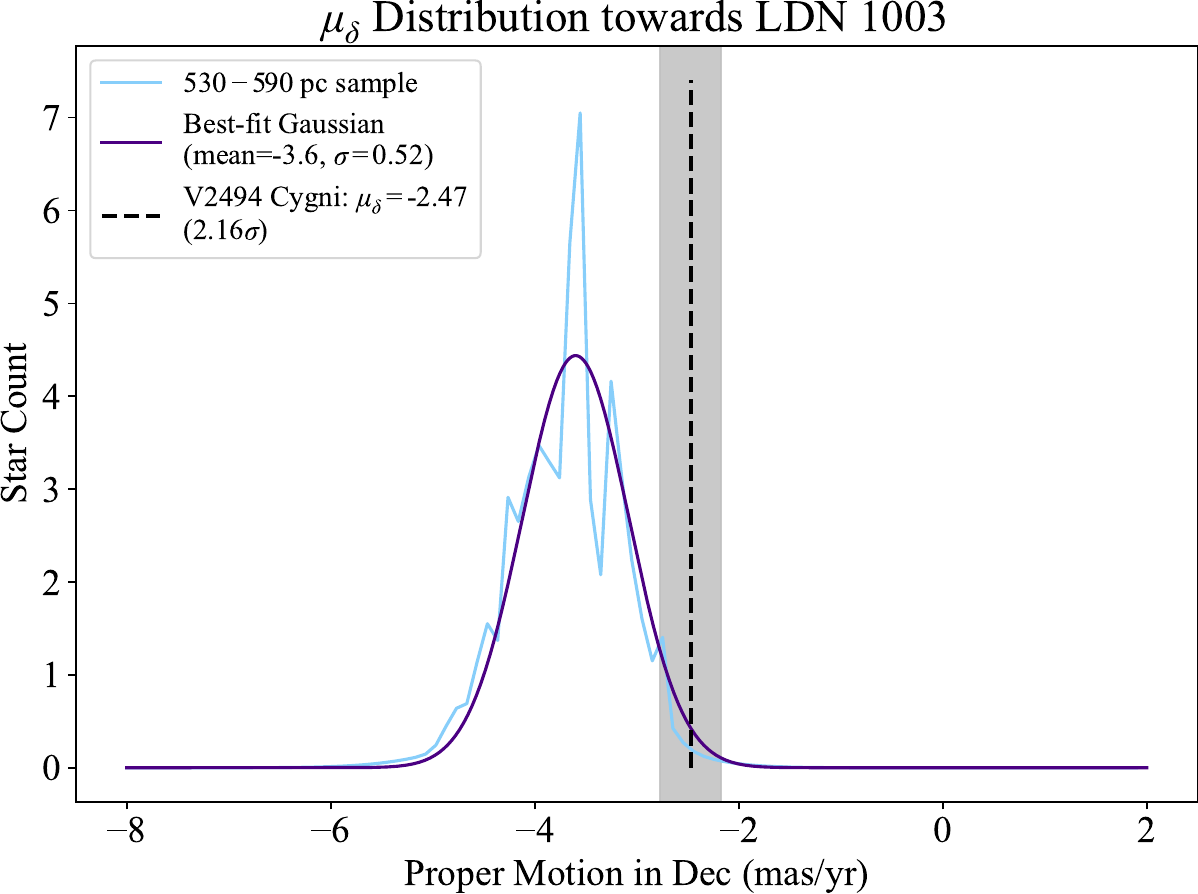}
 \includegraphics[width=6.2cm]{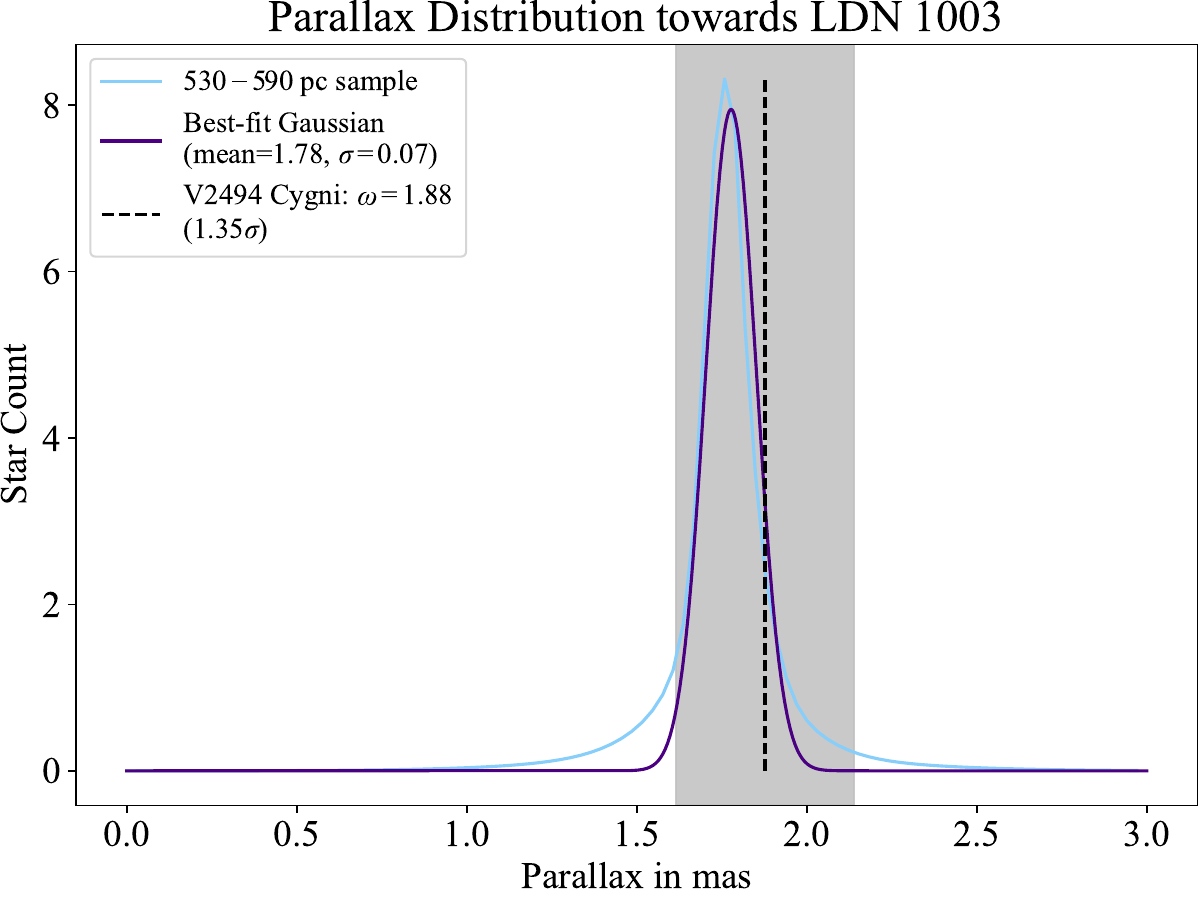} \\
\caption{ Same as \ref{fig:histograms} but for the Braid Nebula region with the full retrieved sample (top row), and the cleaned sample between 530--590 pc and within the kinematic concentration (bottom row), 
compared to V2494 Cyg. 
}

    \label{fig:histograms4}
\end{figure}

For each FU Ori source in our sample,
once a kinematically coherent stellar component has been identified, we summarize its properties in the form of Gaussian-smoothed histograms for the \pmra, \pmdec\ and parallax values. We fit bimodal Gaussians to the full retrieved samples, and unimodal Gaussians to the cleaned sample. Figures~\ref{fig:histograms} to ~\ref{fig:histograms4} shows these distributions along with the values for the respective FU Orionis stars associated with them.  There are five cases.

For V1057 Cygni, V1515 Cygni and HBC 722, the histograms show consistency to within $1.5\sigma$ in all but one of the nine histograms. The one outlier is a $2.43\sigma$ consistency for \pmra\ of V1057 Cygni relative to the NAP stellar component.

For V1735 Cygni, while \pmdec\ agrees very well with the mean of IC 5146, its \pmra\ is consistent only at the $3\sigma$ level, and the parallax is inconsistent at $>3\sigma$.  Better agreement with the \pmra\, \pmdec\ and parallax is found with stars in the identified IC 5146 streamer, to $<1.3\sigma$, indicating kinematic association. 

For LDN 1003/Braid nebula, since the correlation is high in \pmra\ and \pmdec, we additionally fit and check against a bivariate Gaussian in proper motion space. We find a $2.16\sigma$ consistency between the bivariate Gaussian fit and the proper motion of V2494 Cygni as indicated in Table~\ref{tab:n-sigma}. We also fit individual Gaussians to \pmra, \pmdec\ and parallax. We again use a unimodal Gaussian since our restrictions on parallax range and proper motions ensure that only the cluster population is present in the final sample. In these fits too, V2494 Cygni is consistent in all three parameters with LDN 1003 to within $\lesssim 2.2\sigma$.

\subsection{Cluster/Stellar Group Kinematics}

We attempt to crossmatch our derived cluster properties (mean and dispersion of proper motions) to literature catalogues of star clusters analysed using \textit{Gaia}, primarily those of \cite{Hunt_Reffert_2023} and \cite{Cantat-Gaudin2020}. We only find a corresponding entry for NGC 6914 (identified in both catalogues as Dolidze 8). Our derived values for proper motion and parallax (\pmra, \pmdec, $\pi$) = (-2.6, -6.2, 1.08) are consistent with both \cite{Hunt_Reffert_2023} finding (\pmra, \pmdec, $\pi$) = (-2.54, -6.2, 1.02) and \cite{Cantat-Gaudin2020} finding (\pmra, \pmdec, $\pi$) = (-2.63, -5.92, 1.02) to within $1\sigma$.

The absence of the other clusters can be attributed to multiple reasons -- the NAP is likely composed of multiple members \citep{Kuhn_NAP_2020}, which would then either be difficult to resolve or would constitute several separate entries. For NAP, we thus compare our derived properties to those by \cite{Kuhn_NAP_2020} for NAP, and find a good match (Fig~\ref{fig:histograms}). Both the IC 5146 streamer and Braid Nebula are elongated, sparsely populated structures and are not likely to stand out as well as an overdensity if examined in bulk within a more crowded field. Thus any precise astrometry for these two groups is absent in prior literature. 

For consistency, we perform the kinematic analysis for all four groups using our own methods.
However, unlike our previous similar analysis of the Orion FU Ori objects relative to nearby clusters \citep{TR_LAH_2024}, here we do not have abundant reliable radial velocity (RV) data for the cluster stars, either from Gaia DR3 or the APOGEE survey. Our histograms for parallaxes also clearly indicate moderate uncertainties. Additionally, for the cleaned samples, we check our parallax-derived distances with the photogeometric distances (based on star colours constraining the allowed range of absolute magnitudes and hence distances based on stellar evolution models) derived by \citep{Bailer_Jones_2021}. In all cases we find the distribution of photogeometric distances to have a tail towards higher distance ranges ($\sim 2000-3000$ pc), whereas the cleaned sample parallax distances do not extend beyond $~1200$ pc by our selection constraints. While this could be a potential issue in assessing distance consistency, we observe that for both the parallax distance and photogeometric distance, the peak and width of the distributions in the range of interest remain consistent to less than the typical distance uncertainties. We thus use the inverse-parallax distances in our histograms and for $n-\sigma$ calculations.

We can thus ascertain kinematic association reliably only in two dimensions along the proper motions. Due to the absence of RV, we estimate the total velocity dispersion by extrapolating from the dispersions along \pmra\ and \pmdec, which are themselves directly obtained from a Gaussian fit to the cluster component in the \pmra\ and \pmdec\ histograms. 

For NAP (containing V1057 Cyg and HBC 722), these come out to be $0.6$ and $0.69$ mas yr$^{-1}$, nearly equal as would be expected from isotropy. This translates to $2.50$ km/s at an adopted distance of $795$ pc. We take the dispersion along RV to be the same average value, thus giving the total velocity dispersion to be $4.34$ km s$^{-1}$. 

We do not impose any proper motion error cuts anywhere as the final histograms are Gaussian-smoothed i.e. spread out in proportion to the errors. However, we also observe that almost all ($\approx 96\%$) of the stars have \pmra\ and \pmdec\ errors less than $0.68$ mas yr$^{-1}$, which is almost equal to the velocity dispersion itself along that direction.

For NGC 6914 (containing V1515 Cyg), the dispersions along \pmra\ and \pmdec\ are $0.55$ and $0.58$ mas yr$^{-1}$ respectively, again almost equal. Assuming the dispersion along RV to be the root-mean-square of these two values, the total velocity dispersion comes out to be $2.83$ km s$^{-1}$, setting the 2D proper motion difference limit to be $8.52$ km s$^{-1}$. Here, the fraction of stars with errors in both \pmra\ and \pmdec\ less than $0.45$ mas yr$^{-1}$ is $\approx 92\%$.

For IC 5146 and its streamer (containing V1735 Cyg), the dispersions along \pmra\ and \pmdec\ are $0.78$ and $0.45$ mas yr$^{-1}$ respectively which differ slightly as expected from the asymmetric streamer structure. Assuming the dispersion along RV to be the root-mean-square of these two values, the total velocity dispersion comes out to be $2.55$ km s$^{-1}$. About $88\%$ of the stars in the sample have both errors in \pmra\ and \pmdec\ to be less than 0.46 mas yr$^{-1}$. 

For the Braid Nebula region (containing V2494 Cyg), the dispersions along \pmra\ and \pmdec\ are equal to 0.61 and 0.52 mas yr$^{-1}$. About 89\% of the total stars in the sample have their errors in both \pmra\ and \pmdec\ less than this number. However, as can be seen in the VPD of the region in Fig~\ref{fig:braid-vpd-WISE} ~and~\ref{fig:ldn1003-pm-dss}, the distribution is clearly asymmetric elliptical. The proper motions in RA and Dec have a correlation coefficient $\rho = -0.89$. The dispersions of the fitted bivariate Gaussian along the major and minor axes are 0.73 and 0.17 mas yr$^{-1}$ respectively.

\subsection{Colour-Magnitude Diagrams}
\begin{figure}[h!]
	\includegraphics[width=9cm]{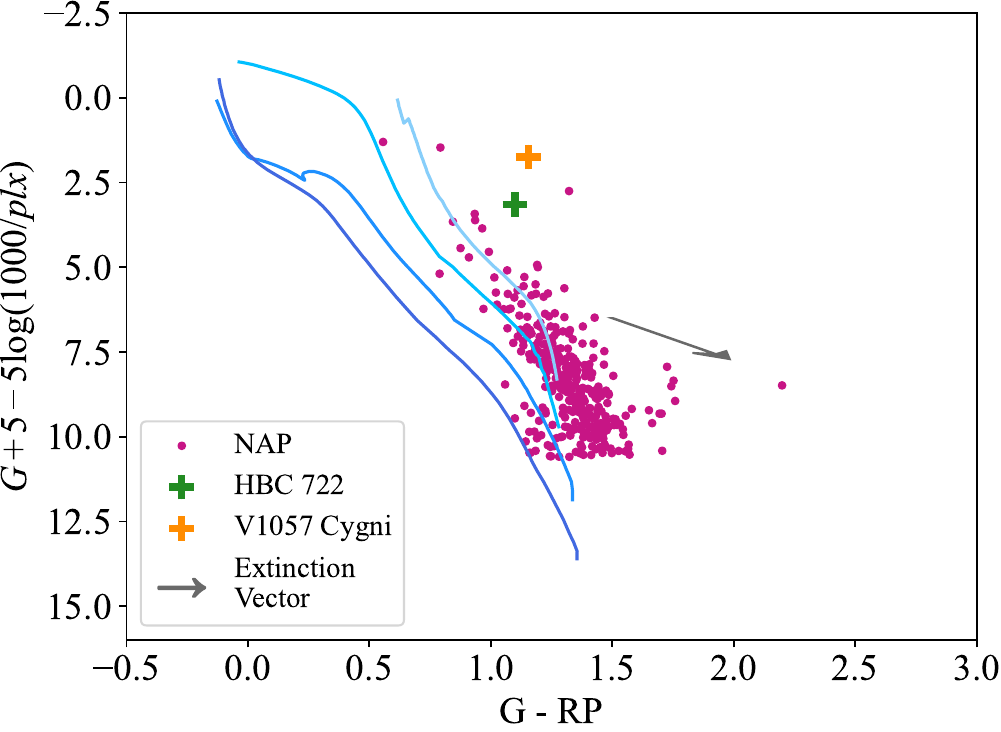}
    \includegraphics[width=9cm]{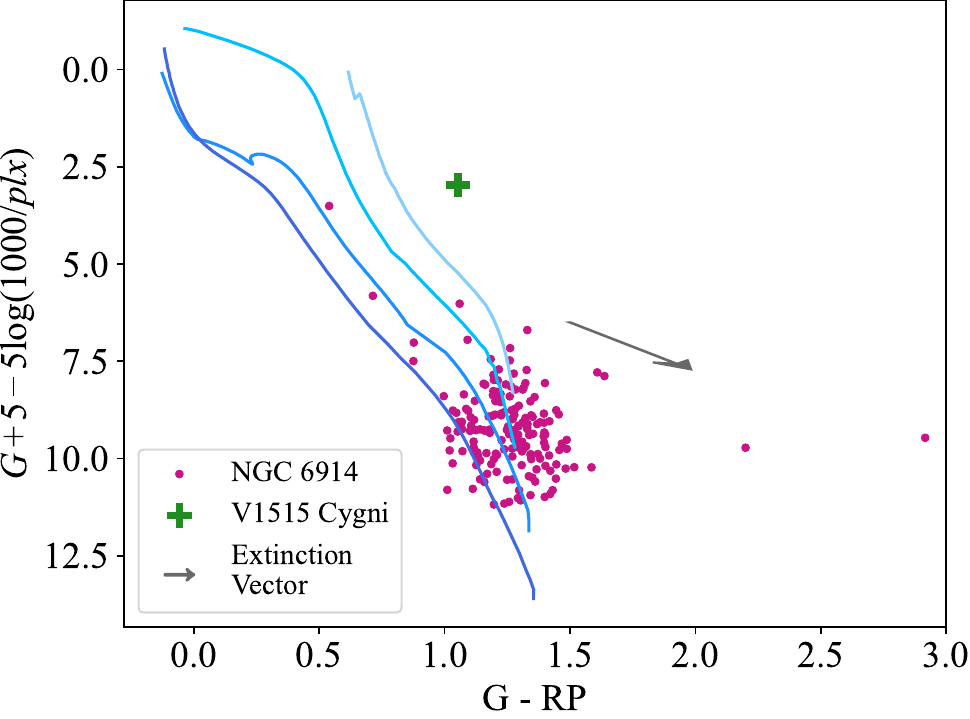} \\
    \includegraphics[width=9cm]{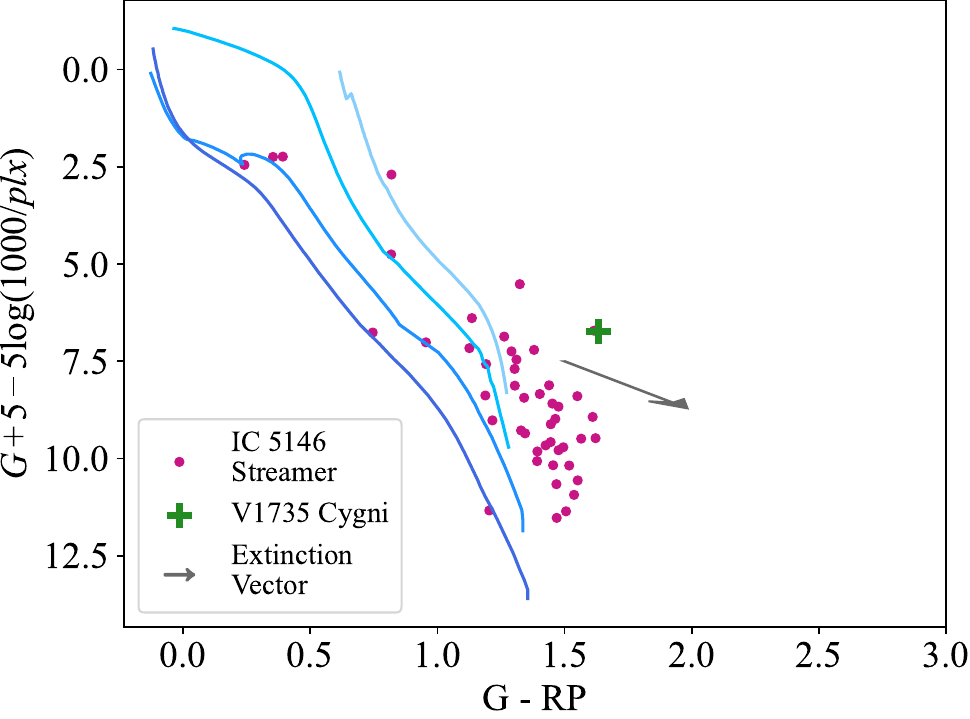}
    \includegraphics[width=9cm]{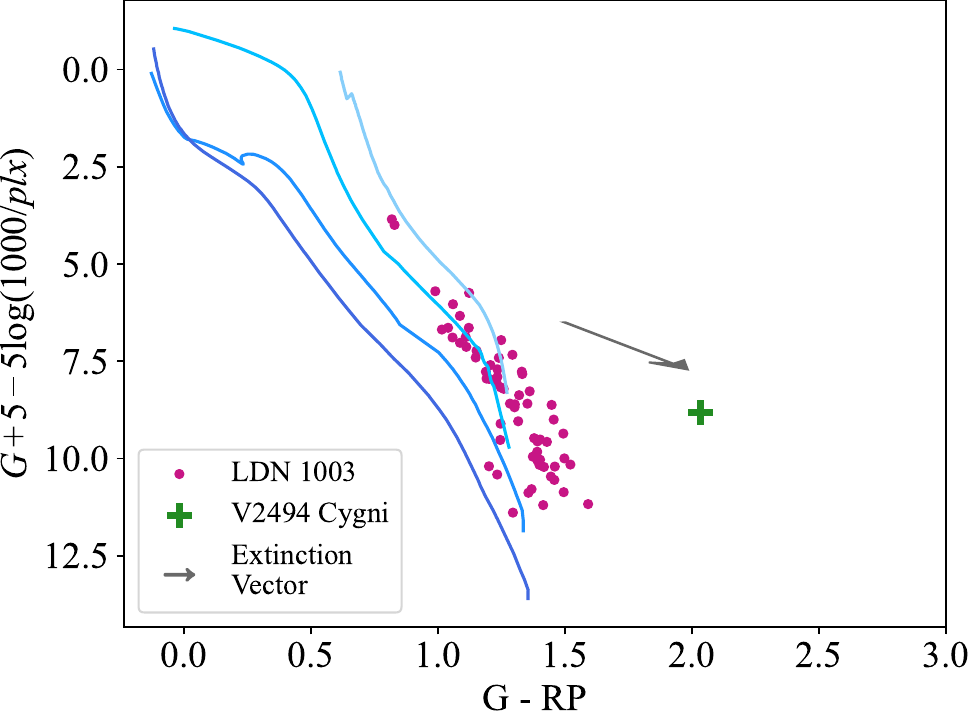}

    \caption{Colour-magnitude diagrams (CMDs) of the four FU Ori targets along with their respective kinematically associated clusters or groups. We also overlay theoretical isochrones from \texttt{PARSEC}, of ages = 0.1, 1, 10 and 100 Myr. The extinction and reddening arrow corresponds to $A_G = 1$ mag, with stars expected to be displaced from young isochrones in the indicated direction. We find all four of the kinematic clusters to have a significant pre-main sequence  component, though NGC 6914 has more main-sequence contamination than the other three regions.
    }
    
    \label{fig:cmd}
\end{figure}

Fig~\ref{fig:cmd} shows the colour-magnitude diagrams for each of the component clusters with their corresponding kinematically associated FU Ori star(s). We also overlay \texttt{PARSEC}\footnote{\url{https://stev.oapd.inaf.it/cgi-bin/cmd}} isochrones \citep{Bressan_2012} of ages  0.1, 1, 10 and 100 Myr on these. In all four clusters, we find a significant pre-main-sequence (PMS) population.  Except for NGC 6914, the kinematically selected stars populate a track parallel to the main sequence but just above it, matched well by isochrones of 0.1--10 Myr; NGC 6914 has somewhat more non-pre-main sequence contamination. The FU Ori targets themselves lie well above the PMS tracks, due to the dominance of accretion luminosity.

\section{Conclusions}
We have examined the evidence for physical association of five known FU Ori stars in the Cygnus region of the sky with stars, stellar clusters, and star-forming regions that are proximal to them.  To do so, we have checked for kinematic consistency in parallax and proper motion space using astrometric information from \textit{Gaia} DR3. We first looked for known clusters in the vicinity of each target, then we performed cluster selection using criteria placed on a vector point diagram (VPD) centered on the coordinates and the literature distance to the cluster. This method is applied to stars associated with the North America Nebula (physically close to HBC 722 and V1057 Cygni) and NGC 6914 (physically close to V1515 Cygni). The search criteria from \textit{Gaia} are adapted in each case to account for uncertainties in parallax and the presence of multiple clusters in the same region of the sky.

We check for kinematic and positional consistency by fitting bivariate Gaussians in RA--Dec and proper motion space to the cluster members alone, as well as by fitting univariate bimodal Gaussians to the cluster + field population for each proper motion component and the parallax. All three among HBC 722, V1057 Cygni and V1515 Cygni are positionally consistent to $<2.5\sigma$, in proper motions to $<1.5\sigma$ and $<1\sigma$ in parallax (which has the broadest distribution due to relatively higher measurement uncertainties).

For V1735 Cygni, we initially identified IC 5146 as a likely physically associated young cluster; however, its physical separation in position ($\gtrsim 100$ pc) renders this unlikely. Instead, we demonstrate the kinematic consistency of V1735 Cygni and the streamer structure to the west of IC 5146, which we astrometrically characterize using the YSOs cataloged by \cite{Harvey_2008}. V1735 Cygni is located within this streamer, as inferred by visual inspection of images from the \textit{Herschel} mission \citep{herschel_overview} and YSO positions, and it has a parallax consistent to within $0.7\sigma$. Furthermore, its proper motions agree to $<1.5\sigma$, supporting its physical association with the streamer structure.

V2494 Cygni did not have any known clusters in its vicinity and further has a high uncertainty in its reported parallax. We thus performed a spatial search in multiple parallax bins over a selected angular radius around it, and looked for kinematic clustering in the VPD. We find a component corresponding to the dark nebula LDN 1003 (also known as the Braid Nebula) at $\sim 575$ pc. The stellar members form an asymmetric structure projected on the plane of the sky, and we show that the proper motion components indicate expansion along the longer axis of the cloud. V2494 Cygni is positionally consistent with LDN 1003 (apparent from visual inspection) with parallax consistent to within $\sim 1.4\sigma$, and proper motion components consistent to $\lesssim 2.2\sigma$. However, this study relies entirely on Gaia-based measurements of proper motions of the FU Ori targets. Targeted radial velocity surveys in these star-forming regions are necessary to build a better 3D picture of the kinematics and ascertain associations.

Finally, by constructing color-magnitude diagrams for the cluster or stellar group associated with each FU Ori star, and comparing it to a main-sequence open cluster, we find that three of the four clusters have a significant population of pre-main-sequence stars, strengthening the possibility of the young FU Ori target being physically associated with the group in each case. 

\begin{acknowledgements}
We thank the referee for comments that helped improve our presentation.
This work has made use of the SIMBAD, Vizier, ADS, IRSA, and SkyView data services.
\end{acknowledgements}

\bibliography{cygnus}{}
\bibliographystyle{aasjournal}

\end{document}